\documentclass[a4paper]{JHEP3}
\usepackage[centertags]{amsmath}
\usepackage{amssymb,amstext}
\usepackage{graphicx}
\usepackage{epsfig}
\usepackage[stable]{footmisc}
\usepackage{appendix}

\newcommand{\half}{\frac{1}{2}}
\newcommand{\Tr}{\text{Tr}}

\def\ie{\begin{equation}\begin{aligned}}
\def\fe{\end{aligned}\end{equation}}

\title{Supersymmetric States in Large $N$ Chern-Simons-Matter Theories}

\author{Shiraz Minwalla$^a$, Prithvi Narayan$^a$, Tarun Sharma$^a$, V. Umesh$^a$
and Xi Yin$^b$\\
$^a$Dept. of Theoretical Physics, Tata Institute of Fundamental Research, Homi Bhabha Rd,
Mumbai 400005, India. \\
$^b$Center for the Fundamental Laws of Nature, Jefferson Physical Laboratory, 
Harvard University,Cambridge, MA 02138 USA.\\
Email:\ \ {\bf  minwalla@theory.tifr.res.in, prithvi@theory.tifr.res.in, 
tarun@theory.tifr.res.in, umesh@theory.tifr.res.in, xiyin@fas.harvard.edu}}

\abstract{In this paper we study the spectrum of BPS operators/states in ${\cal N}=2$ superconformal $U(N)$ Chern-Simons-matter theories with adjoint chiral matter fields, with and without superpotential.
The superconformal indices and conjectures on the full supersymmetric spectrum of the theories in the large $N$ limit with up to two adjoint matter fields are presented. Our results suggest that some of these theories may have supergravity duals at strong coupling, while some others may be dual to higher spin theories of gravity at strong coupling. For the ${\cal N}=2$ theory with no superpotential, we study the renormalization of $R$-charge at finite 't Hooft coupling using ``${\cal Z}$-minimization". A particularly intriguing result is found in the case of one adjoint matter.
}

\begin{document}

\section{Introduction}

The coupling constant of a four dimensional gauge theory coupled to matter 
generically runs under the renormalization group. While it is sometimes 
possible to choose the matter content and couplings of the 
theory so that the gauge $\beta$ function vanishes, such choices 
are very special. In three dimensions, on the other hand, gauge fields 
are naturally self coupled by a Chern-Simons type action. As the coefficient
of the Chern-Simons term in the action is forced by gauge invariance to be 
integrally quantized,  the low energy gauge coupling (inverse of coefficient 
of the Chern-Simons term) cannot be continuously renormalized and so does 
not run under the renormalization group. All these statements are for 
every choice of matter content and couplings of the theory. 
As a consequence CFTs are much easier to construct starting with Chern-Simons coupled gauge fields in $d=3$ than with Yang Mills coupled gauge fields
in $d=4$ \cite{Ivanov:1991fn, Avdeev:1991za, Avdeev:1992jt, Schwarz:2004yj, Gaiotto:2007qi}. 

Precisely because the coefficient, $k$, of the Chern Simons term is an integer, 
the Chern-Simons coupling cannot be varied continuously. The set of 
Chern-Simons CFTs obtained, by varying a given Lagrangian over the allowed
values of $k$, yields a sequence rather than a fixed line of CFTs. Consider, 
however an $SU(N)$ Chern Simons theory at level $k$. Such a theory admits 
a natural 't Hooft limit in which we take  $N \to \infty$, $k \to \infty$ 
with $\lambda =\frac{N}{k}$ held fixed. As explained by 't Hooft, $\lambda$ 
is the true loop counting parameter or coupling constant in this limit.
Several physical quantities - like the spectrum of operators with 
finite scaling dimension-  are smooth functions of 
$\lambda$. Now a unit change in $k$ changes $\lambda$ by 
$-\frac{\lambda^2}{N}$, a quantity that is infinitesimal in the large $N$ 
limit. As a consequence, even though $k$ and $N$ are both integers, $\lambda$
is an effectively continuous parameter in the large $N$ limit. Effectively, 
the discretum of Chern-Simon-matter CFTs at finite $N$ merges into an effective 
fixed line of Chern-Simon-matter theories at large $N$, parameterized by the 
effectively continuous variable $\lambda$.

Lines of fixed points of large $N$ CFTs map to families of 
theories of quantum gravity, under the AdS/CFT correspondence \cite{Maldacena:1997re}. 
CFTs at weak or finite 't Hooft coupling $\lambda$ are generically expected to map to relatively complicated 
higher spin theories of gravity \cite{Mikhailov:2002bp, Sezgin:2002rt, Klebanov:2002ja} or string theories on AdS spaces of string scale radii. In many examples of explicit realization of AdS/CFT, the bulk description simplifies 
in some manner at strong $\lambda$. It is then natural to ask whether the large class of fixed lines of 
Chern-Simons-matter theories admit simple dual descriptions at large $\lambda$ \cite{Gaiotto:2007qi}.
The first explicit realization of the gravity dual of a large $N$ Chern-Simons-matter theory, as a critical string theory, was achieved by ABJM \cite{abjm}.
At infinitely strong coupling the ABJM Chern-Simons-matter theory develops a supergravity dual 
description, which is a considerable simplification over the highly curved 
stringy dual description at finite coupling. A direct field theoretic 
hint for the nature of the dual of ABJM theory \cite{abjm} at strong coupling 
comes from the observation that the set of single trace supersymmetric 
states in ABJM theory have spins $\leq 2$ (and in fact match the spectrum 
of supergravitons of IIA theory on $AdS_4 \times CP^3$). 

While many examples of gravity duals of supersymmetric Chern-Simons-matter theories have been proposed following the work of ABJM (see for instance \cite{Benna:2008zy, Jafferis:2008qz, Klebanov:2008vq, Martelli:2009ga, Benini:2009qs, Jafferis:2009th}), essentially all such proposals involve quiver type matter content in the field theory. The gravity duals of seemingly simpler Chern-Simons-matter fixed points, both with and without 
supersymmetry, remain unknown (and may well be most interesting in the non 
supersymmetric context). On the other hand, it is of significant interest to find the CFT duals of gravity theories in $AdS_4$ with as few four-dimensional bulk fields (apart from gravity itself) as possible, and one may hope that the Chern-Simons theories with simple matter content are good candidates.

In order to maintain a degree of technical control,
however, in this paper we study only supersymmetric 
theories with at least four supercharges. We will consider large $N$ ${\cal N}=2$
and ${\cal N}=3$ Chern Simons theories with a single $U(N)$ gauge 
group and $g$ adjoint chiral multiplets (for all integer $g$). Such theories have been studied perturbatively in \cite{Gaiotto:2007qi, Chang:2010sg} (see also \cite{Niarchos:2009aa}).
We will study theories both with and without superpotentials.
We address and largely answer the following 
question: what is the spectrum of supersymmetric operators as a function 
of the 't Hooft coupling $\lambda$? In the rest of this introduction we 
elaborate on our motivation for asking this question. In the 
next section we briefly summarize our principal results.

In this paper we will 
compute (or present conjectures for) the supersymmetric spectrum of a large 
class of large $N$ Chern-Simons-matter CFTs. As we will now 
describe in some detail, the results we find for the supersymmetric spectrum 
of several fixed lines is quite simple, and has several intriguing features.
As we describe in more detail below, an important difference between some 
theories with ${\cal N}=2$ supersymmetry and theories with ${\cal N}=3$ 
and higher supersymmetry is that the $R$-charge of the chiral multiplets 
of the  ${\cal N}=2$ theories (and hence the charges and dimensions of 
supersymmetric operators) may sometimes be continuously renormalized 
as a function of $\lambda$ \cite{Gaiotto:2007qi} (and sometimes not \cite{Chang:2010sg}). Luckily, Jafferis \cite{jafferis} 
has presented a proposal 
that effectively allows the computation of the $R$-charge as a function 
of $\lambda$ in many of these theories. In the rest of this paper we 
assume the correctness of Jafferis' proposal; we proceed to use a 
combination of analytic and numerical techniques to present a complete 
qualitative picture of this $R$-charge as a function of $\lambda$ and 
the number of chiral multiplets $g$.

While we hope that our results will eventually inspire conjectures for 
relatively simple large $\lambda$ descriptions of some of the theories 
we study, in no case that we have studied have our results proven familiar 
enough to already suggest a concrete conjecture for the dual description of 
large $\lambda$ dynamics. Of the supersymmetric spectra we encounter 
in this paper, the one that appears most familiar is the 
spectrum of the ${\cal N}=3$ theory with two chiral multiplets. As 
we will describe in more detail below, this spectrum includes only states 
of spins $\leq 2$, and so might plausibly agree with the spectrum of 
some supergravity compactification: however a detailed study of 
the spectrum as a function of global charges reveals some unexpected
features that has prevented us (as yet) from finding a supergravity 
compactification with precisely this spectrum. The spectrum of supersymmetric
states in ${\cal N}=2$ deformations of the ${\cal N}=3$ 
theory also has similar features. We hope to return to an investigation
into the possible meanings of these spectra in the future.

In the next section we will present a more detailed description 
of the theories we have studied  and our results. 

\bigskip

Note added in proof: Upon completion of this work, we received \cite{Niarchos:2011sn} which overlaps with
section 4 of this paper.

\section{Summary of results}

\subsection{Theories with a vanishing superpotential}

\subsubsection{$R$-charge as a function of $\lambda$}

The first class 
of theories studied in this paper consists of ${\cal N}=2$  $U(N)$ 
Chern Simons theories at level $k$, coupled to $g$ adjoint 
chiral multiplets with vanishing superpotential. This class of theories was 
studied, and demonstrated to be superconformal (for all $N$, $k$ and $g$) in 
\cite{Gaiotto:2007qi}. In the free limit the $R$-charge of the chiral fields
in this theory equals $\frac{1}{2}$. However, it was demonstrated in 
\cite{Gaiotto:2007qi} that this $R$-charge is renormalized as a function of 
$\lambda$; indeed at first nontrivial order in perturbation theory 
\cite{Gaiotto:2007qi} demonstrated that the $R$-charge of a chiral field 
is given by 
\begin{equation}
h(\lambda)=\frac{1}{2} -(g+1) \lambda^2
\end{equation} 
where $\lambda=\frac{N}{k}$. As the $R$-charge of a supersymmetric 
operator plays a key role in determining its scaling dimension 
(via the BPS formula), the exact characterization of the spectrum of 
supersymmetric states in this theory at large $\lambda$ clearly requires
control over the function $h(\lambda)$ at large $\lambda$. Such control
cannot be achieved by perturbative techniques, but is relatively 
easily obtained by an application of the extremely powerful recent results of Jafferis 
\cite{jafferis} to this problem. 
In \cite{jafferis} Jafferis used localization methods to 
derive a formula (in terms of an integral over $r$ variables, where $r$ is the 
rank of the gauge group) for the partition function on $S^3$ of the CFT in question, as a function of $h_i$ the $R$-charges of all the chiral fields in the 
theory. He then demonstrated that the modulus of this partition function 
is extremized by the values of $h_i$ at the conformal fixed point, assuming the absence of accidental global symmetries. In the large $N$ limit 
of interest to this paper, Jafferis' matrix integral is dominated by a saddle 
point. Using a combination of analytic and numerical techniques, it is not 
difficult to solve the relevant saddle point equations, extremize the 
action with respect to $h$, and thereby evaluate $h(\lambda)$. It turns out that 
$h(\lambda)$ is a monotonically decreasing function for all $g$. In fact
at large $g$ (but all values of $\lambda$) 
\begin{equation} \label{remu}
h(\lambda)=\frac{1}{2}-
\frac{g\pi^{2}}{ \left( {g\pi^{2}\over 2} \right)^{2} + \left( {\pi\over \lambda} \right)^{2} } +{\cal O}\left(\frac{1}{g^2} \right).
\end{equation} 
Note that $h(\lambda)$ tends to a constant value at $\lambda=\infty$. At large 
$g$ this constant value is barely below the free value $\half$; it is 
given by 
$$\frac{1}{2} -\frac{4}{\pi^2 g} +{\cal O}\left(\frac{1}{g^2}\right).$$
Although there is no a priori reason for this formula to apply at  $g$ of order unity, 
we have used numerical techniques to find that it appears to work at the 10-15 percent 
level even down to $g=2$. More specifically, at $g=3$, our numerics indicates 
$h(\infty)\approx 0.35$, and at $g=2$, $h(\infty) \approx 0.27$ (see Figs. \ref{hlmdafigg3} and 
\ref{hlmdafigg2} below); these do not compare badly with $0.365$  and $0.3$ as predicted by \eqref{remu}.
 
 Most interestingly, however, at $g=1$, 
$h(\infty)=0$ (see Fig. \ref{io}  below); i.e. the $R$-charge of the 
chiral multiplet decreases without 
bound in this special (extreme) case (this was anticipated in \cite{Niarchos:2009aa}) , raising several interesting questions 
that we will come back to later.\footnote{We emphasize that these 
results have been obtained using large $N$ saddle point techniques on 
Jafferis's integral. The fact that our results agree with perturbation theory 
at small $\lambda$ give us confidence that we have identified the correct saddle point
at small $\lambda$. It is possible, however, that the integral undergoes 
a large $N$ phase transition to another saddle point at a finite value of 
$\lambda$. In this case the field theory would undergo a phase transition
at that $\lambda$, and our results above apply only below the phase 
transition. Though this seems unlikely given the analysis of \cite{Niarchos:2009aa}, it 
does not seem completely ruled out that the theory develops pathology, and ceases 
to be completely well behaved at a critical $\lambda$. 
If this is the case, the results of this paper apply only below this
critical coupling $\lambda$.}

\subsubsection{Spectrum of single trace operators}

Given the function $h(\lambda)$, it is not difficult to evaluate the
superconformal index \cite{Bhattacharya:2008bja} of these theories 
as a function of $h(\lambda)$. As was already noted in \cite{Bhattacharya:2008bja}, 
this index demonstrates that the spectrum of supersymmetric single trace operators 
grows exponentially with energy for $g \geq 3$. In the absence of a 
superpotential one can actually compute a slightly more refined 
Witten index (adding in a chemical potential that couples to the global 
symmetry generators). Below we demonstrate that this refined index 
implies an exponentially growing density of states for the supersymmetric
spectrum (in the theory without a superpotential) even for $g=2$. 
This immediately suggests that the simplest possible dual description for 
all theories with $g \geq 3$ (and the theory without a superpotential at 
$g=2$) is a string theory, with an exponential growth in {\it supersymmetric} string 
oscillator states. 

However, the index indicates a sub exponential growth of supersymmetric states 
for all theories with $g=1$ and theories with a nontrivial superpotential 
when $g=2$. This leaves open the possibility of a simpler dual (one with a 
field theory's worth of degrees of freedom) in these cases. In this 
subsection we 
focus on theories without a superpotential, and so consider only the case 
$g=1$. Making the assumption that the spectrum 
of supersymmetric states in this theory is isomorphic to the cohomology 
of the {\it classical} action of the susy operator, we 
have computed the full spectrum of single trace 
supersymmetric operators in this theory. Our explicit results are listed
in Table (\ref{tab:1chiralAdjWzero}). While the states listed in this table
do grow in number with energy in a roughly Kaluza-Klein fashion, 
notice that the primaries listed in Table (\ref{tab:1chiralAdjWzero})
include states of arbitrarily high spins, ruling out a possible dual 
supergravity dual description.

The supersymmetric states in Table (\ref{tab:1chiralAdjWzero}) of course 
include the states in the chiral ring ${\rm Tr}\,\phi^n$ for all $n$ where 
$\phi$ is the scalar component of the chiral field. 
The scaling dimension of these chiral ring operators is given by 
$n h(\lambda)$ ($h(\lambda)$ was defined in the previous subsection). 
Unitarity, however, requires that every scalar operator in any 3 d CFT has 
scaling dimension $\geq \half$, and that an operator with dimension $\half$ 
is necessarily free (i.e. decoupled from the rest of the theory). 
Recall that $h(\lambda)$ 
decreases monotonically to zero as $\lambda$ is increased. 
Let $\lambda_n^f$ denote the unique solution to the equation 
$$h(\lambda_n^f)= \frac{1}{2n}.$$ 
For $\lambda> \lambda_n^f$, the scaling dimension of $\Tr \,\phi^n$ descends 
below its unitarity bound $\frac{1}{2}$. 
(For later use we will also find it useful to define 
$$h(\lambda_n^m)=\frac{2}{n}.$$
$\lambda_n^m$ is the value of the coupling at which a superpotential deformation by $\Tr\, \Phi^n$ becomes 
marginal. Note that $\lambda^m_n < \lambda_n^f$.)

Assuming Jafferis' proposal, it follows from unitarity that our theory must 
either cease to exist \footnote{However Niarchos's study of this theory 
in \cite{Niarchos:2009aa} makes this possibility unlikely} or must undergo a phase transition at a critical 
value, $\lambda=\lambda_c \leq \lambda_2^f$. 
While many possibilities 
are logically open, one attractive scenario (which is close to the 
scenario suggested in \cite{Niarchos:2009aa}) is the following. 
As $\lambda$ is increased past $\lambda_2^f$ then 
$\Tr \,\phi^2$ becomes free and decouples from the theory. As $\lambda$ is 
further
increased past $\lambda_3^f$ then $\Tr \,\phi^3$ also becomes free and decouples. 
This process continues ad infinitum, leading to an infinite number of phase 
transitions. 

The picture oulined above for the $g=1$ theory, namely that each of the  $\Tr \phi^{n+1}$
decouple at successively larger values of $n$ can be subject to a consistency 
check. It was demonstrated in \cite{Niarchos:2009aa}, using brane constructions, that the deformation of the zero superpotential 
system by the operator $\Tr \phi^{n+1}$ breaks supersymmetry precisely at $\lambda=n$ (and in particular 
susy is not broken at smaller values of $\lambda$).\footnote{We are grateful to Ofer Aharony and Zohar Komargodski for bringing this to our attention.} However the deformation of a superpotential by a free field always breaks 
supersymmetry. Consistency with the scenario outlined above therefore requires that $\lambda_n^f>n-1$. In other
words 
$$ h(\lambda) \geq \frac{1}{2(\lambda+1)}.$$
Our data (see Fig. \ref{io}) seems consistent with this bound, and moreover suggests (and we 
conjecture that) $h(\lambda)$ asymptotes to $\frac{1}{2(\lambda+1)}$ from above. It would 
be very interesting to establish this conjecture by analytic methods, but we leave that 
for future work. 

If this picture outlined in the last two paragraphs is correct, then, in the limit 
$\lambda \to \infty$ we 
have an effective continuum of chiral primaries, with scaling dimension 
$\geq \half$ (all primaries with lower dimension have decoupled). 
The higher spin fields listed in 
Table (\ref{tab:1chiralAdjWzero})  also reduce to a 
continuum at large $\lambda$. All this suggests that the large $\lambda$ 
behavior of this theory is intriguing, and possibly singular.\footnote{Note, on the other hand, that when $g \geq 2$, $h(\lambda) > \frac{1}{4}$ 
for all $\lambda$. The large $\lambda$ behaviour for these theories
shows no hint of any singular behaviour.}

\subsection{Theories with a superpotential}

Let us now turn to the study of superconformal theories with a single $U(N)$ 
gauge group, only adjoint fields (as above) but with appropriate
superpotentials. 
For theories that reduce to free 
systems as $\lambda \to 0$,  the superconformal index 
is independent of the details of the superpotential, other than the fact that the 
index cannot now be weighted with respect to a chemical potential for any global 
symmetry under which the superpotential is charged, and depends only on the $R$-charge of matter fields which may be renormalized.\footnote{At fixed points that are described by a generic quartic superpotential, or in fact, points that lie on the ``conformal manifold", the $R$-charge is not renormalized \cite{Chang:2010sg}. } So in particular, the index
demonstrates 
the presence of an exponentially growing spectrum of supersymmetric states
for $g\geq 3$, exactly as above. For this reason we focus our study on 
$g \leq 2$. Let us first start with $g=1$. 

\subsubsection{ $\Tr\, \Phi^4$ at $g=1$}

The superpotential deformation 
$\Tr \,\Phi^4$ is marginal at $\lambda=0$, but is relevant at finite $\lambda$ 
(this follows because $h(\lambda)<\frac{1}{2}$ for all finite $\lambda$). 
It has been argued in \cite{Gaiotto:2007qi} that the beta function for this 
superpotential term vanishes when its coefficient is of order $\lambda$ 
(at small $\lambda$) leading to a weakly coupled CFT with a $\Tr\, \Phi^4$ 
superpotential turned on. The presence of the superpotential forces 
the $R$-charge of the field $\phi$ to be fixed at $h=\half$ at all values 
of $\lambda$ in this new 
fixed line. While the superconformal index of this 
theory is blind to the presence of the superpotential, the spectrum of 
single trace supersymmetric operators is not. We present a conjecture for 
this spectrum in Table (\ref{tab:table2adjsupphi4}) below 
(this conjecture is based on the same 
assumption described above, namely that the supersymmetric spectrum is 
accurately captured by the classical supercharge cohomology at all $\lambda$.) 
As in the case of theories without a superpotential, our conjectured supersymmetric 
spectrum grows with energy in a manner expected 
of Kaluza-Klein compactification, but continues to include states 
of arbitrarily high spin. 

\subsubsection{ $\Tr\, \Phi^3$ at $g=1$} 

$\Tr\, \Phi^3$ is another relevant deformation of the weakly coupled theory 
at $g=1$. This deformation leads to a $\phi^4$ term in the scalar potential of the 
theory. At least at weak coupling we would expect the RG flow seeded by this 
operator to end at the supersymmetric large $N$ analogue of a Wilson Fisher 
fixed point. At this fixed point the scaling dimension (and superconformal $R$-charge)
of $\phi$ is fixed to 
be $\frac{2}{3}$. This fixed point can then be continued to large $\lambda$ by 
varying the gauge coupling. For this reason the superconformal index of 
this theory cannot be calculated from a free path integral; however it 
can be computed using the techniques of localization using the results of 
\cite{Imamura:2011su} (following the original work of \cite{Pestun:2007rz, Kim:2009wb}). We have performed 
this computation in section \ref{localization} below. While the computation of this index
is exact at all $N$ and $k$, the final result simplifies 
dramatically in the large $N$ 't Hooft limit of interest to this paper, 
and in fact reduces to the index of the free theory with appropriate charge 
renormalizations (in order to account for the fact that the $R$-charge of 
$\phi$ is $\frac{2}{3}$ rather than $\frac{1}{2}$). We have also computed 
the full spectrum of single trace supersymmetric primaries in this theory 
(using assumptions similar to those described above). Our rather simple 
final results are listed in Table (\ref{tab:table2adjsupphi3}) .  

\subsubsection{ $\Tr\, \Phi^n$ at $g=1$} 

At small $\lambda$, operators of the form $\Tr\, \Phi^n$ are irrelevant for 
$n \geq 5$. However if our description of the $g=1$ theory without a 
superpotential in the previous subsection is indeed correct then 
for each $n$ the operator $\Tr\, \Phi^n$ is in fact relevant for 
$\lambda > \lambda^m_n$. 
It seems likely that the RG flows 
seeded by these operators end in new lines of CFTs in  which the scaling 
dimension of $\phi$ is fixed at $\frac{2}{n}$. We expect the index of these 
theories (in the 't Hooft limit) to once again be given by the formula for non-interacting theory but with renormalized charges for all fields. Using 
methods similar to those described above, it should be easy to compute 
the full spectrum of single trace superconformal primaries in this 
theory. We leave this computation to future work.

\subsubsection{${\cal N}=3$ theory at $g=2$}

Let us now turn to $g=2$ theories with a superpotential. First 
consider superpotentials  of the form $\Tr\, [\Phi_1, \Phi_2]^2$. Like any quartic 
superpotential in this theory, this superpotential is marginal at 
$\lambda=0$, but is relevant at finite $\lambda$ (regarded as a 
deformation about the theory with no superpotential). It was argued in 
\cite{Gaiotto:2007qi} that the RG flow seeded by this operator ends 
with the coefficient  of this superpotential stabilized at that finite 
value (of order $\lambda$) that enhances the supersymmetry of the theory to 
${\cal N}=3$. This ${\cal N}=3$ theory enjoys invariance under an 
enhanced  $SU(2)$ $R$ symmetry group, and also enjoys invariance under 
an $SU(2)$ flavour symmetry group. We have computed the spectrum of 
supersymmetric states in this theory; our results are presented in Table (\ref{tab:prmrycntntshrt2}). 
Interestingly, it turns out that the spins of supersymmetric states 
in this theory grow roughly in the manner one would expect of a Kaluza-Klein compactification of a supergravity theory on $AdS_4 \times S^3$. 
In particular the spins of supersymmetric states in this theory never exceed two. 
We have not, however, managed to identify a specific supergravity 
compactification that could give rise to this spectrum (there seems to be 
a qualitative difficulty in making such an identification, as we describe 
in more detail in section \ref{n3theories}). 

\subsubsection{Superconformal ${\cal N}=2$ deformations of the 
${\cal N}=3$ theory}

There exists a manifold of exactly marginal ${\cal N}=2$ deformations \cite{Chang:2010sg} of the ${\cal N}=3$ 
theory described in the previous paragraph. This manifold can be 
characterized rather precisely in the neighbourhood of the ${\cal N}=3$ 
fixed point using the recent results of \cite{Green:2010da}. We have computed the spectrum 
of supersymmetric states in these deformed theories. Our results are presented
in Tables (\ref{tab:N2U(1)spectrum}), (\ref{tab:N2nosymmetry}) and (\ref{tab:N2spctrm0eig}). Qualitatively, our results for these deformed theories are 
similar to those described in the previous paragraph. The spins of all 
supersymmetric states are less than or equal to two, potentially describing the 
supersymmetric spectrum of a Kaluza-Klein compactification. 

\subsubsection{Other superpotentials at $g=2$}

Finally, as $\lambda$ is increased superpotentials
of up to $7^{th}$ order in the chiral fields eventually turn relevant (Here we use $h(\infty)\simeq0.27$
at explained above).
Just as 
in the case of $g=1$, this suggests the existence of new fixed lines 
with superpotentials of up to $7^{th}$ order in chiral fields turned on.
In addition, at every value of $\lambda$ there probably exist 
superconformal field theories with cubic superpotentials.
We leave the investigation of these theories and their supersymmetric 
spectra to future work.

\section{${\cal N}=2,3$ superconformal algebras and their 
unitary representations\,\footnote{This section was worked out in 
collaboration with Jyotirmoy Bhattacharya.}}

In order to lay out the background (and notation)
for our analysis of supersymmetric states, in 
this section we present a brief review of the structure of the 
${\cal N}=2, 3$ superconformal algebras, their unitary representations 
and their Witten indices. We also explicitly decompose every 
representation of the relevant superconformal algebras into irreducible 
representations of the conformal algebra. The paper \cite{Bhattacharya:2008bja} is 
useful background material for this section. The reader who is familiar with 
the superconformal algebras and their representation theory may wish to skip 
to the next section. 

\subsection{The superconformal algebras and their Witten indices}

In this section we briefly review the representation theory of the 
${\cal N}=2$ and ${\cal N}=3$ superconformal algebras. The bosonic 
subalgebras of these Lie super algebras is given by 
$SO(3,2) \times SO(2)$ (for ${\cal N}=2$) and $SO(3,2)\times SO(3)$ (for 
${\cal N}=3$). Primary states of these algebras are labeled by 
$(\Delta, j, h)$ where $\Delta$ is the scaling dimension of the primaries, 
$j$ is its spin and $h$ is its $R$-charge (or $R$ symmetry highest weight).
$h$ can be any positive or negative real number for ${\cal N}=2$, but is a 
positive half integer for ${\cal N}=3$. 

The labels of unitary representations of the superconformal algebra obey 
the inequalities forced by unitarity. When $j \neq 0$
the condition 
$$\Delta \geq |h|+j+1$$ 
is necessary and sufficient for unitarity. When $j=0$ unitary representations 
occur when 
$$\Delta=|h|, ~~~(|h| \geq \half)$$
or when 
$$\Delta \geq |h|+1$$ 
The isolated representations, and those that saturate this bound, are all 
short. 

The Witten index $ {\rm Tr} (-1)^F x^{\Delta+j}$ vanishes on all long representations
of the supersymmetry algebra but is nonzero on short representations. This 
index captures information about the state content of a conformal field 
theory. The only way that the Witten index of a CFT can change 
under continuous variations of parameters (like the parameter $\lambda$ in 
our theory), is for the $R$-charge to be renormalized as a function of 
that parameter. Note that the $R$-charge is fixed to be a half integer 
at ${\cal N}=3$, but can in principle be continuously renormalized 
at ${\cal N}=2$. 

In the case of ${\cal N}=2$ theories, we have glossed over a detail. 
At the purely algebraic level, in this case, there are really 
two independent Witten indices; ${\cal I}_+$ and 
${\cal I}_-$. These are defined as 
$$ {\cal I}_+={\rm Tr} (-1)^F x^{H+J} e^{-\beta (H -J-R)}$$
and
$$ {\cal I}_-={\rm Tr} (-1)^F x^{H+J} e^{-\beta (H -J+R)}$$
respectively. We used the notation $H$ for the dilatation operator, $J$ the third component of the spin,
and $R$ the $R$ symmetry generator. The indices above are distinct, even though they 
both evaluate to quantities independent of $\beta$. The first index receives
contributions only from states with $\Delta=j+h$ ; all such states are 
annihilated by and lie in the cohomology of the supercharge with charges 
$(\half, -\half, 1)$. The second index
receives contributions only from states with $\Delta=j-h$; all such 
states are annihilated by and lie in the cohomology of the supercharge with charges $(\half, \half, 1)$. The existence of two algebraically independent
Witten Indices is less useful than it might, at first seem, in the study 
of quantum field theories, as the two indices are closely linked by the 
requirement of CPT invariance.

\subsection{State content of all 
unitary representations of the ${\cal N}=2$ superconformal algebra}

In the rest of this section we will 
list the conformal representation content of all unitary representations 
of the superconformal algebra, and use our listing to compute the 
index of all short representations of this algebra. 

To start with, we present a group theoretic 
listing of the state content of an antisymmetrized product of supersymmetries. 
This is given in Table (\ref{tab:susytab}).
\begin{table}
\centering
\begin{tabular}{|c|c|}
\hline
Operator & States \\
\hline
$I$ & $(0,0,0)$ \\
\hline
$Q$ & $(\half, \half, 1)$, $(\half, \half, -1)$ \\
\hline
$Q^{2}$ & $(1,0, 2)$, $(1,0,-2)$, $(1, 0, 0)$, $(1,1,0)$ \\
\hline
$Q^{3}$ &  $(\frac{3}{2}, \half, 1)$, $(\frac{3}{2}, \half, -1)$\\
\hline
$Q^{4}$ & $(2,0,0)$\\
\hline
\end{tabular}
\caption{A decomposition of the antisymmetrized 
products of supersymmetries into irreducible representations of the maximal
compact bosonic subgroup of the relevant superalgebras. Representations 
are labeled by $(\Delta,j,h)$ where $\Delta$ is the scaling dimension, 
$j$ the angular momentum (a positive half integer) and $h$ the $R$-charge
of the representation. The same labeling convention is used in all 
tables in this section. }\label{tab:susytab}
\end{table}

The conformal primary content of a long representation of the superconformal 
algebra is given by the Clebsh Gordon product of the state content of the 
product of susy generators above with that of the primary. 
We list the conformal primary content of an arbitrary long representation in 
Table (\ref{tab:prmrycontent}).
\begin{table}
\centering
\begin{tabular}{|c|c|c|}
\hline
Primary & Conformal content & Index \\
\hline
$(\Delta, j, h)$ & $(\Delta, j, h)$, & $0$ \\
                 & $(\Delta + \half, j + \half, h+1)$, $(\Delta + \half, j - \half, h+1)$, & \\
                 & $(\Delta + \half, j + \half, h-1)$, $(\Delta + \half, j - \half, h-1)$, & \\
                 & $(\Delta + 1, j, h+2)$, $(\Delta + 1, j, h-2)$, $(\Delta + 1, j+1, h)$, & \\
                 & $2(\Delta + 1, j, h)$, $(\Delta + 1, j-1, h)$, & \\
                 & $(\Delta + \frac{3}{2}, j + \half, h+1)$, $(\Delta + \frac{3}{2}, j - \half, h+1)$, & \\
                 & $(\Delta + \frac{3}{2}, j + \half, h-1)$, $(\Delta + \frac{3}{2}, j - \half, h-1)$, & \\
                 & $(\Delta + 2, j, h)$ & \\
\hline
$(\Delta, 0, h)$ & $(\Delta, 0, h)$, & $0$ \\
               & $(\Delta + \half, \half, h+1)$, $(\Delta + \half, \half, h-1)$, & \\
               & $(\Delta + 1, 0, h+2)$, $(\Delta + 1, 0, h-2)$, $(\Delta + 1, 1, h)$, & \\
               & $2(\Delta + 1, 0, h)$, $(\Delta + \frac{3}{2}, \half, h+1)$, & \\
               & $(\Delta + \frac{3}{2}, \half, h-1)$, $(\Delta + 2, 0, h)$ & \\
\hline
\end{tabular}
\caption{Conformal primary content of long representations. 
 Representations 
are labeled by $(\Delta,j,h)$ where $\Delta$ is the scaling dimension, 
$j$ the angular momentum (a positive half integer) and $h$ the $R$-charge
of the representation.}\label{tab:prmrycontent}
\end{table}

Note that a long representation of the susy algebra decomposes into $16$ 
long representations of the conformal algebra when $j\neq 0$; when $j=0$
we must delete the representations with negative values for the $SO(3)$ highest
weight ($j-\half$ and $j-1$) from the generic $j$ result 
leaving us with a total of $11$ conformal representations. The Witten index of 
all long representations automatically 
vanishes. 

Let us now turn to the short representations. To start with consider 
representations with $h \neq 0$,  $j \neq 0$ and $\Delta=|h|+j+1$. 
These representations are short because they include a family of 
null states. These null states themselves transform in a short 
representation of the superconformal algebra, with quantum numbers 
$(|h|+j+\frac{3}{2}, j-\half, h+1)$ (when $h$ is positive) and 
$(|h|+j+\frac{3}{2}, j-\half, h-1)$ (when $h$ is negative). It is not 
too difficult to verify that the conformal primary content of such a 
short representation (represented by $\chi_S(j, h)$) and the Witten index 
of these representations is as given in 
Table (\ref{tab:prmrycntntshrt1}) 
(we list the result for positive $h$; the result for negative $h$ follows from symmetry).\footnote{The Witten index 
of these representations may be evaluated as follows. When $h>0$  
there are no states with $\Delta=j-h$ and so ${\cal I}_-=0$. 
States with $\Delta=j+h$ occur only in the representation 
$(j+h +\frac{3}{2}, j+\half, h+1)$ and we find 
${\cal I}_+=(-1)^{2j+1}\frac{x^{2j+h+2}}{1-x^2}$ where we have used 
the spin statistics theorem to assert that the fermion number of a primary 
of angular momentum $j$ is $(-1)^{2j}$. Similarly when $h<0$ we have 
 ${\cal I}_+=0$ and ${\cal I}_-=(-1)^{2j+1}\frac{x^{2j-h+2}}{1-x^2}$.}
Note that all $8$ conformal primaries that occur in this 
decomposition are long (recall we have assumed $h \neq 0$). 

\begin{table}
\centering
\begin{tabular}{|c|c|c|}
\hline
Primary & Conformal content & Indices ($h>0$; $h<0$)\\
\hline
$(j + h + 1, j, h)$          & $(j+h+1, j, h)$, &  ${\cal I}_-=0$, \\
($j \neq 0$, $h \neq 0$)   & $(j + h + \frac{3}{2}, j + \half, h + 1)$, $(j + h + \frac{3}{2}, j + \half, h-1)$, &
                                                   ${\cal I}_+=(-1)^{2j+1}\frac{x^{2j+h+2}}{1-x^2}$ ;\\
               & $(j + h + \frac{3}{2}, j + \half, h-1)$, $(j + h + 2, j, h-2)$, & ${\cal I}_+=0$, \\
               & $(j + h + 2, j + 1, h)$, $(j + h + 2, j, h)$, & ${\cal I}_-=(-1)^{2j+1}\frac{x^{2j-h+2}}{1-x^2}$ \\
                       & $(j + h + \frac{5}{2}, j + \half, h -1)$ & \\
\hline
\end{tabular}
\caption{Conformal primary content and index of generic short representation. 
 Representations 
are labeled by $(\Delta,j,h)$ where $\Delta$ is the scaling dimension, 
$j$ the angular momentum (a positive half integer) and $h$ the $R$-charge
of the representation.}\label{tab:prmrycntntshrt1}
\end{table}

It is not difficult to verify that 
$\chi_L(h+j+1, j, h)=\chi_S(j, h)+\chi_S(j-\half, h+1)$. 
This expresses the fact that the state content of a long 
representation just above unitarity is equal to the sum of the state 
content of the short representation it descends to plus the state 
content of the short representation of null states.

Let us now turn to the special case of short representation 
$(j+1, j, 0)$. The null states of this representation consist of a 
sum of two irreducible representations with quantum numbers 
$(j+\frac{3}{2}, j-\half, 1)$ and $(j-\frac{3}{2}, j-\half, -1)$. 
It is not too difficult to convince oneself that the primary content  
of such a short representation is as given in Table (\ref{tab:prmrycntntshrt2}).
Note that all $4$ conformal representations that appear in this split are 
short.\footnote{The Witten index of these representations may be evaluated 
as follows. States with 
$\Delta=j+h$ occur only in the representation  $(j+\frac{3}{2}, j+\half, 1)$
while states with $\Delta=j-h$ occur only in the representation  
$(j+\frac{3}{2}, j+\half, -1)$. The Witten indices of this representation 
are given by ${\cal I}_+={\cal I}_- =(-1)^{2j+1}\frac{x^{2j+2}}{1-x^2}$.  }
It may be verified\footnote{In order to perform this verification, it is important to recall that 
$\chi_L(j+1, j, 0)$ is the sum of $16$ {\it long} characters of the 
conformal group, $4$ of which are at the unitarity threshold. Equivalently 
we may write this as the sum of $12 + 4 = 16$ long conformal characters 
and $4$ short conformal characters (where we have used the fact that 
a long conformal character, at its unitarity bound, decomposes into the 
sum of a short and a long character). The $4$ short characters in this 
decomposition simply yield $\chi_S(j+1, j, 0)$ above, while the $16$ long 
characters constitute 
$\chi_S(j+\frac{3}{2}, j-\half, 1)+ \chi_S(j+\frac{3}{2}, j-\half, 1)$.} that $\chi_L(j+1, j, 0)=\chi_S(j+1, j, 0)
+\chi_S(j+\frac{3}{2}, j-\half, 1)+ \chi_S(j+\frac{3}{2}, j-\half, -1)$.
\begin{table}
\centering
\begin{tabular}{|c|c|c|}
\hline
Primary & Conformal content & Index \\
\hline
$(j+1, j, 0)$ & $(j+1, j, 0)$, & $\cal{I}_{+}$= $\cal{I}_{-}$= $(-1)^{2j+1}\frac{x^{2j+2}}{1-x^2}$ \\
            & $(j + \frac{3}{2}, j + \half, 1)$, $(j + \frac{3}{2}, j + \half, -1)$, & \\
            & $(j+2, j+1, 0)$ & \\
\hline
\end{tabular}
\caption{Conformal primary content and index for $(j+1, j, 0)$ representation. 
 Representations 
are labeled by $(\Delta,j,h)$ where $\Delta$ is the scaling dimension, 
$j$ the angular momentum (a positive half integer) and $h$ the $R$-charge
of the representation.}\label{tab:prmrycntntshrt2}
\end{table}

Let us next turn to the special case of a short representation with $j=0$ 
and with quantum numbers $(h+1, 0, h)$. Such representations are often referred
to as semi short, to distinguish them from the `short' $j=0$ representations 
we will deal with next. We will deal with the case $h>0$ 
(the results for $h<0$ can then be deduced from symmetry). The primary 
for the null states of this representation has quantum numbers 
$(h+2, 0, h+2)$. Note that the null states transform in an isolated short
representation. The state content and Witten index of a semishort $j=0$ 
representation are listed in  
Table (\ref{tab:prmrycntntshrt5}). Note that $\chi_L(h+1, 0, h)=\chi_S(h+1, 0, h)+ \chi_S(h+2, 0, h+2)$; this formula
captures the split of a long representation into the 
short representation and null states.
\begin{table}
\centering
\begin{tabular}{|c|c|c|}
\hline
Primary & Conformal content & Index($h>0$; $h<0$) \\
\hline
$(h+1, 0, h)$ & $(h+1, 0, h)$, &  ${\cal I}_{-} = 0$, ${\cal I}_+=-\frac{x^{h+2}}{1-x^2}$ ; \\
     & $(h + \frac{3}{2}, \half, h+1)$, $(h+ \frac{3}{2}, \half, h-1)$, & \\
     & $(h+2, 0 ,h-2)$, $(h+2, 1, h)$, $(h+2, 0, h)$, & ${\cal I}_+=0$, ${\cal I}_-= -\frac{x^{-h+2}}{1-x^2}$\\
     & $(h + \frac{5}{2}, \half, h-1)$ & \\
\hline
\end{tabular}
\caption{Conformal content and index of representation $(h+1,0,h)$.  
Representations 
are labeled by $(\Delta,j,h)$ where $\Delta$ is the scaling dimension, 
$j$ the angular momentum (a positive half integer) and $h$ the $R$-charge
of the representation.}\label{tab:prmrycntntshrt5}
\end{table}

The short representation with primary labels $(1, 0, 0)$  is a bit 
special; its null states have primaries with quantum numbers $(2, 0, 2)$ 
and $(2,0, -2)$ (these are isolated short representations, see below). 
The state content and Witten index of this representation are given in Table (\ref{tab:prmrycntntshrt6}).
Of the $5$ conformal primaries that appear in this split,
only the representation with quantum numbers $(2,1,0)$ is short.\footnote{The conformal representation 
$(1,0,0)$ and $(\frac{3}{2}, \half, \pm 1)$ 
are not short as spin $0$ and spin $\half$ are exceptions to the general rule.}
Using the results we present below, it is possible to verify the 
character decomposition rule:\footnote{The character on the LHS is a sum of $10$ long conformal representations or 
$11 =5+3 +3$ 
short  conformal representations (the extra representation is the conformal shortening vector of 
$(2,1,0)$ and is given by $(3,0,0)$). States with 
$\Delta=j+h$ occur only in the representation  $(\frac{3}{2}, \half, 1)$
while states with $\Delta=j-h$ occur only in the representation  
$(\frac{3}{2}, \half, -1)$.}
$$\chi_L(1,0,0)=\chi_S(1,0,0)+\chi_S(2,0,2) +\chi_S(2,0,-2).$$

\begin{table}
\centering
\begin{tabular}{|c|c|c|}
\hline
Primary & Conformal content & Index \\
\hline
$(1,0,0)$ & $(1,0,0)$ & ${\cal I}_+={\cal I}_- = -\frac{x^{2}}{1-x^2}$ \\
        & $(\frac{3}{2}, \half, 1)$, $(\frac{3}{2}, \half, -1)$ & \\
        & $(2,0,0)$, $(2,1,0)$ & \\
\hline
\end{tabular}
\caption{Conformal content and index for representation $(1,0,0)$. 
 Representations 
are labeled by $(\Delta,j,h)$ where $\Delta$ is the scaling dimension, 
$j$ the angular momentum (a positive half integer) and $h$ the $R$-charge
of the representation.}\label{tab:prmrycntntshrt6}
\end{table}

Now let us turn to the isolated short representations $(h, 0, h)$ for $|h| \geq 1$.
The primaries for the null states of these representations have quantum numbers 
$(h+\half, \half, h+1)$. The null states transform in a (short) non-unitary
representation, reflecting the fact that the isolated representations 
cannot be regarded as a limit of unitary long representations but can 
be regarded as the limit of non-unitary long reps. The conformal 
primary content and Witten indices for these representations are given 
in Table (\ref{tab:prmrycntntshrt3}). Here we have written conformal content for $h$ positive; 
the result for negative $h$ is given by symmetry.
\begin{table}
\centering
\begin{tabular}{|c|c|c|}
\hline
Primary & Conformal content & Index($h>0$; $h<0$) \\
\hline
$(h, 0, h)$ & $(h, 0, h)$, & ${\cal I}_-=0$, ${\cal I}_+=\frac{x^{h}}{1-x^2}$ ; \\
         & $(h + \half, \half, h-1)$, &  ${\cal I}_+=0$, ${\cal I}_-=\frac{x^{-h}}{1-x^2}$ \\
         & $(h+1, 0, h-2)$ & \\
\hline
\end{tabular}
\caption{Conformal content and index for representation $(h,0,h)$.}\label{tab:prmrycntntshrt3}
\end{table}
Recall that $h \geq \half$ for the  representations we have just discussed. 
The lower bound of this inequality, $h=\half$, is a special case. 
The conformal decomposition and index for the 
$h=\half$ and $-\half$ cases are given in Table (\ref{tab:prmrycntntshrt4}).
\begin{table}
\centering
\begin{tabular}{|c|c|c|}
\hline
Primary & Conformal content & Index \\
\hline
$(\half, 0, \half)$ & $(\half, 0, \half)$, & ${\cal I}_{-}=0$, ${\cal I}_{+} = \frac{x^{\half}}{1-x^2}$ \\
                    & $(1, \half, -\half)$ & \\
\hline
$(\half, 0, -\half)$ & $(\half, 0, -\half)$ & ${\cal I}_{+}=0$, ${\cal I}_{-} = \frac{x^{\half}}{1-x^2}$ \\
                      & $(1, \half,  \half)$ & \\
\hline
\end{tabular}
\caption{Conformal content and index for representations $(\half, 0, \half)$ and $(\half, 0, -\half)$}\label{tab:prmrycntntshrt4}
\end{table}


\subsection{Decomposition of all unitary representations of the 
${\cal N}=3$ algebra into ${\cal N}=2$ representations}

In this subsection we record the decomposition of all ${\cal N}=3$ representations 
into ${\cal N}=2$ representations. 
Representations of the ${\cal N}=3$ algebra are labeled as $(\Delta,j,h)$, 
where $h$ is the highest weight under the Cartan of the $SO(3)$ $R$ symmetry 
(normalized to be a half integer). 
A generic ${\cal N}=3$ long representation with $j \neq 0$ breaks as follows
\begin{equation}\label{long}
(\Delta,j,h)_3=\bigoplus_{m=-h}^{h}\bigg[ (\Delta,j,m)_2 \oplus (\Delta+\frac{1}{2},j+{1 \over 2},m)_2 \oplus
                   (\Delta+\frac{1}{2},j-{1 \over 2},m)_2 \oplus (\Delta+1,j,m)_2 \bigg].
\end{equation}
where $\oplus$ denotes a direct sum and $\bigoplus_{m=-h}^{h}$ denotes the
direct sum of representations.

Here the subscript denotes $\cal{N}$ of the algebra; i.e. $()_3$ 
denotes a representation of the ${\cal N}=3$ algebra, while $()_2$ 
denotes a representation of the ${\cal N}=2$ algebra.

The summation outside the brackets on the RHS of \eqref{long} reflects the 
fact that a primary that transforms in a given irreducible $SO(3)$
($R$ symmetry in ${\cal N}=3$ algebra) representation consists of several 
different $SO(2)$ primaries (with distinct $R$-charges). The four terms 
in the bracket on the RHS of \eqref{long} represent the states obtained 
by acting on the ${\cal N}=3$ primary with supercharges that belong to the 
${\cal N}=3$ algebra, but are absent in the ${\cal N}=2$ algebra. 

The decomposition \eqref{long} may be rewritten as follows:
\begin{eqnarray} \label{longp}
\nonumber
&&(j+h+1 +\epsilon ,j,h)_3=\bigoplus_{m=-h}^{h}\bigg[ (j+h+1+\epsilon,j,m)_2 \oplus 
(j+h+\frac{3}{2}+\epsilon,j+{1 \over 2},m)_2\bigg]  \\
&&~~~~~~~~~~~~\oplus\bigoplus_{m=-(h+1)}^{h+1}\bigg[ (j+h+\frac{3}{2} +\epsilon, 
j-{1 \over 2},m)_2 \oplus (j+h+2 +\epsilon,j,m)_2 \bigg].
\end{eqnarray}
In this equation we have 
grouped together terms on the RHS for the following reason. 
Recall that the decomposition of a long representation - 
 with $j\neq 0$  - at the unitarity bound, into short unitary representations
of the superconformal algebra, is given both at ${\cal N}=3$ and 
at ${\cal N}=2$ by  
\begin{equation}\label{deco}
(j+h+1+\epsilon,j,h)~ \longrightarrow^{\hspace*{-6.5mm}\epsilon \to 0} ~(j+h+1,j,h) \oplus (j+h+\frac{3}{2},j-\frac{1}{2},h+1)
\end{equation}
This formula should apply to \eqref{longp}. Comparing \eqref{deco} and 
\eqref{longp}, it is plausible (and correct) that the generic short 
${\cal N}=3$ representation decomposes into ${\cal N}=2$ representations 
according to the formula 
\begin{equation} \label{shortgen}
(j+h+1,j,h)_3=\bigoplus_{m=-h}^{h}\bigg[ (j+h+1,j,m)_2 \oplus 
(j+h+\frac{3}{2},j+{1 \over 2},m)_2\bigg]
\end{equation}
where all representations that saturate the unitarity bound, on the 
RHS of \eqref{shortgen}, are short.

We may deduce the split of a generic $j=0$  short ${\cal N}=3$ representation
into representations of the ${\cal N}=2$ algebra using identical reasoning. 
To start with we note that  ${\cal N}=3$ long representation at $j=0$ splits 
up into  long ${\cal N}=2$ representations as follows
\begin{equation}
(\Delta,0,h)_3=\bigoplus_{m=-h}^{h}\bigg[ (\Delta,0,m)_2 \oplus (\Delta+\frac{1}{2},{1 \over 2},m)_2 \oplus 
                       (\Delta+1,0,m)_2 \bigg]
\label{eq:j=0breakup}
\end{equation}
We next note that, both in the ${\cal N}=3$ and the ${\cal N}=2$ algebras,
\begin{eqnarray}
\nonumber
&&(h+1+\epsilon,0,h)~ \longrightarrow^{\hspace*{-6.5mm}\epsilon \to 0}~ (h+1,0,h) \oplus (h+2,0,h+2) \\
&&(h+\frac{3}{2}+\epsilon,\frac{1}{2},h) \longrightarrow^{\hspace*{-6.5mm}\epsilon \to 0} ~
(h+\frac{3}{2}+\epsilon,\frac{1}{2},h) \oplus (h+2,0,h+1)
\end{eqnarray}
These two equations allow us to deduce that, for short representations,  
\begin{equation}
(h+1,0,h)_3 = \bigoplus_{m=-h}^{h}\bigg[ (h+1,0,m)_2 \oplus (h+\frac{3}{2},\frac{1}{2},m)_2 \bigg]
               \oplus \bigoplus_{m=-(h+2)}^{h+2} (h+2,0,m)_2
\label{eq:shortjeq0breakup}
\end{equation}
In the equation above, representations that saturate the BPS bound are short.

The breakup of the ${\cal N}=3$ isolated short representations needs slightly
more indirect reasoning to deduce; we simply present the final result:
\begin{equation}
(h,0,h)_3 = \bigoplus_{m=-h}^{h} (h,0,m)_2
\label{eq:Isoshrtbrkup}
\end{equation}

The complete decomposition of the ${\cal N}=3$ algebra into ${\cal N}=2$ representations is given in Table (\ref{tab:decomposition1})
for long representations and Table (\ref{tab:decomposition2}) for short representations.

\begin{table}
\centering
\begin{tabular}{|c|c|c|}
\hline
Spin $j$ & ${\cal N}=3$ primary & ${\cal N}=2$ primaries  \\
\hline
$j\neq0$ &  $(\Delta, j, h)$    & $\bigoplus_{m=-h}^{h}\bigg[ (\Delta,j,m) \oplus (\Delta+\frac{1}{2},j+{1 \over 2},m) \oplus$ \\
         &                       & $(\Delta+\frac{1}{2},j-{1 \over 2},m) \oplus (\Delta+1,j,m) \bigg]$ \\
\hline                      
$j=0$    &  $(\Delta, 0, h)$    & $\bigoplus_{m=-h}^{h}\bigg[ (\Delta,0,m) \oplus (\Delta+\frac{1}{2},{1 \over 2},m) \oplus$ \\ 
         &                      & $(\Delta+1,0,m) \bigg]$  \\
\hline
\end{tabular}
\caption{Decomposition of long ${\cal N}=3$ representations into ${\cal N}=2$ representations.}\label{tab:decomposition1}
\end{table}

\begin{table}
\centering
\begin{tabular}{|c|c|c|}
\hline
Spin $j$ & ${\cal N}=3$ primary & ${\cal N}=2$ primaries \\
\hline
$j\neq0$ & $(j+h+1, j, h)$      & $\bigoplus_{m=-h}^{h}\bigg[ (j+h+1,j,m) \oplus$ \\
         &                      & $(j+h+\frac{3}{2},j+{1 \over 2},m)\bigg]$  \\
         &                      &                                                  \\
\hline 
Generic short & $(h+1, 0, h)$   &  $\bigoplus_{m=-h}^{h}\bigg[ (h+1,0,m) \oplus (h+\frac{3}{2},\frac{1}{2},m) \bigg]$\\
       $j=0$       &                 & $\oplus \bigoplus_{m=-(h+2)}^{h+2} (h+2,0,m) $     \\ 
               &                 &                                                \\ 
\hline 
Isolated short &  $(h,0,h)$     & $\bigoplus_{m=-h}^{h} (h,0,m)$                                         \\
  $j=0$        &                &                                                    \\
\hline
\end{tabular}
\caption{Decomposition of short ${\cal N}=3$ representations into ${\cal N}=2$ representations.}\label{tab:decomposition2}
\end{table}

\section{The $R$-charge as a function of $\lambda$ in the absence of 
a superpotential}

It was explained in \cite{Gaiotto:2007qi} (see the introduction) that 
the ${\cal N}=2$ $U(N)$ Chern Simons theory with $g$ chiral multiplets 
and  no superpotential, 
is superconformally invariant at every value of $N$ and $k$, and so 
at every value of $\lambda$, in the large $N$ limit. In the free limit
the $R$-charge of each of the chiral multiplets in this theory is equal 
to half. As was explained in  \cite{Gaiotto:2007qi}, however, 
this $R$-charge is renormalized as a function of $\lambda$. As the 
$R$-charge of an operator appears in the BPS formula that determines its 
scaling dimension, the determination of the charge of a chiral field, as 
a function of $\lambda$, is perhaps the most elementary characteristic 
of the supersymmetric spectrum of the theory. In this section we will 
adopt a proposal by Jafferis \cite{jafferis} to perform this determination.

\subsection{The large $N$ saddle point equations}

According to the prescription of \cite{jafferis}, the 
superconformal $R$-charge of the theories we study is determined by extremizing 
the magnitude of their partition function on $S^{3}$ with respect to 
the trial $R$-charge,\footnote{More precisely, as shown in \cite{jafferis}, a supersymmetric theory on $S^3$ can be defined with an arbitrary choice of $R$-charge $h$, and the partition function of this theory on $S^3$ is ${\cal Z}(h)$. The superconformal $R$-charge is such that $|{\cal Z}(h)|^2$ is minimized.} $h$, assigned to a chiral multiplet. The partition 
function itself is determined by the method of supersymmetric localization to 
be given by the finite dimensional integral 
\begin{equation} \label{pf}
{\cal Z}(h) = \int \prod_{i=1}^{N} du_i\, \exp\left\lbrace N^2 \left[ {i \pi \over \lambda} {1\over N}
{\sum_i u_i^2 } +  {1\over N^2} \sum_{i\ne j}  \log \sinh{(\pi u_{ij} )} + {g\over N^2}
\sum_{i,j} \ell(1 - h + i u_{ij}  ) \right]  \right\rbrace 
\end{equation}
where $\lambda={N / k}$ is the 't Hooft coupling, 
$u_{i}$  ($i =1 \ldots N$) are real numbers (and the integration range is from 
$-\infty$ to $\infty$), $u_{ij}\equiv u_i-u_j$, and the function $\ell(z)$ is given by 
\begin{equation}\label{lfunc}
\ell(z) = - z \log{\left(1 - e^{2\pi i z}\right)} + {i \over 2} \left[ \pi z^2 + {1 \over \pi} 
          {\rm Li}_2(e^{2 \pi i z}) \right] - {i \pi \over 12}
\end{equation}
where ${\rm Li}_2$ is the dilogarithm function. While the function $\ell(z)$ is complicated looking, 
its derivative is elementary
$$\partial_z \ell(z) = -\pi z\cot(\pi z)$$
and is all we will need in this paper. 

According to \cite{jafferis}, once the partition function ${\cal Z}$ 
is obtained by performing the integral in \eqref{pf}, the $R$-charge of 
the chiral fields is determined (up to caveats we will revisit below) 
by solving the equation
$\partial_h |{\cal Z}(h)|^2 = 0$.
This gives the exact superconformal $R$-charge.

In the large $N$ limit the integral in \eqref{pf} may be determined 
by saddle point techniques. The saddle point equations, together 
with the equation $\partial_h|{\cal{Z}}(h)|^2 =0$ 
(which determines $h$, given the saddle point) are given by 
\begin{eqnarray}
\label{eq:saddlept}
0 &=& {i u_k\over\lambda} + {1\over N} \sum_{j(\ne k)}^N\left\{ \coth{(\pi u_{kj})} 
-{i\over 2} g \bigg[ (1-h + iu_{kj}) \cot{\pi(1-h+i u_{kj})} \right. \\ \nonumber
 && \left. -(1-h-iu_{kj} ) \cot{\pi(1-h-i u_{kj} )} \bigg] \right\}, \\
0 &=& {\rm Re} \left[ \sum_{i, j=1}^N (1 - h + i u_{ij}) \cot{\pi(1 - h + i u_{ij})} \right].
\label{eq:deltaeqn}
\end{eqnarray}

\subsection{Perturbative solution at small $\lambda$}

While we have been unable to solve the equations \eqref{eq:saddlept}
in general even at large $N$, it is not difficult to solve these equations 
either at small $\lambda$ (at all $g$) or at large $g$ (for all $\lambda$). 
In this subsection we describe the perturbative solution to these equations 
at small $\lambda$ (for all $g$). In the next subsection we will outline the 
perturbative procedure that determines $h(\lambda)$ at all $\lambda$ but 
large $g$. 

It is apparent from a cursory inspection of \eqref{eq:saddlept} that 
the eigenvalues $u_i$ must become small in magnitude (in fact must scale 
like $\sqrt{\lambda}$) at small $\lambda$. It follows that complicated functions
of $u$ simplify to their Taylor series expansion in a power expansion in 
$\lambda$. This is the basis of the perturbative technique described in this
subsection.

More quantitatively, at small $\lambda$ we expand $u_i$ and $h$ as
\begin{align}\label{upertexp}
u_k =& \sqrt{\lambda} \left( u_k^{(0)} + \lambda  u_k^{(1)} + \cdots \right), \\
h =& h^{(0)} +\lambda h^{(1)} + \lambda^2 h^{(2)}+\cdots ,
\end{align}
and attempt to solve our equations order by order in $\lambda$. 
At leading nontrivial order, ${\cal O} ( \lambda^0)$, 
equation \eqref{eq:deltaeqn} reduces to 
\begin{equation}
(1 - h^{(0)}) \cot{\pi(1 - h^{(0)})}=0 \Rightarrow h^{(0)}={1 \over 2}
\end{equation}
which tells us that $h=\frac{1}{2}$ at leading order in $\lambda$. 
On the other hand, equation \eqref{eq:saddlept} at its leading nontrivial order, 
namely ${\cal O}({1 \over \sqrt{\lambda}})$,
reduces to 
\begin{equation}\label{0theqn}
i u_i^{(0)} = - {1 \over \pi} {1 \over N} \sum_{j (\ne i)} {1 \over u_i^{(0)} - u_j^{(0)} } .
\end{equation}
Apart from an unusual factor of $i$, this is precisely the large $N$ saddle 
point equations of the Wigner model. The extra factor of $i$ may be dealt 
with by working with the rescaled variable 
$$ y_j= e^{-{\pi i\over 4}}u_j$$
in terms of which 
\begin{equation}\label{0theqn1}
y_i^{(0)} =  {1 \over \pi} {1 \over N} \sum_{j (\ne i)} 
{1 \over y_i^{(0)} - y_j^{(0)} } .
\end{equation}
The solution to this equation is well known in the large $N$ limit. 
The eigenvalues $y_i^{(0)}$ cluster themselves into a ``cut" along the interval 
$(-a, a)$ with 
$$a=\sqrt{2\over\pi}~.$$
The density of eigenvalues, $\rho(y)= \sum_{i=1}^N \delta(y-y_i^i)$, in this 
interval is given by 
\begin{equation}\label{eigdistrib0}
\rho(y)=\frac{2}{\pi a^{2}}\sqrt{a^{2}-y^{2}}.
\end{equation}
Using the fact that $u\approx e^{\pi i\over 4}\sqrt{\lambda}  y$, we see that, at 
leading order in $\lambda$, the saddle point is given by the eigenvalues 
$u_i$ clustering along a straight line of length of order $\sqrt{\lambda}$, 
oriented at $45$ degrees in the complex plane. 

Note that the distribution of eigenvalues has $u \to -u$ symmetry 
and in particular the average value of eigenvalues is zero. The $u \to -u$ 
symmetry is an exact symmetry of the saddle point equations, and we will 
assume that it is preserved in the solution (i.e. not spontaneously broken)
in the rest of this paper. 

Let us now proceed beyond the leading order. \eqref{eq:deltaeqn}
is automatically satisfied at ${\cal O}(\sqrt{\lambda})$ 
(this is because ${\rm Im} \left[\sum_{i \ne j} ( u_i^{(0)} -  u_j^{(0)}) \right] = 0$). 
 At order $\lambda$ the same equation reduces to 
\begin{equation} \label{oab}
h^{(1)} = -2\, {\rm Re} \left[{1 \over N^2} \sum_{i \ne j} ( u_i^{(0)} -  u_i^{(0)} )^2 \right]
\end{equation}
Now recall that the phase of $u_i$ is $e^{\pi i\over 4}$. As a consequence 
the real part vanishes and hence, from \eqref{oab},  $h^{(1)}=0$.

In order to compute the correction to $h(\lambda)$ at ${\cal O}(\lambda^2)$ 
we need to find the first correction $u^{1}_i$ to the eigenvalue distribution. 
We now turn to that task. At order ${\cal O}(\lambda^0)$, 
\eqref{eq:saddlept} reduces to 
\begin{equation}\label{1storder}
i u_k^{(1)} - {1 \over N} \sum_{j(\ne k)} \left\lbrace { \pi \over 6} (3g-2) (u_k^{(0)}-u_j^{(0)}) + { u_k^{(1)}-u_j^{(1)} \over \pi (u_k^{(0)}-u_j^{(0)})^2 } \right\rbrace  = 0
\end{equation}
Now, part of the second term on the RHS of \eqref{eq:saddlept} is easily
simplified. It follows immediately from \eqref{0theqn} (differentiating 
that equation with respect to $u_i^{(0)}$)  that 
$${1 \over \pi N}\sum_{j (\ne k)} {1 \over (u_k^{(0)}-u_j^{(0)})^2} = i. $$
Inserting this relation into part of the RHS of \eqref{1storder} gives 
us a piece that cancels the LHS, and \eqref{1storder} simplifies to 
\begin{equation}
{1 \over \pi N} \sum_{j (\ne k)} {1 \over (u_k^{(0)}-u_j^{(0)})^2} u_j^{(1)} = {\pi \over 6  N } (3g-2) \sum_{j \ne k}  ( u_k^{(0)}-u_j^{(0)} ).
\end{equation}
The RHS of this equation may be further simplified using 
$ \sum_j u_j=0.$
Retaining only terms that contribute at leading order in $N$, we find 
\begin{equation}
{1 \over \pi N} \sum_{j (\ne k)} {1 \over (u_k^{(0)}-u_j^{(0)})^2} u_j^{(1)} 
= {\pi \over 6} (3g-2) u_k^{(0)}.
\end{equation}
In order to solve this equation we once again move to the ``real'' variable 
$y$. That is we define $u^{(0)}_j=e^{\pi i\over 4} y_j$ as above. Let us also define 
 $u^{(1)}_j = e^{\pi i\over 4} v_j(y_j)$. In the large $N$ limit $u^{(0)}_i$ is effectively 
a continuous variable on the $45$ degree cut on the complex plane, and 
$$u^{(1)}=e^{\pi i\over 4} v(y)$$ 
for a continuous function $v(y)$ that we now determine. The equation for 
$v(y)$ is given by   
\begin{equation}\label{contn}
{1 \over \pi} {\cal P} \int {v(y) \rho(y) \over (y_1 - y)^2}  dy  = 
{i \pi \over 6} (3g-2) y_1 .
\end{equation}
Integrating both sides of this equation with respect to $y_{1}$ we find
\begin{equation} \label{inteq}
{\cal P} \int {v(y) \rho(y) \over (y_1 - y)}  dy  = -{i\pi^{2} \over 12} (3g-2) y_{1}^{2} + k_1 ,
\end{equation}
where $k_1$ is the constant of integration. 

In order to proceed we must solve the integral equation \eqref{inteq}. We will 
now explain how, in more generality, it is possible to solve the equation
\begin{equation} \label{foreq}
{\cal P} \int {z(y) \over (y_1 - y)}  dy  = g_{n}(y_{1})
\end{equation}
where $g_n(y)$ is a complex polynomial in $y$, and $z(y)$ is a function defined 
over the range $(-a, a)$. The solution may be found by constructing an 
analytic function $P(y)$ whose only singularities on the real axis are a 
cut in the range $(-a, a)$, and whose real part on this interval is given 
by $g_n(y)$. An obvious ansatz for such a function is
\begin{equation}
P(y)=g_{n}(y)-h_{n}(y)\sqrt{y^{2}-\frac{2}{\pi}}
\end{equation}
where $h_n(y)$ is a yet to be determined polynomial in $y$. Let us now choose 
$h_n(y)$ to ensure that the leading behaviour of $P(y)$ at infinity is
$P(y) \sim \frac{1}{y}$. If we manage to achieve this then it follows 
from an application of Cauchy's theorem that, for any complex valued $w$, 
$$P(w) =-\frac{1}{2 \pi i} \int_{-a}^a \frac{{\rm disc} \,P(x)}{x-w}dx$$
where ${\rm disc}P(x)$ is the discontinuity of $P(x)$ across the branch cut
which is on the real axis.
We can now apply this equation to $w=y+i\epsilon$ and also to $w=y-i\epsilon$ 
(where $y \in (a, -a)$), take the average of the two, and take the limit 
$\epsilon \to 0$. This gives the equation 
$$g_n(y) =-\frac{1}{2 \pi i} {\cal P} \int_{-a}^a \frac{{\rm disc} \,P(x)}{x-w}dx$$
where we have used the fact that the real part of $P(y)$ is $g_n(y)$
in the range $y \in (-a, a)$. Comparing with \eqref{foreq} we conclude 
that 
\begin{eqnarray}\nonumber
z(y)&=& \frac{1}{2\pi i}{\rm disc}\, P(y) 
               =\frac{1}{\pi}h_{n}(y)\sqrt{\frac{2}{\pi}-y^{2}} .
\end{eqnarray}
If, as is the case in our situation, that 
$$z(y)= v_n(y) \rho(y)$$
then the solution above for $z(y)$ implies that 
\begin{equation} \label{finvsol}
v_{n}(y)=\frac{1}{\pi}h_{n}(y)
\end{equation}

Applying this general formalism to the equation \eqref{inteq} is 
straightforward. We have
\begin{equation}
g_{1}(y)=-\frac{i\pi^{2}}{12} (3g-2) y^{2}+ k_1
\end{equation}
where $k_1$ is an as yet arbitrary constant. In order to be able to 
find a suitable function $P(y)$ we need to choose $k_1=\frac{i \pi}{12} (3g-2)$ 
and 
$$h_1(y)= -\frac{i \pi^2 y}{12} (3g-2).$$
These choices lead to the analytic function 
\begin{equation}
P(y)=-\frac{i\pi^{2}}{12} (3g-2) \left(y^{2}-\frac{1}{\pi}-y\sqrt{y^{2}-\frac{2}{\pi}}\right)
\end{equation}
and yield\footnote{This correction to the eigenvalue distribution tilts the eigenvalue 
cut - originally at $45$ degrees in the complex plane - a little nearer 
to the real axis.}
\begin{equation}
v(y)=-\frac{i\pi}{12} (3g-2) y.
\end{equation}

With the first nontrivial correction to the eigenvalue distribution in hand, 
it is now a simple matter to compute the shift in the scaling dimension 
$h(\lambda)$ at leading order nontrivial order in $\lambda$. 
\eqref{eq:deltaeqn} is automatically obeyed at 
${\cal O} (\lambda^{3 \over 2})$.\footnote{The identity in question is 
\begin{equation}
{\rm Im} \sum_{i \ne j} \left[ \pi^3 (u_i^{(0)} - u_j^{(0)})^3 - 3 \pi (u_i^{(1)} - u_j^{(1)}) \right] = 0
\end{equation} }
However at ${\cal O}(\lambda^2)$, the same equation yields 
\begin{equation}
{\rm Re} \left[ \sum_{k \ne j} - 2 \pi^3 (u_i^{(0)} - u_j^{(0)})^4 + 
3 \pi \left( 4 (u_i^{(0)} - 
u_j^{(0)}) (u_i^{(1)} - u_j^{(1)}) + h^{(2)} \right) \right]=0
\end{equation}
In other words 
\begin{equation}
\begin{split}
3 h^{(2)} =& {\rm Re} \left[ \int dy_1 dy_2 \rho(y_1) \rho(y_2) \left\lbrace -2 \pi^2 (y_1 - y_2)^4  - 
12 i (y_1 - y_2) (v(y_1)- v(y_2)) \right\rbrace \right]\\
=&  {\rm Re} \left[ -2  \pi^2 \left( 2 \langle y^4 \rangle + 
6  \langle y^2 \rangle^2 \right)  - 24 i  \langle y \ v(y) \rangle \right]\\
\end{split}
\end{equation}
where we have defined
$$\langle O(y) \rangle \equiv  \int O(y) \rho(y) dy .$$ 
Evaluating the integrals we find
\begin{equation}
h^{(2)}=-(1+g)
\end{equation}
This exactly matches the explicit perturbative result of 
Gaiotto and Yin \cite{Gaiotto:2007qi}. Similar agreement was
also found with \cite{Amariti:2011}.

The method presented here is easily iterated to higher orders in 
$\lambda$. It turns out that at each order the correction to the eigenvalue 
distribution is determined by the solution to an equation of the form 
\eqref{foreq}. This solution may be obtained, order by order, using the 
method described above. The new correction to the eigenvalue 
distribution yields a new term in the correction to the anomalous dimension. 

We have explicitly implemented this perturbative procedure to a few 
orders in perturbation theory. At $g=1$ we find 
\begin{equation}\label{hresultpert}
h=\frac{1}{2}-2\lambda^{2}+\frac{13\pi^{2}}{3}\lambda^{4}-\left(\frac{207\pi^{4}}{10}-32\pi^{2}\right)\lambda^{6}+
   \left(\frac{339019\pi^{6}}{2520}-\frac{3355\pi^{4}}{9}+\frac{160\pi^{2}}{3}\right)\lambda^{8} + \cdots
\end{equation}
while for general $g$ we have 
\begin{eqnarray} \nonumber
h&=&\frac{1}{2} - (1+g)\lambda^{2} + \frac{1}{12}(1+g)\left[-24(-1+g)+\pi^{2}(3g^{2}+15g+8)\right] \lambda^{4}  \\  \nonumber & &
+\bigg[-8 - 25 {\pi^2 \over 3} - {61 \pi^4\over60} +  g (-{4 \pi^2 \over 3}- {637 \pi^4 \over 120} )+
 g^2 (8 + {64 \pi^2 \over 3} - {395 \pi^4 \over 48} ) \\ \nonumber & & 
 + 
 g^3 ({52 \pi^2 \over 3} - {239 \pi^4 \over 48}) + 
 g^4 (3 \pi^2 - {53 \pi^4 \over 48}) - g^5 {\pi^4 \over 16} \bigg] \lambda^{6} + \cdots
\end{eqnarray}

\subsection{Perturbative solution at large $g$} \label{pertlargeg}

As we have described in the previous subsection, in the limit of small 
't Hooft coupling the solution to the saddle point equation, 
\eqref{eq:saddlept}, is obtained by balancing the first term (large because 
of the inverse power of $\lambda$) against the second (large because of the 
singularity at small $u$), and treating the third term as a perturbation. 

Let us now consider another limit; one in which the number of flavours 
$g$ becomes large, without making any assumptions on $\lambda$. In this 
case the small $u$ singularity of the second term in \eqref{eq:saddlept}
balances against either the largeness of $g$ (in the third term) or the 
smallness of $\lambda$ (in the first term). The important point is that
$u$ is necessarily small for this balance to work, so perturbative techniques
apply. The most interesting regime
is one in which $\lambda= {\cal O} (1/g)$. In order to focus in on this 
regime we formally scale 
 $\lambda \rightarrow \lambda x^{2}$, $g \rightarrow \frac{g}{x^{2}}$ 
and work in a power series expansion in $x$. 
We make the expansion
$$u_i=x \left( u_i^{(0)}+ x^2 u_i^{(1)} + \ldots \right)$$
$$h=\frac{1}{2}+x^{2}h^{(1)} + x^4 h^{(2)} + \ldots $$
and solve the equations order by order in the formal expansion 
parameter $x$. The perturbative procedure proceeds exactly as in the previous
subsection. At lowest order we obtain the equation
\begin{equation}
\frac{1}{N}\sum_{j\neq k}\frac{1}{(u_{k}^{(0)}-u_{j}^{(0)})} = u_{k}^{(0)}\bigg( \frac{-i\pi}{\lambda}+\frac{g\pi^{2}}{2} \bigg)
\end{equation}
so that, once again, the eigenvalue distribution is given by a Wigner type 
cut at leading nontrivial order. More specifically, 
the eigenvalue distribution for this is given by 
\begin{equation}
\rho(u)=\frac{2}{\pi a^{2}}\sqrt{a^{2}-u^{2}};~~~a^{2}=\frac{2}{\frac{-i\pi}{\lambda}+\frac{g\pi^{2}}{2}}
\end{equation}
The lowest order correction to $R$-charge can be calculated as before by expanding the ${\cal Z}$ extremization equation 
(\ref{eq:deltaeqn}) to lowest order in $x$ . This gives 
\begin{equation}
h^{(1)}=-4 \, {\rm Re}\langle u^{2}\rangle
\end{equation}
where 
$\langle ~~\rangle$ denotes expectation value as earlier. Using the zeroth order eigenvalue distribution this gives 
\begin{equation}
h^{(1)}=-{\rm Re}\langle a^2\rangle=-\frac{g\pi^{2}}{ \left( \frac{g\pi^{2}}{2} \right)^{2} + \left( \frac{\pi}{\lambda} \right)^{2} }
\end{equation}

In summary, we have, to leading order
\begin{equation} \label{sum}
h(\lambda)= \frac{1}{2} -\frac{g\pi^{2}}{ \left( \frac{g\pi^{2}}{2} \right)^{2} + \left( \frac{\pi}{\lambda} \right)^{2} }\, +\, {\rm higher~order}.
\end{equation}

\subsection{Numerical study of $R$-charge for small $g$ and large $\lambda$}\label{smallglargelmda}

In the previous subsections we used perturbative techniques to establish 
the following. The function $h(\lambda)$ starts out at the value $\frac{1}{2}$ 
at $\lambda=0$. It always decreases at small $\lambda$; at leading order 
$h(\lambda)= \frac{1}{2} -(g+1) \lambda^2$. What happens at larger 
values of $\lambda$ ? This question turned out to be easy to answer at large 
$g$. In this limit the decrease of $h(\lambda)$ as a function of $\lambda$ 
stops at $\lambda \sim \frac{1}{g}$, after which $h(\lambda)$ settles down 
at its asymptotic value 
\begin{equation} \label{ha}
h(\lambda)=\frac{1}{2}-\frac{4}{\pi^2 g} + {\cal O}(\frac{1}{g^{2}}).
\end{equation}

Neither of the analyses we have performed, however, reliably predict the 
behaviour of $h(\lambda)$ at large $\lambda$ when $g$ is of order unity. 
An unjustified extrapolation of \eqref{ha} suggests that $h(\lambda)$ 
always monotonically decreases, asymptoting to a constant value at 
$\lambda=\infty$. In order to check whether all this is really
the case we have numerically solved a discretized version of the saddle 
point equations with $Ne$ eigenvalues using Mathematica. \footnote{In the rest of 
this paper $Ne$ denotes the number of eigenvalues used for the purposes of discretized 
numerics. $Ne$ is a composite symbol - (it is not equal to product of $N$ with $e$).}

. We give details of our numerical 
procedure in Appendix \ref{numerics}. In this section we merely present our 
results. 

\begin{figure}
\begin{center}
\begin{tabular}{cc}
\includegraphics[width=70mm]{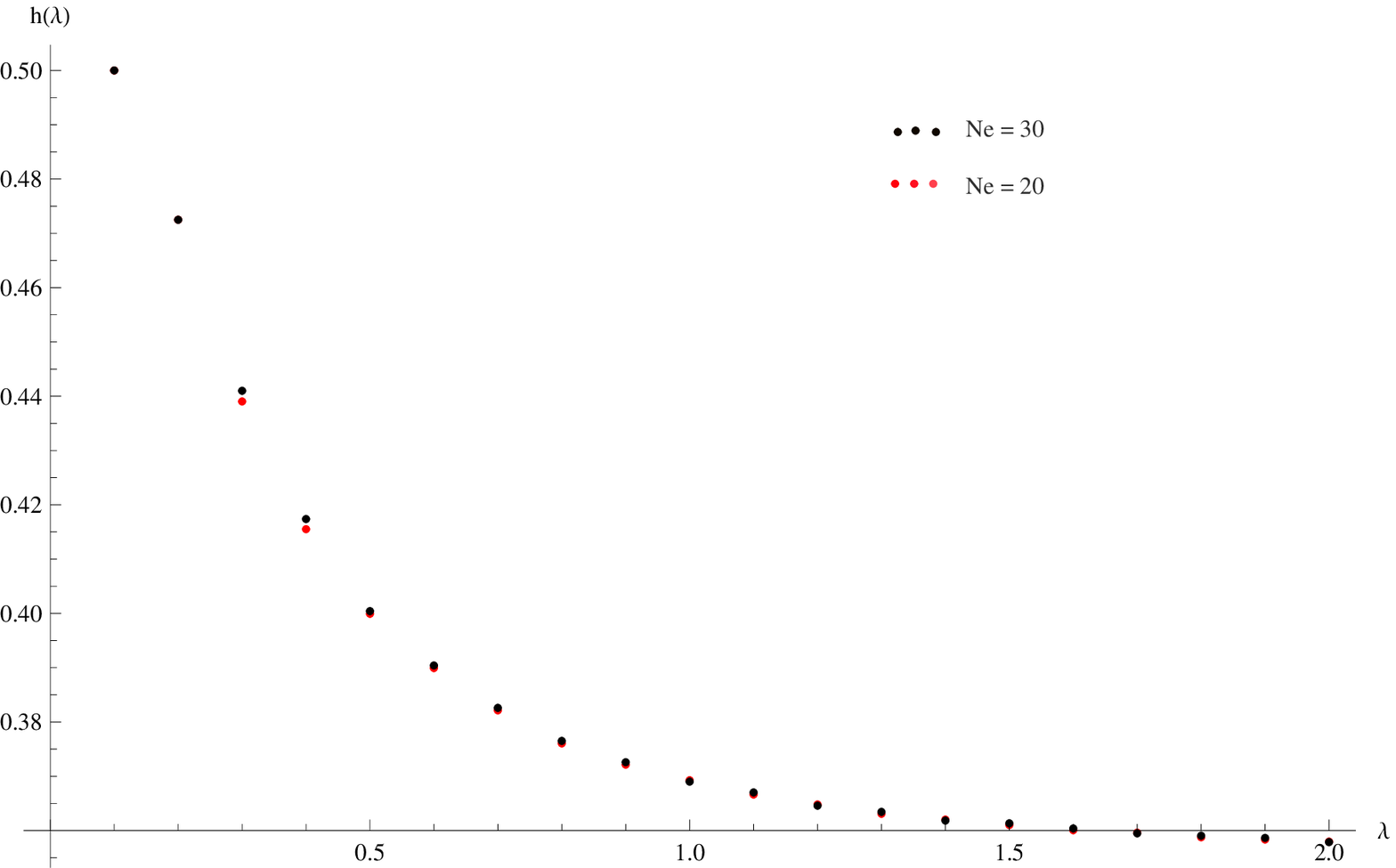} & \includegraphics[width=70mm]{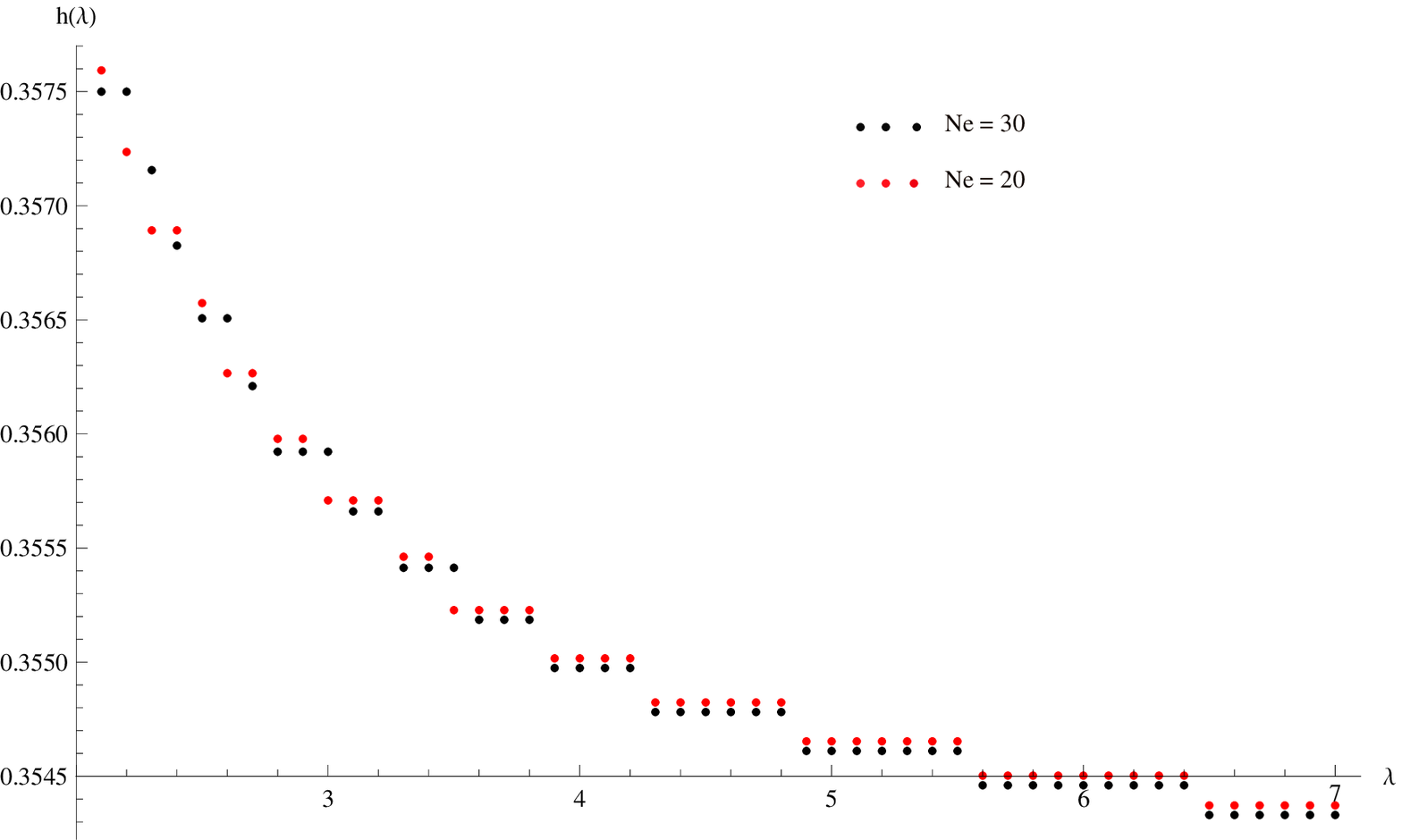} \\
\end{tabular}
\end{center}
\caption{$h$ vs. $\lambda$ for $g=3$, $Ne=20$ and $Ne=30$. In the figure on the left, $\lambda$ varies from $0$ to $2$. In the figure on the 
right $\lambda$ varies from 2 to 7. Note that $h(\lambda)$ scale is different in the two figures. $R$-charge saturates to around $0.354$. While we have not performed a serious error
estimate, it seems unlikely to us that the error in this asymptote value exceeds $\pm 0.01$.}
\label{hlmdafigg3}
\end{figure}

At $g=3$ the function is as given by the graph in Fig. \ref{hlmdafigg3}. Note that $h(\lambda)$ asymptotes to almost a constant 
value by $\lambda=2$, and varies only slightly in the $\lambda$ range 2 to 7 (this constant is approximately 0.354). 

\begin{figure}
\begin{center}
\begin{tabular}{cc}
\includegraphics[width=70mm]{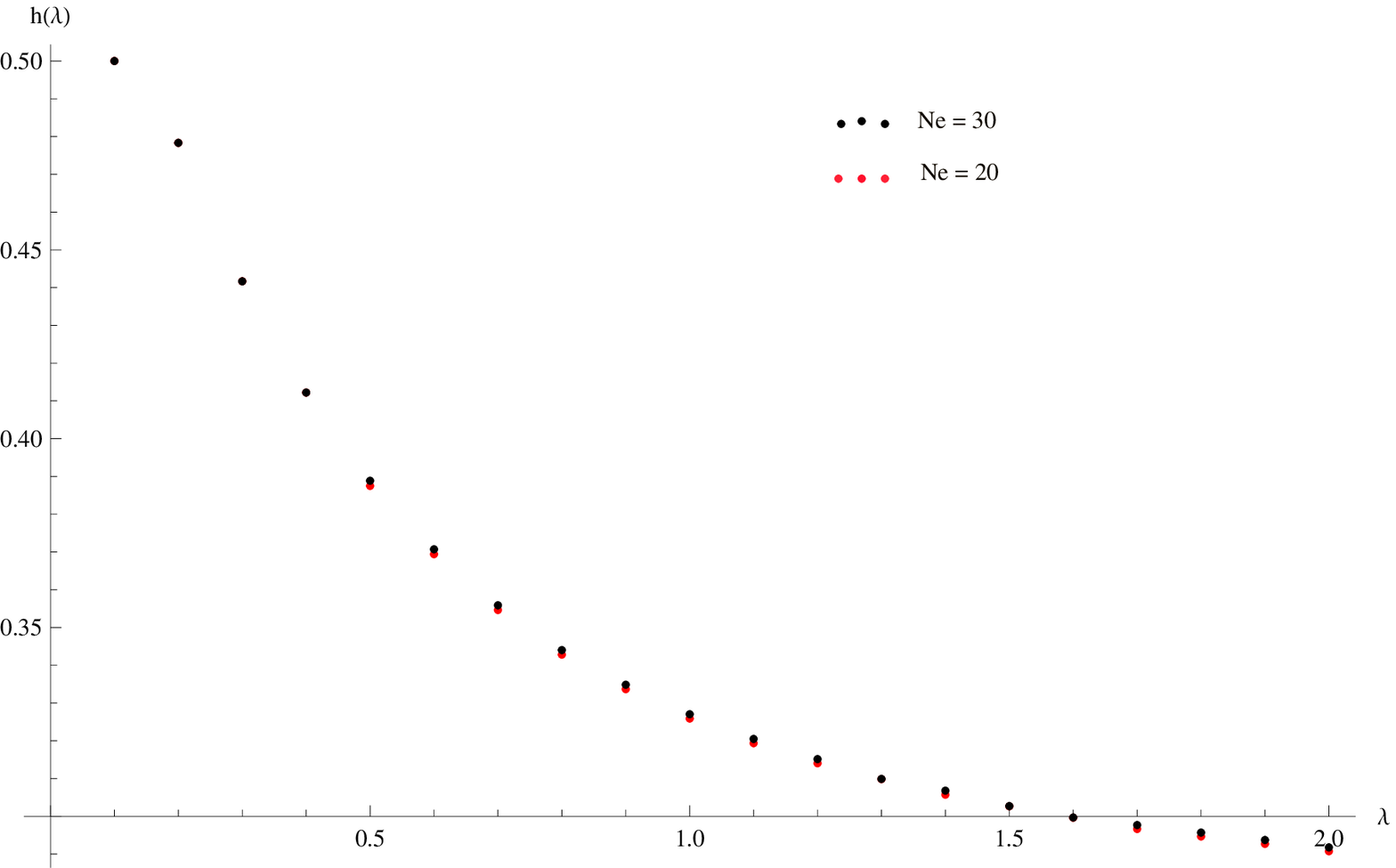} & \includegraphics[width=70mm]{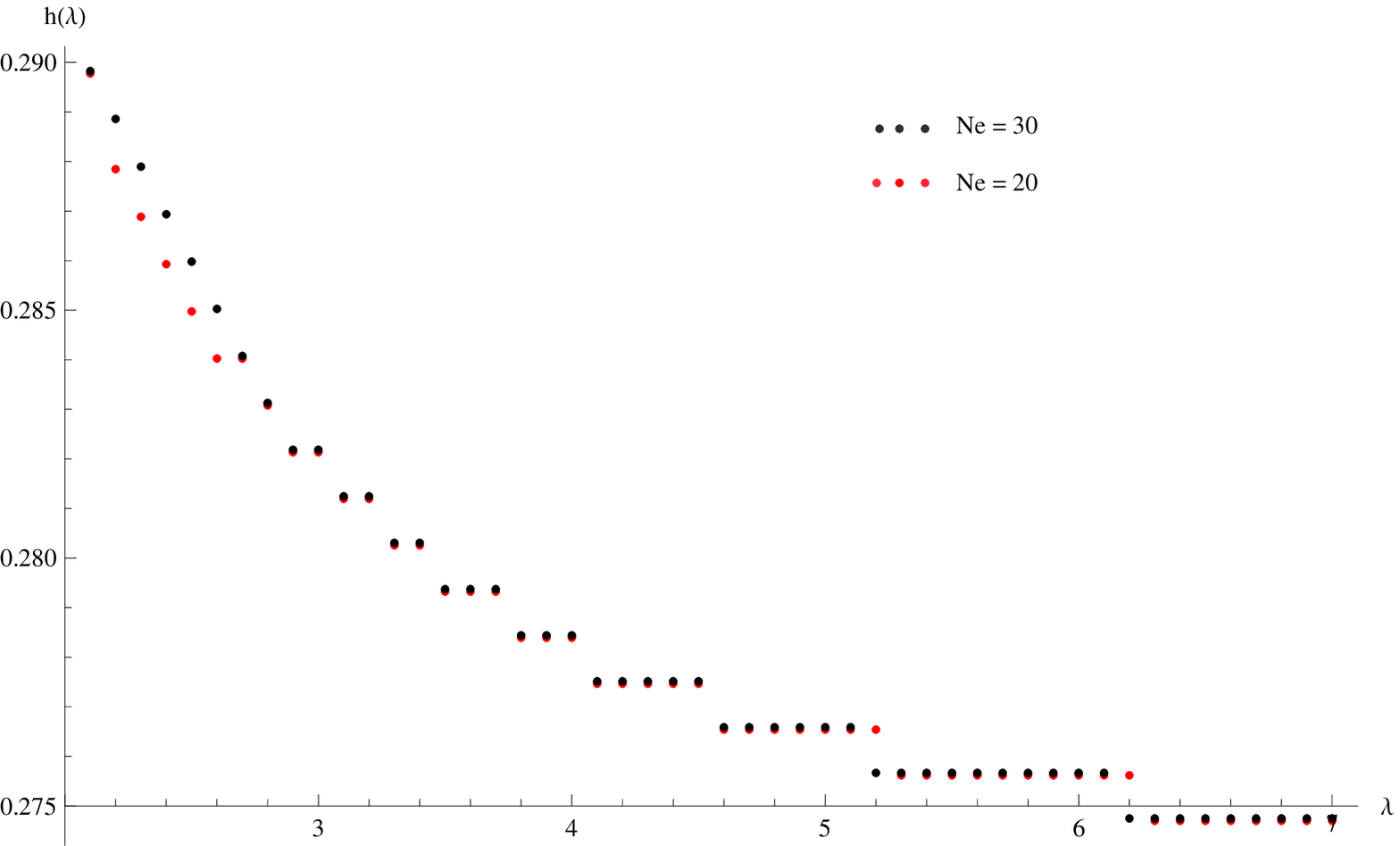} \\
\end{tabular}
\end{center}
\caption{$h$ vs. $\lambda$ for $g=2$, $Ne=20$ and $Ne=30$. In the figure on the left, $\lambda$ varies from $0$ to $2$. In the figure on the 
right $\lambda$ varies from $2$ to $7$. Note that $h(\lambda)$ scale is different in the two figures. $R$-charge saturates to around $0.274$.While we have not performed a serious error
estimate, it seems unlikely to us that the error in this asymptote value exceeds $\pm 0.01$.}
\label{hlmdafigg2}
\end{figure}

At $g=2$ we find the function $h(\lambda)$ given by Fig.  \ref{hlmdafigg2}. Note that $h(\lambda)$ asymptotes to almost a constant 
value by $\lambda=2$, and varies only slightly in the $\lambda$ range $2$ to $7$ (the asymptote value is approximately 0.274).

\begin{figure}
\begin{center}
\includegraphics[width=70mm]{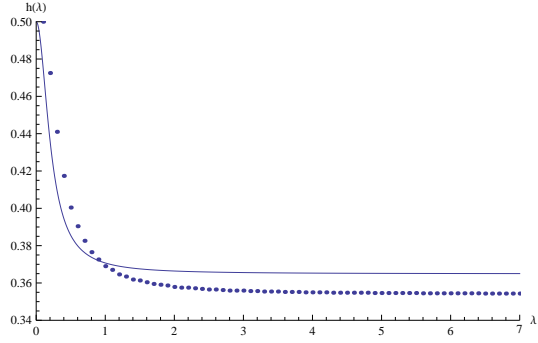}
\end{center}
\caption{$h$ vs. $\lambda$ for $g=3$ and $Ne=30$. The blue line is large g perturbation theory prediction. $R$-charge saturates to around $0.354$. While we have not performed a serious error
estimate, it seems unlikely to us that the error in this asymptote value exceeds $\pm 0.01$.}
\label{hlmdafigg3part2}
\end{figure}

As is apparent, in both these cases the function $h(\lambda)$ displays
the qualitative behaviour predicted by the large $g$ formula \eqref{sum}; 
$h(\lambda)$ monotonically decreases from $h=\half$ at $\lambda=0$ 
to a finite value of $h$ (greater than $\frac{1}{4}$) at $\lambda=\infty$. 
Notice that at $g=3$, $\frac{1}{2}-h(\infty) =\frac{1}{2}-.354 \approx 0.146$. This shift 
from $\frac{1}{2}$ agrees to 10 percent with the first order prediction of large $g$ 
perturbation theory \eqref{ha},  $\frac{4}{3\pi^2 }=0.135$. At $g=2$, $1-h(\infty)=\frac{1}{2}-.274=0.226$
which agrees to 12 percent with the first order prediction of large $g$ perturbation theory \eqref{ha}, 
$\frac{4}{2\pi^2 }=0.203$. It thus appears that, even quantitatively,  the results of large $g$ perturbation theory
are not too far off the mark from the correct answer  all the way down to $g=2$. In order to see this more clearly, 
in Fig \ref{hlmdafigg3part2} we replot our results for $h(\lambda)$ versus $\lambda$ at $g=3$ and compare with 
the predictions of large $g$ perturbation theory at first order in the perturbative expansion. 
Note the semi quantitative agreement between the curves.

\begin{figure}
\begin{center}
\begin{tabular}{cc}
\includegraphics[width=70mm]{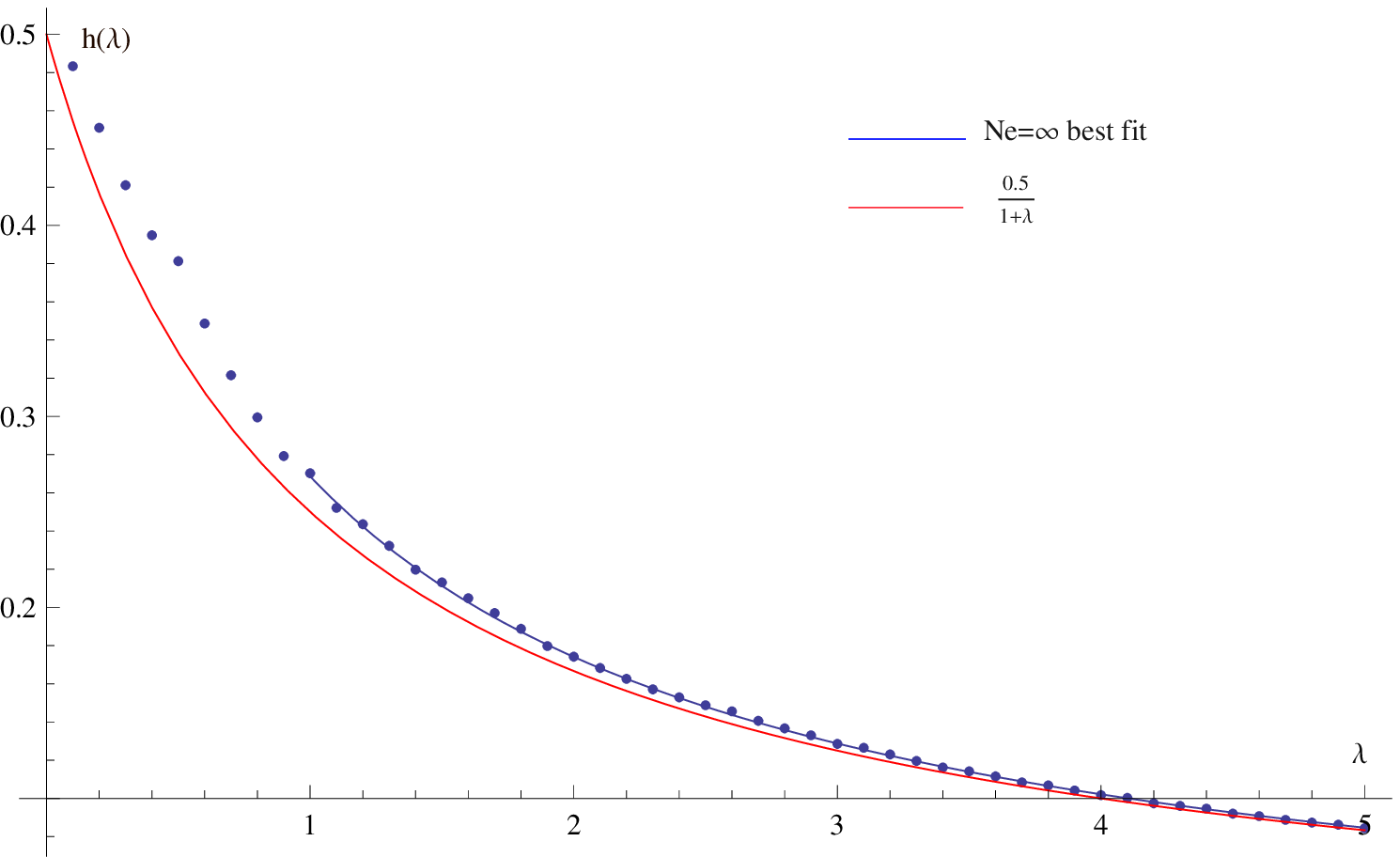} & \includegraphics[width=70mm]{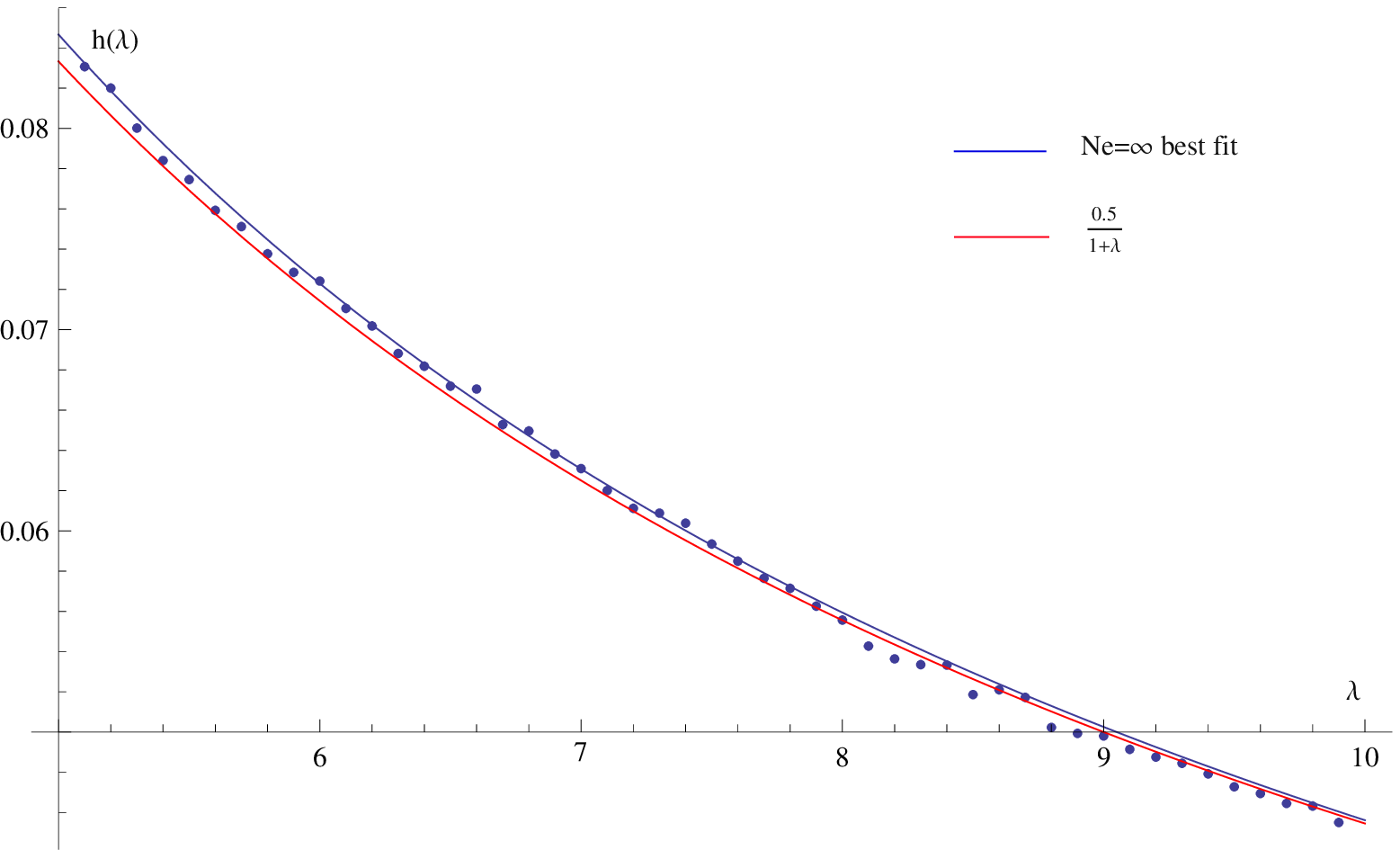}\\
\end{tabular}
\end{center}
\caption{$h$ vs. $\lambda$ for $g=1$.  Blue line is best fit of the data points to the form $\frac{\alpha}{\beta+\lambda}$. The red curve 
is $\frac{1}{2(1+\lambda)}$. }
\label{io}
\end{figure}

On the other hand our numerical result for the function $h(\lambda)$ 
at $g=1$ is presented in Fig. \ref{io} for $\lambda \in (0, 10)$ \footnote{It would of course be possible 
to obtain data at larger values of $\lambda$ as well. However this process becomes computationally increasingly 
expensive, as $\frac{1}{Ne}$ errors appear to increase upon increasing $\lambda$. In order to 
generate reliable data at larger and larger $\lambda$ requires solving the equations at 
larger and larger $Ne$.} As we explain in 
more detail in Appendix \ref{numerics}, at every $\lambda$ we have solved the saddle point 
equations at $Ne=20, 30, \ldots 100$ and bestfitted our results 
to the form 
\begin{equation} \label{oo} 
h(\lambda)=h(\lambda) + \frac{b(\lambda)}{c^2(\lambda)+(Ne)^2}.
\end{equation}

In the plot in Fig. \ref{io} the data points represent the values of $h(\lambda)$ (obtained 
out of this best fit procedure) versus $\lambda$. In Appendix \ref{numerics} we have also 
presented a very crude estimate of the likely magnitude of the error in $h(\lambda)$; 
we estimate that this error is not larger that a few (conservatively, say,  5) percent. In Appendix \ref{numerics} 
we also demonstrate that our curve of $h(\lambda)$ versus $\lambda$ agrees quite well at small $\lambda$ with 
the perturbative predictions of previous subsections.

In Fig. \ref{io} we have also presented two curves. The blue (upper) curve represents the 
bestfit of $h(\lambda)$ versus $\lambda$ to the form $h(\lambda)=\frac{\alpha }{ \beta + \lambda}$. 
The bestfit values turn out to be $\alpha=0.495$ and $\beta=0.841$. The red (lower) curve in Fig. \ref{io} is simply a graph of the 
function $f(\lambda)=\frac{1}{2(\lambda+1)}$.
Notice that our data (the blue curve) always lies above the red curve, but appears to asymptote rather 
accurately to the later at large $\lambda$. As explained in the introduction, the red curve is 
a theoretical lower bound for $h(\lambda)$. Based on our data we conjectrure that $h(\lambda)$ 
asymptotes to from above to the curve $\frac{1}{2(\lambda+1)}$ at large $\lambda$. It would be interesting 
(and may be possible) to establish this fact by a direct analytic study of the saddle point equations at 
large $\lambda$; however we leave this exercise for future work.

\begin{table}
\centering
\begin{tabular}{|c|c|c|c|c|c|c|c|c|c|}
\hline
    & $\lambda_2^f$&$\lambda_3^f$ &$\lambda_4^f$ &$\lambda_5^f$ & $\lambda_6f$ &$\lambda_7^f$ &$\lambda_8^f$ &$\lambda_{9}^f$ & $\lambda_{10}^f$ \\
\hline
$N=100$ corrected(linear interpolation) & 1.14 & 2.12 & 3.17 & 4.13 & 5.17 & 6.14 & 7.08 & 8.02 & 8.97 \\
\hline
$N=\infty$(linear interpolation) & 1.14 & 2.13 & 3.14 & 4.09 & 5.09 & 6.08 & 7.06 & 7.98 & 8.91 \\
\hline
$N=100$ corrected(fit to $\frac{\alpha}{\beta + \lambda}$) & 1.15 & 2.14 & 3.14 & 4.13 & 5.13 & 6.12 & 7.12 & 8.11 & 9.10 \\
\hline
$N=\infty$ (fit to $\frac{\alpha}{\beta + \lambda}$) & 1.14 & 2.13 & 3.12 & 4.10 & 5.09 & 6.08 & 7.07 & 8.06 &9.05 \\
\hline
\end{tabular}\label{tablelmdanf}
\caption{Values of $\lambda_n^f$. $Ne=100$ corrected is $Ne=100$ data corrected for $\frac{b(\lambda)}{c^2(\lambda) + (Ne)^2}$ }
\end{table}

\begin{figure}
\begin{center}
\includegraphics[width=100mm]{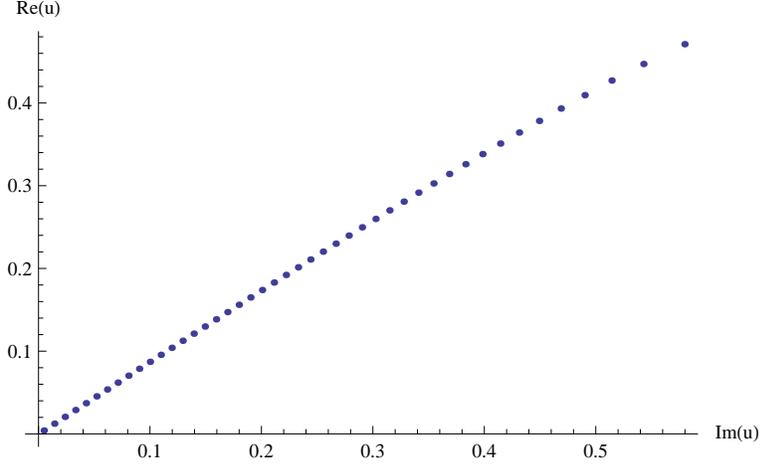}
\end{center}
\caption{Eigenvalue distribution for Ne=90 at $\lambda=11$.}
\label{fig:N90lmda11EVdist}
\end{figure}

As we have explained above, in order to obtain $h(\lambda)$ above we have to solve a saddle point 
eigenvalue equation. In Fig \ref{fig:N90lmda11EVdist} we present 
a scatter plot of the saddle point value of the eigenvalues at $\lambda=11$ and $Ne=90$. Note that
the eigenvalues appear to orient themselves along a curve that does not deviate too far from a straight 
line (in the complex plane). The magnitude of this `cut' is approximately 0.7469 and its angle with the 
real axis in the complex plane is approximately 39.69 degrees.

\begin{figure}
\begin{tabular}{cc}
\includegraphics[width=70mm]{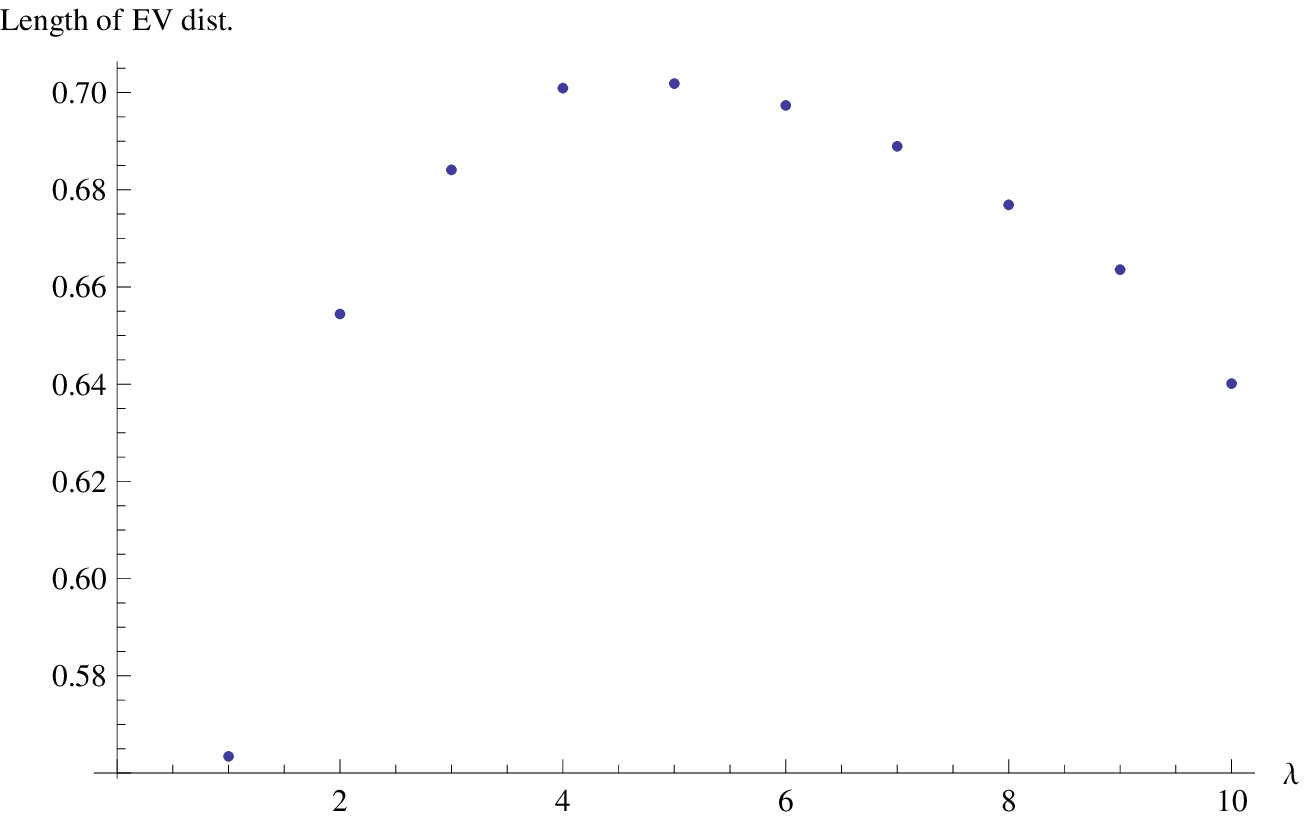} & \includegraphics[width=70mm]{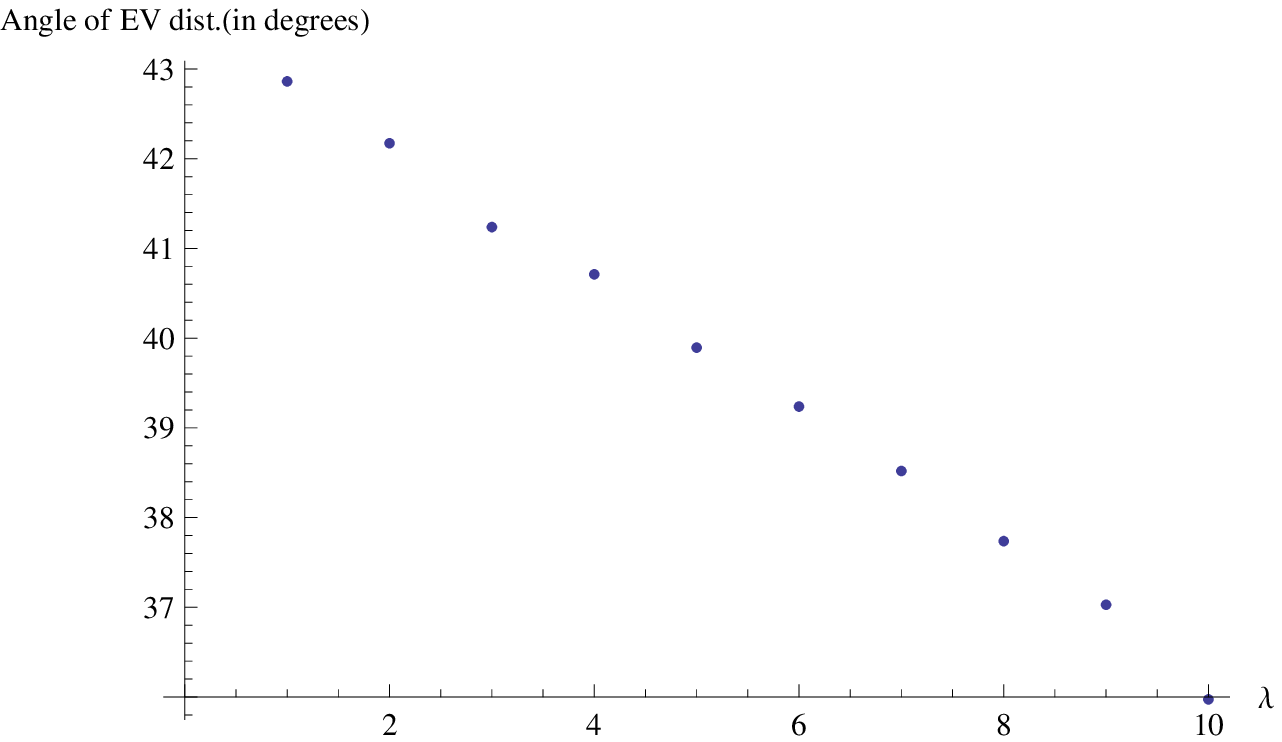}
\end{tabular}
\caption{Variation of size and angle(with real axis) of eigenvalue distribution for Ne=50 from $\lambda$=1 to 10.}
\label{fig:EVlengthangleN50}
\end{figure}

To study the variation of eigenvalue distribution with $\lambda$ we plot the length and angle of the 
eigenvalue distribution for $Ne=50$ from $\lambda$=1 to 10 in Fig. \ref{fig:EVlengthangleN50}. 
These plot show that the length of the eigenvalue distribution first increases with increasing $\lambda$ 
(we know from small $\lambda$ perturbation theory that this length scales like $\sqrt{\lambda}$ at extremely small $\lambda$)
but then reaches a maximum at $\lambda$ somewhere between 5 and 6, and then decreases again.  On the other hand the angle made by the 
cut continues to decrease upon increasing $\lambda$ (see the second graph in \ref{fig:EVlengthangleN50}. It would be interesting 
to continue this analysis to larger $\lambda$, but we leave that for future work. 

In the introduction we had defined $\lambda_n^f$ as the root of the equation
$$h(\lambda_n^f)=\frac{1}{2n}$$
Recall that consistency with known results appears to require that 
$$\lambda_n^f\geq n-1.$$ Note also that, were to be given by $h(\lambda)=\frac{1}{2(\lambda+1)}$ then $\lambda_n^f$ would saturate that inequality. 
In Table \ref{tablelmdanf} we present a table of $\lambda_{n}^{f}$ for $n \in (2, ...10)$ as obtained from our numerical data. In the second 
row we have computed $\lambda_n^f$ using the numerical function $h(\lambda)$ obtained by extrapolating our results to large $Ne$ as explained 
above equation \eqref{oo}. In the first row of Table \ref{tablelmdanf} we have computed $\lambda_n^f$ using $h(\lambda)$ obtained 
out of corrected  $Ne=100$ data, as explained around equation \eqref{corrected} in Appendix \ref{numerics}. The difference between these two 
results gives a crude estimate of the likely order of error in our results. Note that $\lambda_n^f$ starts approaching $\lambda_n^f\approx n-1$ at large $n$.
 This is, of course,
a restatement of the fact that the graph of $h(\lambda)$ appears to asymptote to 
$\frac{1}{2(\lambda+1)}$ from above at large $\lambda$.

To end this section let us recall the significance of $\lambda_n^f$. At $\lambda_2^f \approx 1.13$, the $R$-charge of $\phi$ is renormalized to $1/4$; 
$\Tr\,\phi^2$ then saturates the unitarity bound $1/2$ and becomes a free field 
and (presumably) decouples from the theory. There is a newly emergent $U(1)$ global symmetry which
rotates this free field. When $\lambda>\lambda_2^f$, there seems to be no clear argument
that ${\cal Z}$-minimization should give the correct superconformal $R$-charge due to possible
mixing of the $R$-charge with this accidental $U(1)$. It is nonetheless plausible that 
the naive result obtained by applying ${\cal Z}$-minimization still holds at large $N$, 
since only one field decouples from the theory at a time (on the other hand, there 
could also be decoupled topological sectors, which may well contribute to the ${\cal Z}$ function
at leading order in $N$). If we assume this, then at each $\lambda_n^f$ an operator $\Tr\,\phi^n$ becomes
free and decouples from the theory, at outlined in the introduction.

\section{Supersymmetric states of a theory with a single adjoint 
in the absence of a superpotential}

\subsection{Superconformal index}

\subsubsection{The free theory}\label{indexC}

We begin by analyzing the index of the free theory of a single ${\cal N}=2$ $U(N)$ adjoint chiral multiplet $\Phi$ (the $U(N)$ is gauged), in the large $N$ limit. Let the generator of the global flavour symmetry of this theory - corresponding 
to the rephasing of $\Phi$ - be denoted by $G$.   
$G$ is normalized so that the field $\phi$ has charge $\half$ under $G$. 
In this case the refined Witten index 
\begin{equation}
{\cal I}_+= \Tr \left[ (-1)^F x^{H+J}e^{-\beta(H-J-R)} y^G \right]
\end{equation} 
is easily computed in the free theory.
The letter index relevant to this computation is 
$${\cal I}_{L}= { x^{1\over 2} y^{1\over 2} -x^{3\over 2} y^{-{1\over 2}}\over1-x^2}.$$
Note that the two supersymmetric letters in the basic multiplet are 
$\phi$ and ${\bar \psi_+}$(from now on denoted as $\bar \psi$) and that the flavour charge of ${\bar \psi}$ is 
opposite to that of $\phi$. It follows that the refined index is given by 
\begin{equation}
{\cal I}_{+}=\prod_{n=1}^\infty \frac{1-x^{2n}}{(1-x^{n\over 2}y^{n\over 2})(1+x^{\frac{3n}{2}}
y^{-\frac{n}{2}}) }.
\end{equation} 
Once again it is possible to rewrite this index in terms of an index over 
single trace primaries as
\begin{equation}
{\cal I}_{+}
= \exp\left[\sum_{n=1}^\infty \frac{{\cal I}_{st}(x^n, y^n)}{n(1-x^{2n})}
\right],
\end{equation}
where ${\cal I}_{st}$ is found to be \footnote{Similar techniques were used to compute the Index of ${\cal
N}=4$ Yang Mills theory in \cite{Kinney:2005ej}, ABJM theory in \cite{Bhattacharya:2008bja}, and the
partition function of free gauge theories
in, for instance, \cite{Aharony:2003sx} }
\begin{equation}\begin{split} \label{compone}
&{\cal I}_{st}= (xy)^{1\over 2}+ xy + (xy)^{\frac{3}{2}}-
x^{\frac{3}{2}}y^{-{1\over 2}} + \frac{1}{1-(xy)^3}\bigg[ (xy)^2 -x^2 + (xy)^{\frac{5}{2}}-
x^{\frac{5}{2}} y^{1\over 2} + (xy)^3 - x^3 y   \\
&~~~~~~~+ (xy)^{\frac{7}{2}} - 
x^{7\over 2}y^{3\over 2} + 
(xy)^4-x^4 y^2+ (xy)^{\frac{9}{2}}
- x^{9\over 2}y^{5\over 2} \bigg] + \frac{1}{1-x^3y^{-1}} \left( x^{7 \over 2}y^{-{1 \over 2}} - x^{9 \over 2}y^{-{3 \over 2}}  \right).
\end{split}
\end{equation}
Note that \eqref{compone} receives contributions from (effectively) 
either two or three states at each energy level\footnote{Note that \eqref{compone} reduces to \eqref{simpone} below upon setting $y=1$.}. 
At every energy level 
we see the contribution of the chiral ring (in the form of $(xy)^E$ for 
every $E$). At every level we also, however, see the contribution of either
one or two additional ``particles''.  

\subsubsection{The large $N$ theory at finite 't Hooft coupling}

Now consider the ${\cal N}=2$ $U(N)$ CS theory at level $k$ with one adjoint matter field and no superpotential, in the 't Hooft limit.
As we have seen previously, the $R$-charge of this theory is renormalized 
as a function of $\lambda$. The renormalization of the $R$-charge is, 
really, more accurately a mixing of the $R$-charge and the flavour charge; 
the extent of this mixing varies with $\lambda$. 

The effect of this mixing on the superconformal index may be dealt with 
very simply. If we perform the replacement 
$$ y \rightarrow y x^{2h(\lambda)-1}$$
on all the formulas for the (gauged) free theory in the previous subsubsection, then they apply
to the interacting theory at arbitrary $\lambda$ (and infinite $N$).

\subsection{Conjecture for the supersymmetric spectrum at all couplings}

In this section we (conjecturally) enumerate all single trace primary 
operators that are annihilated by the supercharge $Q$ that was preserved 
in the superconformal index. This enumeration will allow us to 
enumerate the full single trace supersymmetric operator spectrum of our 
theory. We perform our enumeration of operators in the classical (though
nonlinear) theory; and assume without proof that the same result continues to 
hold in the full quantum theory. We initially ignore the effect of the 
renormalization of $R$-charge in this theory as a function of coupling;
as in the previous section, it will prove extremely easy to insert
this effect into our final answer right at the end.\footnote{The reason that the index is insufficient to completely characterize
the supersymmetric spectrum is that it is blind to Bose-Fermi pairs of 
supersymmetric states, whose contribution to the index cancel. The reader
may feel that such conspiratorial cancellations are unlikely, but that 
is far from the truth. In fact it is immediately clear on general grounds 
that our result for the index in the previous section must include some 
important cancellations. In order to see this recall
 that the $R$ current and stress tensor 
appear in the supersymmetry multiplet with quantum numbers $(\Delta=2, J=1, 
R=0, G=0)$. It follows that the contribution of this multiplet to the 
index is $-x^4 y^0$ : however such a term is absent in \eqref{compone}. 
It must be that the full susy spectrum of the theory includes a bosonic 
state whose contribution to the index cancels that of the stress tensor
multiplet. Below we will see in some detail how this works.}

We now describe the procedure we will use for the enumeration in this 
subsection in more detail. 
Every supersymmetric operator of the sort we seek must 
be built out of $\phi$, ${\bar \psi_+}$ and $D_{++}$ where 
$\phi$ and ${\bar \psi_+}$ both have $R$-charge $\half$ and the $+$ subscript 
denotes $SO(3)$ charge (these are the only letters with $\Delta=h+j$).
As we have explained above, the letters $\phi$ and ${\bar \psi_+}$ have 
flavour charge $\pm \half$ respectively. 
 
In the free theory any operator constructed out of these elements is 
annihilated by our special supercharge $Q$ (with $R$-charge unity and $SO(3)$ 
charge $-\half$). In the interacting theory, on the other hand,
while it continues to be true that $[Q, \phi]=\{Q, {\bar \psi} \}=0$
we now have   
$$[Q, D_{++} O] \sim [[\phi, {\bar \psi_+}], O]$$
As illustrated by this equation, the derivative $D$ carries the same 
$x$ and $y$ charges as the combination of letters $\phi {\bar \psi_+}$. 
Consequently a given term in the index counts an infinite number of distinct 
possible operator structures. For instance, at order $x^5y$ the 
cohomology potentially has operators of the form $\Tr\, (\phi ^4 {\bar \psi}^2)$,
$\Tr\, (\phi^3 D {\bar \psi})$ and $\Tr\, (\phi D^2 \phi)$. In any given charge sector we 
refer to the {\it number of derivatives} in the operator as its {\it level}. As we see
from the equation above, the operator $Q$ preserves charge but maps 
states of level $l$ to states of level $l-1$. At any given fixed value 
of the charge let the number of states 
in the free theory at level $l$ be $n(l)$. Let the number of states in the 
kernel of the action of $Q$ (those that are annihilated by $Q$) at level 
$l$ be denoted by $c(l)$. It follows that the number of states in $Q$ 
cohomology at level $l$ is given by $c(l)-(n(l+1) -c(l+1))$ (because 
$n(l+1) -c(l+1)$ gives the number of exact states at level $l$). 

The precise value of this cohomology, as defined above, contains some redundant
information. This is because a total $D_{++}$ derivative (outside the trace)
of any member of cohomology obviously itself belongs to the cohomology.  Such 
descendant elements of cohomology are trivial (they do not map to new 
particles under the AdS/CFT map) and should be removed from the analysis.
This is easy to do in a self consistent manner.

We have written a Mathematica
routine that computes the numbers $c(l)$ and $n(l)$ and hence the cohomology
at low orders. Our routine then proceeds to eliminate descendant. 
All the results we have obtained so far are consistent with the following 
conjectures for the structure of primary states in cohomology: 
\begin{itemize} 
\item The cohomology (obviously) contains only states with $x$ and $y$ 
quantum numbers of $\Tr\, (\phi^m {\bar \psi}^n)$ for positive $m$ and $n$. 
\item We conjecture that, at level zero, all such charges admit a unique
(conformal primary)
state in cohomology unless $m=0$ and $n$ is even. This state in cohomology 
may be thought of as the trace of the completely symmetric combination of 
${\bar \psi}$ and $\phi$. The exception ($m=0$ and $n$ even) is a consequence 
of the fact that $\Tr\, {\bar \psi}^{2k}$ vanishes because of the 
fermionic nature of ${\bar \psi}$.
\item We conjecture that, at level one, there also always exists a unique 
(conformal primary)
state in the cohomology subject to the following exceptions. There are no 
states at level one when $n=0$ or when $n=1$.\footnote{This statement is 
obvious when $n=0$ because then there do not exist any level one states 
with the right quantum numbers. Moreover when $n=1$ the unique 
level one state with the right quantum numbers is a descendant of 
$\Tr\, \phi^{m-1}$.} There are also no level one states when $m=0$ or 
when $m=1$ and $n$ is even.\footnote{When $m=0$ there are no level one 
states with the right quantum numbers. When $m=1$ and $n$ is even, the 
unique such state is a descendant of  $\Tr\, {\bar \psi}^{n-1}$. Of course 
no such statement can be (or is) true when $n$ is odd because 
$\Tr\, {\bar \psi}^{n-1}$ vanishes.}
\item We conjecture that the cohomology has no (conformal primary) 
states at levels higher than one.
\item That our conjecture is consistent with the index listed above as may 
be seen as follows. A pair of a level zero and level one state gives 
a vanishing contribution to the index. It follows that we have a contribution 
to the index only in those cases in which level zero and level one states 
are unpaired. According to our conjecture, unparied states occur 
at charges at which there exists a level one state but no level zero state. 
This occurs for states of the charges $\Tr\, \phi^m $, $\Tr\, (\phi^m {\bar \psi})$, 
$\Tr\, {\bar \psi}^{2m+1}$  and $\Tr\, ({\bar \psi}^{2m} \phi)$. This precisely accounts 
for the index computed above. 
\end{itemize} 

We summarize the conjecture described above in 
Table (\ref{tab:1chiralAdjWzero}). This is found to agree with the
index calculated for this theory as calculated in section \ref{indexC}. In that table we have also used the 
fact that every short representation of the ${\cal N}=2$ superconformal 
algebra has a unique (conformal primary) state in $Q$ cohomology
to read off the full spectrum of the short (or supersymmetric) single 
trace primary operators of the theory implied by our conjecture 
for $Q$ cohomology.
\begin{table}
\centering
\begin{tabular}{|c|c|c|c|}
\hline
Cohomology state & \# & ${\cal N}=2$ Primary & Allowed $h$ and $j$ \\
\hline
$(h,0,h,h)$ & $1$ & $(h,0,h,h)$ & $h \in{1\over 2}\mathbb{Z}^{+}$ \\
\hline
$(j+h,j,h,h-2j)$ & $1$ & $(j+h-\frac{1}{2},j-\frac{1}{2},h-1,h-2j)$ & $j,h\in {1\over 2}\mathbb{Z}^{+}, ~h\geq j$ \\
 & & & $h=j\in \mathbb{Z}$ not allowed. \\
\hline
$(j+h,j,h,h-2j+2)$ & $1$ & $(j+h-\frac{1}{2},j-\frac{1}{2},h-1,h-2j+2)$ & $h,j\in {1\over 2}\mathbb{Z}^{+},~ j\geq\frac{3}{2},~ h\geq j-\frac{1}{2}$ \\
\hline
$(2h+1,h+1,h,-2h)$ & $1$ &$(2h+\frac{1}{2},h+\frac{1}{2},h-1,-2h)$ & $h\in {1\over 2}\mathbb{Z}^{+}$ \\
\hline
\end{tabular}
\caption{${\cal N}=2$ primary content of single adjoint matter theory with zero superpotential. 
The second column stands for the multiplicity of the cohomology states.
The notation is $(\Delta,j,h,g)$ which respectively are scaling dimension,
spin, $R$-charge and $U(1)$ flavor charge. The flavor charges are normalized 
to be $\frac{1}{2}$ and $-\frac{1}{2}$ respectively. The results above 
apply when $h(\lambda)=\frac{1}{2}$. The results for the general case 
are obtained from the results of the table above by the replacement 
$\Delta \rightarrow \Delta -(1-2h(\lambda))g$, $ h \rightarrow 
h-(1-2h(\lambda))g$ for every primary in the table.}\label{tab:1chiralAdjWzero}
\end{table}

\section{Supersymmetric states of theories with a single adjoint field with 
nonzero superpotential}

\subsection{Space of theories}

One may construct several  superconformal field theories with a single 
chiral multiplet by perturbing the theory with no superpotential. 
The simplest theory of this sort may be constructed by perturbing 
the superpotential of the theory of the previous subsection with a ${\rm Tr}\,\Phi^4$ 
term. While this perturbation is marginal at zero $\lambda$, as we have 
explained above it is relevant at every finite $\lambda$ (this follows 
as $h(\lambda)< \half$ for all finite $\lambda$). 
The RG flow seeded by this operator may be argued to terminate 
at a new fixed point at which the coefficient $c$ of $\Tr\,\Phi^4$ in the 
superpotential is of order unity, in units in which a uniform factor of 
$k=\frac{N}{\lambda}$ sits outside the whole action.\footnote{The argument that the RG flow ends 
at a finite value of $c$ is simple. When $c\gg 1$, the gauge interaction 
in the theory is negligibly weak compared to the $\Tr \, \Phi^4$ interaction and 
may be ignored. The model is then effectively the Wess Zumino theory whose 
$\beta$ function towards the IR is known to be negative. Consequently 
the sign of the $\beta$ function flips from positive for small $c$ to 
negative at large $c$, and so must have a zero at $c$ of order unity.} 
This line of CFTs also reduces to the free theory as $\lambda \to 0$.
The $\Tr\,\Phi^4$ superpotential in this theory breaks the flavour symmetry 
of the zero superpotential theory down to $Z_4$. The requirement of the 
invariance of the superpotential under $R$ symmetry transformations 
forces the $R$-charge of this system to equal $\half$ at every value 
of the coupling constant $\lambda$. 

There exist no other lines of CFTs with this matter content that reduce 
to the free CFT in the limit $\lambda \to 0$. However there plausibly 
exist many other lines of superconformal fixed points that cannot be deformed
to the free theory. To start with the superpotential term $\Tr\, \Phi^3$ is relevant at all values 
of $\lambda$. At small $\lambda$ the  RG flow seeded by this term presumably 
terminates at a large $N$ and supersymmetric analogue of the Wilson Fisher 
fixed point. It thus seems plausible that the new fixed point exists at 
every value of $\lambda$. The $R$-charge of $\phi$ is fixed at $\frac{2}{3}$
at this fixed point. 

More exotically, as we have explained in the introduction, the fact that 
$h(\lambda)$ decreases without bound (as $\lambda$ is increased) plausibly 
suggests the existence of fixed points seeded by $\Tr\, \Phi^n$ induced 
RG flows, for all values of $n$, at sufficiently large $\lambda$. 
The $R$-charge of the operator $\phi$ is fixed at $\frac{2}{n}$ 
along these fixed lines of theories.

\subsection{Superconformal index of the theory with a $\Tr\, \Phi^4$ superpotential}\label{indexB}

Let us compute the Witten index 
 $$ \Tr (-1)^F x^{H+J}$$
for the theory with the $\Tr\, \Phi^4$ 
superpotential. 
The letter index of this 
theory is the Witten index of the sum of the $(\half, 0, \half)$ 
and $(\half, 0, -\half)$ representations. We find the single letter contribution\footnote{Note that this letter partition function is finite (and 
in fact evaluates to half) in the limit $x=1$.}  
$${\cal I}_{L}=\frac{x^\half}{1+x}.$$
It follows that the Witten index of the 
theory - in the large $N$ limit - is given by 

$${\cal I}_{+}= \prod_{n=1}^{\infty} \frac{1}{1-{\cal I}_{L}(x^n)} = 
                   \prod_{n=1}^\infty \frac{1+x^n}{1-x^{\frac{n}{2}}+x^n}.$$

The formula above gives the Witten index of the full theory (including 
all multi trace operators) in the large $N$ limit. It is interesting 
to inquire about the single trace index for the same theory. Now the 
full index is obtained from the single trace index by Bose exponentiation. 
Now the single trace index actually receives contributions both from 
single trace conformal primaries and single trace conformal descendants. 
It is of most interest to isolate the index over all single trace primaries
(as this gives the index over the particle spectrum of the dual theory). 
Let us define the index over single trace primaries as ${\cal I}_{st}$. 
It then follows that 
$${\cal I}_{+}= \prod_{n=1}^\infty \frac{1+x^n}{1-x^{\frac{n}{2}}+x^n}=
\exp\left[\sum_{n=1}^\infty \frac{{\cal I}_{st}(x^n)}{n(1-x^{2n})}
\right]$$
This equation completely determines ${\cal I}_{st}$. It is possible to 
show that 
\begin{equation}\begin{split} \label{simpone}
 {\cal I}_{st}&=x^{1\over 2} +x + x^{\frac{7}{2}}-x^{\frac{9}{2}}
+x^{\frac{13}{2}}-x^{\frac{15}{2}}+ x^{\frac{21}{2}} -x^{\frac{23}{2}}
+\ldots \\
&=x^{1\over 2}+x+\frac{x^{7\over 2}}{1+x+x^2}. \\
\end{split}
\end{equation}
Note that the spectrum of single traces is periodic at large enough energies, 
with one new boson and one new fermion being added at energy intervals of 
three.

\subsection{Conjecture for the supersymmetric spectrum of the 
theory with a $\Tr\, \Phi^4$ superpotential}\label{superphi4}

In this subsection we present a conjecture for the partition function 
over single trace supersymmetric operators in the theory with a 
$\Tr\,\Phi^4$ term in the superpotential. Our method is the same as that 
employed in the previous section: we list the classical 
cohomology of the special supercharge $Q$ (the supercharge that annihilates 
the superconformal index). 

The only difference between the cohomology of $Q$ in this section, and the 
cohomology of $Q$ in the absence of a superpotential (computed in the 
previous section) is that $Q$ no longer annihilates 
$\bar{\psi}$, but we instead have 
$$Q \bar{\psi} \sim \phi^3$$. 

As in the previous section the cohomology of interest can be calculated 
separately for operators with distinct values of $\Delta+j$. At any given 
value of $\Delta+j$, we grade operators by their {\it level}, defined to be 
{\it twice the angular momentum} of the operator in question. As in the 
previous section, we work with the full set of single trace operators 
constructed out of $\phi$, ${\bar \psi}$ and $D_{++}$. States at level $0$ are
constructed entirely out of $\phi$'s, and are all annihilated by 
$Q$. There is a unique level zero state at every value of $\Delta+j$. States at level $1$ are built out 
of several $\phi$'s but one ${\bar \psi}$. There is also a unique 
state at level $1$ at every value of $\Delta+j$. 

The situation is more complicated at higher levels. As in the previous section,
we have written a Mathematica code to compute the cohomology for all states
with $\Delta+j \leq \frac{23}{2}$. As in the previous subsection, our 
Mathematica routine automatically removes the contributions of descendants. 
Our results are consistent 
with following conjecture. 
\begin{itemize}
 \item  At every $\Delta+j$ the number of (conformal primary) 
states in cohomology is either $0$ or $1$.
 \item  When $\Delta + j = {1 \over 2}$ there exists a single conformal primary 
state in cohomology at level $0$ and no states at any other level.
\item When $\Delta + j = 1$ there exists a single conformal primary state 
in cohomology at level $0$ and no state at any other level.
 \item When $ \frac{3}{2} \geq \Delta +j \geq 3$ there are no conformal primary states in cohomology.
\item When $\Delta + j = {6k + 7 \over 2}$ for $k=0,1 \ldots \infty$
there is a single conformal primary state in cohomology at 
level $2(k+1)$
\item When $\Delta + j = {6k + 8 \over 2}$ for $k =0,1 \ldots \infty$ 
there are two conformal primary states in cohomology. The first occurs at 
level $2(k+1)$ while the second occurs at level $2(k+1)+1$.
\item When $\Delta + j = {6k + 9 \over 2}$ for $k=0,1 \ldots \infty$ 
there is a single conformal primary state in cohomology at level $2(k+1)+1$
\item There are no conformal primary states in cohomology when 
$\Delta+j =\frac{6k +10}{2}$ or $\frac{6k +11}{2}$
or $\frac{6k +12}{2}$ (for $k = 0,1 \ldots \infty$)
\end{itemize}

This conjecture has been summarized in Table (\ref{tab:table2adjsupphi4}) in 
terms of charges $(\Delta,j,h)$ of the cohomology states. This is found to agree with the
index calculated for this theory in section \ref{indexB}. As states 
of the form $(j+h,j,h)$ (for $j>0$) are descendants of the superconformal 
primary $(j+h-{1 \over 2},j-{1 \over 2},h-1)$, the corresponding superconformal 
primaries are also listed out in the table. 
\begin{table}
\centering
\begin{tabular}{|c|c|c|c|}
\hline
Cohomology states & Multiplicity  & Protected primaries \\
\hline
$({1 \over 2},0,{1 \over 2})$ & $1$ & $({1 \over 2},0,{1 \over 2})$	 \\
\hline
$(1,0,1)$ & $1$ & $(1,0,1)$\\
\hline
$(2k+{5 \over 2},k+1, k + {3 \over 2})$ & $1$ & $(2k + 2,k+{1 \over 2}, k+ {1 \over 2})$\\
\hline
$(2k+3,k+ 1,k+2) $ & $1$ & $(2k+{5 \over 2},k+{1 \over 2},k+1)$ \\
\hline
$(2k+{5 \over 2},k + {3 \over 2},k+1 )$ & $1$ & $(2k+2,k+1,k)$ \\
\hline
$(2k+3,k + {3 \over 2},k + {3 \over 2 } )$ & $1$ & $(2k+{5 \over 2},k+1,k+ {1 \over 2})$\\
\hline
\end{tabular}
\caption{Supersymmetric spectrum for ${\cal N}=2$ single adjoint with superpotential 
$\Tr\,\Phi^4$, $k \in 0,\mathbb{Z}^+ $. The notation again is $(\Delta,j,h)$.}\label{tab:table2adjsupphi4}
\end{table}

\subsection{Conjecture for the supersymmetric spectrum of the
theory with a $\Tr\,\Phi^3$ superpotential}\label{superphi3}

In this subsection we present a conjecture for the classical
cohomology of the particular supercharge $Q$ for the theory with a $\Tr\,\Phi^3$ 
superpotential. A theory with such a superpotential is always strongly 
coupled, and so we cannot use a free calculation to compute its superconformal
index. However, the computation of $Q$ cohomology in such a theory 
is easily performed (under the same assumption of the previous subsection, i.e.
that it is sufficient to use the classical supersymmetry transformation rules), 
using the same method as in the previous subsection. 

The difference between the computation of this subsection and that of the 
previous one is as follows. In this case the action of $Q$ on $\bar \psi$ 
is given by $Q \bar{\psi} \sim \phi^2$. Further the fact that we have a 
superconformal theory with a $\Phi^3$ superpotential forces the following
charge assignments: the $R$-charges of $\phi,\bar{\psi}$ are 
${2 \over 3},{1 \over 3}$ respectively, while the scaling dimensions 
of $\phi,\bar{\psi}$ as determined by the BPS relation  $\Delta = j +h$, 
are ${2 \over 3},{5 \over 6}$ respectively.

As in the previous subsection, the cohomology must be computed 
separately at every value of  $\Delta+j$. At any given
value of $\Delta+j$, we grade operators by their {\it level} defined to be
{\it twice the angular momentum} of the operator in question. As in the
previous subsection, we work with the full set of single trace operators
constructed out of $\phi$, ${\bar \psi}$ and $D_{++}$. States at level $0$ are
constructed entirely out of $\phi$'s, and are all annihilated by
$Q$. 

The calculation is formulated exactly as in the previous section.
This calculation has been done up to $\Delta+j=12$ on Mathematica. Our 
results are consistent with following conjecture: 
\begin{itemize}
 \item  At $\Delta+j={2 \over 3}$, there is a single (conformal primary) state in the cohomology, and it occurs at level $0$.
 \item  For every allowed value of $\Delta+j$ and at any given level, the 
number of conformal primary states in the cohomology is either $0$ or $1$.
 \item  The full conformal primary cohomology content of the theory is 
summarized as follows:
\begin{itemize}
\item[a.] $\Delta + j = {4 \over 3}(2k+1)$ one state at level $2k+1$, 
for $k=1,2,3\ldots$
\item[b.] $\Delta + j = {4 \over 3}(2k+{1 \over 2})$ one state at level 
$2k$ for $k=1,2,3\ldots$
\end{itemize}
\end{itemize}

This conjecture has been summarized in Table (\ref{tab:table2adjsupphi3}) in
terms of charges $(\Delta,j,h)$ of the cohomology states. This is found to agree with the
index calculated for this theory in section \ref{localization}, by making the substitutions 
as given in (\ref{relation}). As states
of the form $(j+h,j,h)$ (for $j>0$) are descendants of the superconformal
primary $(j+h-{1 \over 2},j-{1 \over 2},h-1)$, the corresponding superconformal
primaries are also listed out in the table.
\begin{table}
\centering
\begin{tabular}{|c|c|c|c|}
\hline
Cohomology states & Multiplicity  & Protected primaries \\
\hline
$({2 \over 3},0,{2 \over 3})$ & $1$ & $({2 \over 3},0,{2 \over 3})$     \\
\hline
$(\frac{5k+2}{3},k, \frac{2}{3}(k+1))$ & $1$ & $(\frac{5k}{3}+\frac{1}{6},k-{1 \over 2}, \frac{2k-1}{3})$\\
\hline
$(\frac{5}{6}(2k+1),k+\frac{1}{2},\frac{2k+1}{3}) $ & $1$ & $(\frac{5k+1}{3},k,\frac{2}{3}(k-1))$ \\
\hline
\end{tabular}
\caption{Supersymmetric spectrum for ${\cal N}=2$ single adjoint with superpotential
$\Phi^3$, $k \in \mathbb{Z}^+ $. The notation is $(\Delta,j,h)$}\label{tab:table2adjsupphi3}
\end{table}

\section{Supersymmetric states in the theory with 
two adjoint fields and vanishing superpotential}

The superconformal theory with two adjoints and no superpotential 
has a $U(2)$ flavour symmetry, realized as the rotation of the chiral 
multiplets $\Phi_1$ and $\Phi_2$ as a doublet. We denote the two 
$U(1)$ Cartan charges of this $U(2)$ by $G_1$ and $G_2$. 
Our conventions are that the field $\phi_1$ has 
charges $(G_1, G_2)=(1,0)$ while the charges of $\phi_2 $ are 
$(0, 1)$. We compute the refined Witten index defined by 
\begin{equation}
{\cal I}= \Tr \left[ (-1)^F x^{H+J} y_1^{G_1} y_2^{G_2} \right].
\end{equation} 
The letter index is given by 
$${\cal I}_{L}=\frac{x^{\frac{1}{2}}(y_1+y_2)-x^{\frac{3}{2}}(y_1^{-1}
+y_2^{-1})}{1-x^2}. $$
As usual the full multitrace index of the free theory is given by 
$${\cal I}_{+}=\prod_{n=1}^\infty \frac{1}{1-{\cal I}_L(x^n, y^n)}.$$

This index captures an exponential growth of density of supersymmetric states.  In order to see that this must be the case,
note that the number of chiral primaries at level $L$ is of order 
$2^L$. As there are no fermionic states with the quantum numbers of 
the chiral primaries, it is impossible for this contribution to be cancelled 
in the full index. It follows that the index, in this state, receives 
contributions from an exponentially growing number of states. 

As an exponentially growing spectrum of supersymmetric states is presumably
rather difficult to characterize more precisely, we postpone further 
analysis of this case to future work, and turn, instead, to the study 
of theories with two chiral multiplets and a superpotential.

\section{Supersymmetric states in the ${\cal N}=3$ theory with 
two adjoint chiral multiplets (i.e. one adjoint hypermultiplet)}

Of all the possible superpotential deformations of the zero superpotential 
theory with two adjoint chiral multiplets, the deformation  
$W=\alpha \Tr\, [\Phi_1,\Phi_2]^{2}$ has a special role. As we have explained 
above, this deformation is relevant at nonzero $\lambda$. The RG flow 
seeded by this deformation has a fixed point at $\alpha=\frac{2\pi}{k}$; 
at this value of the coefficient, the theory is conformally invariant; 
further its supersymmetry is enhanced to ${\cal N}=3$ and the theory 
enjoys invariance under the full ${\cal N}=3$ superconformal algebra. 

${\cal N}=3$ superconformal symmetry forces the theory to have an $SU(2)$ $R$ symmetry; 
$\phi_1$ and $\phi_2^*$ transform as a doublet under this symmetry.  
$\phi_1$ and $\phi_2$ each have $R$-charge $\half$ under the canonical $U(1)$ 
subgroup of this $R$ symmetry group. Note of course that the value 
of this $R$-charge is protected by $SU(2)$ representation theory, and cannot
be renormalized as a function of $\lambda$.  

In addition the fact that the superpotential is proportional to 
$\Tr\, (\epsilon^{ij} \Phi_i \Phi_j)^2$ reveals that the superpotential 
deformation preserves an $SU(2)$ flavour subgroup of the $U(2)$ 
flavour isometry group of the theory without a superpotential. 

In this section we will compute the supersymmetric spectrum of this theory
at large $N$. 

\subsection{Superconformal index}\label{sec:indexA}

As the ${\cal N}=3$ theory reduces to a free theory at $\lambda=0$, its 
superconformal index is easily computed. The superconformal index is defined 
by 
\begin{equation}
{\cal I}= \Tr \left[ (-1)^F x^{H+J} y^G \right]
\end{equation} 
where $G$ is the $U(1)$ component of the $SU(2)$ flavor group (under which 
$\phi_1$ has charge $\frac{1}{2}$ and $\phi_2$ has charge $-\frac{1}{2}$). 
The relevant letter index is given by 
$${\cal I}_{L}=\frac{x^{\half}y^{\half}-x^{\frac{3}{2}}y^{-\half}+x^{\half}y^{-\half}-
x^{\frac{3}{2}}y^{\half}}{1-x^2}=
\frac{x^{\half}(y^{\half} +y^{-\half})}{1+x},$$
so that the index over the theory is 
$${\cal I}_{+}=\prod_{n=1}^\infty \frac{1+x^n}{(1-x^{\frac{n}{2}}y^{\frac{n}{2}})
(1-x^{\frac{n}{2}}y^{-\frac{n}{2}})}.$$
As in previous sections, 
it is possible to rewrite this index in terms of an index over 
single trace primaries as
\begin{equation}
{\cal I}_{+}=\prod_{n=1}^\infty \frac{1+x^n}{(1-x^{\frac{n}{2}}y^{\frac{n}{2}})
(1-x^{\frac{n}{2}}y^{-\frac{n}{2}})}=
\exp\left[\sum_{n=1}^\infty \frac{{\cal I}_{st}(x^n, y^n)}{n(1-x^{2n})}
\right].
\end{equation}
We find 
\begin{equation}\begin{split} \label{comonen}
 &{\cal I}_{st}= x^{1\over 2} (y+\frac{1}{y}) +
x(1+y^2 +\frac{1}{y^2}) + x^{\frac{3}{2}}(y^3+\frac{1}{y^3})
+ x^2(y^4+\frac{1}{y^4}) \\
&~~~~~~+ \sum_{n=5}^\infty x^{\frac{n}{2}}\left(y^n+\frac{1}{y^n} 
-y^{n-4}-\frac{1}{y^{n-4}} \right).
\end{split}
\end{equation}
This simple result describes a spectrum with $4$ states - two bosons and two 
fermions - at every energy higher than a minimum value. Note that 
\eqref{comonen} reduces to \eqref{indcru} upon setting $y$ to unity. 

\subsection{Supersymmetric cohomology}\label{n3theories}

We now proceed to compute the single trace supersymmetric cohomology of 
the ${\cal N}=3$ theory.\footnote{As in the previous subsubsection, we know that there must exist 
states that contribute to the partition function but are invisible in the 
index simply from the observation that the contribution of the stress tensor
multiplet -$x^4 y^0$ - is not visible in \eqref{comonen}.}
We are instructed to count traces built out of $D_+^n \phi_i$ and 
$D_+^n {\bar \psi}_i$ ($i = 1,2$). We are only interested in $Q$ 
cohomology, where the action of $Q$ on the basic fields is given by 
\begin{equation} \label {Qcohom} \begin{split}
&Q \phi_i=0, \\
&Q {\bar \psi_i} = [\phi_i , [\phi_1, \phi_2]], \\
&Q [D_{++}, \, \cdot\, ] = \left[ [\phi_1,{\bar \psi_1}]+ [\phi_2, {\bar \psi_2}], \,\cdot \,\right].
\end{split}
\end{equation}
As in previous subsections, we have explicitly enumerated this cohomology 
(with the help of Mathematica) at low quantum numbers, and used the 
results of this numerical experiment to suggest a relatively simple conjecture
for the conformal primary content of this cohomology. As in previous 
subsections, each conformal primary member of cohomology implies the existence
of a single short ${\cal N}=2$ superconformal representation. Unlike the 
situation in previous sections, however, this spectrum has to satisfy an 
additional consistency check, as ${\cal N}=2$ representations must group 
together into ${\cal N}=3$ representations. Our conjecture for $Q$ 
cohomology passes this consistency check, and leads us to 
conjecture that the short supersymmetric operator content of the theory 
is given as in Table (\ref{tab:N3spctrm}). This is found to agree with the
index calculated for this theory in section \ref{sec:indexA}. (Our notation for quantum numbers of 
states as well as primaries is 
$(\Delta, j, h, g)$ where $g$ is the $SU(2)$ flavour charge.)
\begin{table}
\centering
\begin{tabular}{|c|c|c|c|c|}
\hline
Cohomology states & Multiplicity & ${\cal N}=2$ Primary & ${\cal N}=3$ Primary & Allowed $h$ \\
(${\cal N}=2$ quantum numbers) & & & & \\
\hline
$(h,0,h,h)$ & $1$ & $(h,0,h,h)$ & $(h,0,h,h)$ & $h \in {1\over 2}\mathbb{Z}^+$ \\
\hline
$(h+\half,\half,h,h)$ & $1$ & $(h,0,h-1,h)$ &  &$h \in {1\over 2}\mathbb{Z}^+$ \\
\hline
$(h+\frac{3}{2},\half,h+1,h)$ & $1$ & $(h+1,0,h,h)$ & $(h+1,0,h,h)$ & $h \in {1\over 2}\mathbb{Z}^+$ \\
\hline
$(h+2,1,h+1,h)$ & $1$ & $(h+\frac{3}{2},\half,h,h)$ &  & $h \in {1\over 2}\mathbb{Z}^+$ \\
\hline
$(h+1,1,h,0)$ & $1$ & $(h+\half,\half,h-1,0)$ & $(h+\half,\half,h-1,0)$ & $h \in \mathbb{Z}^+$ \\
\hline
$(h+\frac{3}{2},\frac{3}{2},h,0)$ & $1$ & $(h+1,1,h-1,0)$ &  & $h \in \mathbb{Z}^+$ \\
\hline
\end{tabular}
\caption{Supersymmetric spectrum for $2$ chiral adjoints at ${\cal N=3}$ fixed point with $SU(2)$ 
flavor symmetry. The flavor charges for $\phi$ and $\bar{\psi}$ are normalized to be $\half$ and 
$-\half$ respectively}\label{tab:N3spctrm}
\end{table}

A striking feature of this conjectured supersymmetric spectrum is that it 
includes no states with spin $\geq 2$. This is unlike all the other 
supersymmetric spectra we have computed in this paper, and suggests that 
the ${\cal N}=3$ theory might admit a supergravity-like dual description 
at large $\lambda$. 

As the full global symmetry group of our theory is $SU(2) \times SU(2)$ it 
is tempting to conjecture that the supergravity description in question 
is obtained by a compactification of a $7$-dimensional supergravity on $S^3$. 
Indeed the states in the first $2$ rows (of the second last column) 
of Table (\ref{tab:N3spctrm}) 
have $SU(2) \times SU(2)$ quantum numbers that are strongly reminiscent 
of scalar and vector spherical harmonics on $S^3$. However the 
states in the last line of this table do not appear to fit well into this 
pattern; as the difference between the two $SU(2)$ quantum numbers of these 
states grows without bound; this never happens for $S^3$ spherical harmonics 
for states with a fixed (or bounded) value of spin. For this reason 
we are unsure whether our results for the supersymmetric spectrum of this 
theory are consistent with a possible dual description in terms of a higher 
dimensional supergravity theory. We leave further discussion of this question
to future work.

\section{Marginal ${\cal N}=2$ deformations of the ${\cal N}=3$ theory}

It was demonstrated in \cite{Green:2010da} that the infinitesimal 
manifold of exactly marginal deformations of a given SCFT has a very 
simple characterization.\footnote{In the perturbative regime, a global characterization of the ``conformal manifold"
was given in \cite{Chang:2010sg}.} This space is simply given by modding out the 
space of marginal (but not necessarily exactly marginal) classical 
deformations by the complexified action of the global (non $R$) symmetry group 
${\cal G}_{C}$ of the theory. 

The space of marginal scalar deformations of the ${\cal N}=2$ theory 
was worked out in the previous section. At the level of the superpotential
it is given by operators $\Tr (\Phi^{a_1}\Phi^{a_2}\Phi^{a_3}\Phi^{a_4})$
with the indices $a_1 \ldots a_4$ completely symmetrized. These operators 
transform in the 5 dimensional (spin $2$) representation of the global 
symmetry group $SU(2) \sim SO(3)$. In colloquial terms they constitute 
a complex traceless symmetric $3 \times 3$  matrix $M$ on which 
complexified $SO(3)$ transformations $O$ act according to the law 
$$M \rightarrow O M O^T$$
The classification of all $3 \times 3 $ matrices $M$ that are inequivalent 
under this transformation law is well studied \footnote{We thank A. Mukherjee
for discussions on this topic}. Generic complex symmetric 
matrices $M$ can be diagonalized by the action of $O$; consequently 
generic exactly marginal deformations of the ${\cal N}=3$ theory 
are in one to one correspondence with complex diagonal traceless 
$3 \times 3 $ matrices and are labeled by two complex eigenvalues. 
In addition to the generic case, however, there exist two special 
classes of matrices $M$ that cannot be diagonalized. Instead, one of them can be 
put in the form \cite{book}
\begin{equation}
\left( \begin{array}{ccc} \lambda_1 + \frac{i}{2} &    \frac{1}{2}      &    0 \\
                            \frac{1}{2}       & \lambda_1-\frac{i}{2} &   0 \\
                            0       &                    0      & -2\lambda_1
       \end{array}
\right)\label{nongeneric1}
\end{equation}
This gives a new one parameter set of exactly marginal deformations of the 
${\cal N}=3$ theory. The second possible form is 
\begin{equation}
\left( \begin{array}{ccc} 0 &          {1+i \over 2}     &      0 \\
                        \frac{1+i}{2}       &0 &            {1-i \over 2} \\
                            0       &      {1-i \over 2}       & 0
       \end{array}
\right)\label{nongeneric2}
\end{equation}

Let us first focus on the generic marginal deformations of the 
${\cal N}=3$ theory. The generic deformation can be put in the form 
\begin{equation}
W=\lambda_{1}\Tr\, (\Phi_1^2 + \Phi_2^2)^2 - \lambda_2 \Tr\, (\Phi_1^2 - \Phi_2^2)^2 - 
(\lambda_1 + \lambda_2)\Tr\, (\Phi_1\Phi_2 + \Phi_2\Phi_1)^2
\label{eq:N=3deformW}
\end{equation}
which is better written as 
\begin{equation}\label{genericsup}
W=\lambda_1 \Tr\, \left[ (\Phi_1 +\Phi_2)^2(\Phi_1 -\Phi_2)^2\right] -\lambda_2 \Tr\, \left[(\Phi_1 +i\Phi_2)^2(\Phi_1 -i\Phi_2)^2\right]
\end{equation}
At a generic point on the conformal manifold the flavor symmetry is completely
broken. For special values of $\lambda_{1,2}$, listed in 
Table (\ref{tab:conformalmanifold}), a $U(1)$ flavor symmetry is restored. 
\begin{table}
\centering
\begin{tabular}{|c|c|c|}
\hline
Points on conformal manifold & Flavour symmetry & Charges \\
\hline
$\lambda_1 =0=  \lambda_2$ & $SU(2)$ & $\Phi_1 ,\Phi_2$ form a doublet \\
\hline
$\lambda_1 = \lambda_2 \neq 0$ & $U(1)$ & $\Phi_1$: $1$, $\Phi_2$: $-1$ \\
\hline
$\lambda_1=0, \lambda_2 \neq 0$ & $U(1)$ & $\Phi_1 +i \Phi_2$: 1, $\Phi_1 - i \Phi_2$: $-1$ \\
\hline
$\lambda_1 \neq 0, \lambda_2 = 0$ & $U(1)$ & $\Phi_1 + \Phi_2$: 1, $\Phi_1 - \Phi_2$: $-1$ \\
\hline
\end{tabular}
\caption{Symmetries on the conformal manifold}\label{tab:conformalmanifold}
\end{table}
The space of generic exactly marginal deformations of the ${\cal N}=3$ 
theory is a two complex dimensional manifold.

Next, for the nongeneric deformation as parametrized by \eqref{nongeneric1}, the 
superpotential deformation is 
\begin{equation}\label{nongenericsup1}
 W = \lambda_2 \left(  \Tr\,( \Phi_2^4 )- \lambda_1 \Tr\,( \Phi_1 \Phi_2 \Phi_1 \Phi_2 )\right)
\end{equation}
There is no flavour symmetry in this case.

As for the nongeneric deformation as parametrized by \eqref{nongeneric2}, the 
superpotential deformation is 
\begin{equation}\label{nongenericsup2}
 W = \lambda_2 \left( (1+i) \Tr\, ( \Phi_1^4 - \Phi_2^4 ) + 2 (1-i) \Tr\, ( \Phi_1^3 \Phi_2 
-\Phi_1 \Phi_2^3 )\right)
\end{equation}
There is no flavour symmetry in this case also.

The $U(1)$ $R$-charge of chiral multiplets in these theories is fixed to 
$\frac{1}{2}$ (merely by the observation that each of these theories has 
a space of exactly marginal deformations, labeled by different quartic 
superpotential deformations). 

\subsection{Superconformal index of these theories}

The Witten index of the deformed theories that preserve $U(1)$ flavor 
symmetry is simply identical to the index of the ${\cal N}=3$ theory 
determined in the previous section.

The Witten index of deformed theories that break the $U(1)$ symmetry 
is also equal to to that of the ${\cal N}=3$ theory, but with $y=1$ 
(as there is no flavor charge with respect to which states can be weighted in this 
case). We find the remarkably simple result 
\begin{equation}\label{indcru}
 {\cal I}_{st} =2 x^{1\over 2} +3x + 2 x^{\frac{3}{2}} + 2x^2.
\end{equation}
Thus the single trace index sees a total of only $9$ conformal primaries!

\subsection{Conjecture for the supersymmetric cohomology}
For cohomology calculations we consider the general marginal superpotential 
as given in 
(\ref{eq:N=3deformW}) along with the original ${\cal N}=3$ superpotential with a coefficient 
normalized to one. 
With this superpotential the action of the special supercharge on the 
basic letters is as follows
\begin{eqnarray}
Q(\bar{\psi_1})&=&-\phi_1\phi_2^2 - \phi_2^2\phi_1 + 2\phi_2\phi_1\phi_2 + (\lambda_1-\lambda_2)\phi_1^3 - 
4(\lambda_1+\lambda_2)(\phi_2\phi_1\phi_2) \\
Q(\bar{\psi_2})&=&-\phi_2\phi_1^2 - \phi_1^2\phi_2 + 2\phi_1\phi_2\phi_1 + (\lambda_1-\lambda_2)\phi_2^3 - 
4(\lambda_1+\lambda_2)(\phi_1\phi_2\phi_1) \\
Q[D_{++}, \,\cdot\,] &=& \left[ [\phi_1,{\bar \psi_1}]+ [\phi_2, {\bar \psi_2}],  \,\cdot\, \right]
\label{eq:QactionN=3Wdeform}
\end{eqnarray}

Although it is not obvious, it (experimentally) appears that the cohomology
is largely independent of the complex ratio $\frac{\lambda_1}{\lambda_2}$
but instead depends only on whether the flavour symmetry of the theory 
is broken or restored. Using the methods described in earlier sections 
we have generated data that suggests that the cohomology of these 
theories takes the following form.

For the generic ${\cal N}=2$ deformations \eqref{genericsup}, the conformal primary states
in the cohomology, and the corresponding ${\cal N}=2$ superconformal representations,
are given, in the case that a $U(1)$ flavor symmetry is preserved 
(($\lambda_1=0, \lambda_2\neq 0$), ($\lambda_1\neq 0, \lambda_2 = 0$),
($\lambda_1 = \lambda_2 \neq 0$)) in Table (\ref{tab:N2U(1)spectrum}).

\begin{table}
\centering
\begin{tabular}{|c|c|c|c|}
\hline
Cohomology states & Multiplicity & ${\cal N}=2$ Primary & Allowed $h$ \\
(${\cal N}=2$ quantum numbers) & & & \\
\hline
$(h,0,h,h)$ & $1$ & $(h,0,h,h)$ &  \\
$(h,0,h,-h)$ & $1$ & $(h,0,h,-h)$ & $h \in {1\over 2}\mathbb{Z}^+$ \\
$(h,0,h,0)$ & $1$ & $(h,0,h,0)$ &  \\
\hline
$(\frac{3}{2},\half,1,0)$ & $1$ & $(1,0,0,0)$ & \\
\hline
$(h+\frac{3}{2},\half,h+1,h)$ & $1$ & $(h+1,0,h,h)$ &  \\
$(h+\frac{3}{2},\half,h+1,-h)$ & $1$ & $(h+1,0,h,-h)$ & $h \in {1\over 2}\mathbb{Z}^+$ \\
$(h+\frac{3}{2},\half,h+1,0)$ & $1$ & $(h+1,0,h,0)$ &  \\
\hline
$(h+1,1,h,0)$ & $1$ & $(h+\frac{1}{2},\half,h-1,0)$ &  $h \in {1\over 2}\mathbb{Z}^+$ \\
\hline
$(h+\frac{1}{2},\frac{3}{2},h-1,0)$ & $1$ & $(h,1,h-2,0)$ & $h \in 2\mathbb{Z}^+$ \\
\hline
\end{tabular}
\caption{Supersymmetric spectrum for $2$ chiral adjoints at ${\cal N}=2$ fixed point with $U(1)$ flavor symmetry. 
Notation is $(\Delta,j,h,g)$ with $g$ normalized to be $\half$ for $\phi$.}\label{tab:N2U(1)spectrum}
\end{table}

On the other hand the conformal primary cohomology and the 
superconformal  primary content of ${\cal N}=2$ cases when there 
is no flavour symmetry i.e. when $\lambda_1\neq\lambda_2$ and $\lambda_1,\lambda_2\neq 0$ 
is given in Table (\ref{tab:N2nosymmetry}). Also for the first non generic ${\cal N}=2$ deformation \eqref{nongenericsup1} with
$\lambda_1 \ne 0$ and for the second nongeneric ${\cal N}=2$ deformation 
\eqref{nongenericsup2} the superconformal  primary content is the same as in 
Table (\ref{tab:N2nosymmetry}). 

\begin{table}
\centering
\begin{tabular}{|c|c|c|c|}
\hline
Cohomology state & Primary & Multiplicity & Allowed values \\
\hline
$(\frac{k}{2},0,\frac{k}{2})$ & $(\frac{k}{2},0,\frac{k}{2})$ & $3$ if $\frac{k}{2}$ is odd, & $k \in \mathbb{Z}^+$ \\
 & & 2 if else & \\
\hline
$(\frac{k+1}{2},\half,\frac{k}{2})$ & $(\frac{k}{2},0,\frac{k}{2}-1)$ & $3$ if $\frac{k}{2}$ is even, & $k \in \mathbb{Z}^+$ \\
 & & $1$ if $\frac{k}{2}$ is odd, & \\
 & & $2$ if $\frac{k}{2}$ is half an odd integer &  \\
\hline
$(2k+\half,\frac{3}{2},2k-1)$ & $(2k,1,2k-2)$ & $1$ & $k \in \mathbb{Z}^+$ \\
\hline
\end{tabular}
\caption{Cohomology and primary content of theories with no flavor symmetry. The notation 
is ($\Delta,j,h$), since there is no flavor symmetry.}\label{tab:N2nosymmetry}
\end{table}

For the first non generic ${\cal N}=2$ deformation \eqref{nongenericsup1}, 
if $\lambda_1 = 0$,  the superconformal  primary content is given in Table (\ref{tab:N2spctrm0eig})

\begin{table}
\centering
\begin{tabular}{|c|c|c|c|c|}
\hline
Cohomology states & Multiplicity & ${\cal N}=2$ Primary & Allowed $k$\\
\hline
 $({1 \over 2},0,{1 \over 2})$& $2$ & $({1 \over 2},0,{1 \over 2})$&\\
$(1,\half,\half)$ & $1$ & $(\half,0,-\half)$& \\
$({3 \over 2},\half,1)$ & $1$ & $(1,0,0)$&\\
 $(2,\half,{3 \over 2})$ & $3$ & $({3 \over 2},0,\half)$&\\
\hline
$(k,0,k)$ & $3$ & $(k,0,k)$ & $k \in {Z^+ \over 2}$,$k \ge 1$\\
\hline
$(k+\half,\half,k)$ & $3$ , if $k = \text{even}$& $(k,0,k-1)$ & $k \in {Z^+ \over 2}$, $k \ge 2$\\
 & $4$ ,  if $k \ne \text{even}$ & & \\
\hline
$(2 k+1,1,2 k )$ & $2$ & $(2 k + \half,\half,2 k-1)$ & $k \in {Z^+}$\\
$( k+{3 \over 2},1,k + \half )$ & $1$ & $(k +1,\half,k-\half)$ & $k \in {Z^+}$ \\
\hline
$(2 k+\half,{3 \over 2},2 k -1 )$ & $1$ & $(2 k,1,2 k-2)$ & $k \in {Z^+}$\\
\hline
\end{tabular}
\caption{Supersymmetric spectrum for ${\cal N}=2$ nongeneric deformation with $\lambda_1=0$}
\label{tab:N2spctrm0eig}
\end{table}

Note that in each case the supersymmetric spectrum has no states with 
spins greater than two, suggesting again the possibility of a 
dual supergravity description for these theories at strong coupling.

\subsection{Theories with three or more chiral multiplets}

In this case the letter partition function equals unity at a value of 
$x<1$. It follows that the Witten index undergoes a Hagedorn transition at 
finite `temperature'. In other words the number of supersymmetric operators 
protected by susy grows exponentially with energy in these theories. 
Restated, our system has a stringy growth in its degrees of freedom; the 
effective string scale is the $AdS$ scale (unity in our units). 
It is clearly impossible for such theories to have a gravitational 
description (in any dimension).

Note that in these theories the index undergoes a phase transition at a finite
value of the chemical potential. In the `high temperature' (more accurately
small $x$) phase the logarithm of the index is of the order $N^2$. It seems possible that
this index captures the entropy of supersymmetric black holes in the as yet mysterious
bulk dual of these theories.

\section{Superconformal index at finite $k$ and $N$ from localization}\label{localization}

In the 't Hooft limit where $\lambda=N/k$ can be treated as a continuous parameter, the superconformal index is expected to be independent of $\lambda$, apart from possible renormalization of $R$-charge, as discussed in previous sections. There, the index was computed from the free limit of the theory. This is no longer the case at finite $k$, as $k$ is quantized and generally the index can jump as $k$ varies. Nonetheless, the exact superconformal index at finite $k$ and $N$ can be computed using the powerful technique of supersymmetric localization \cite{Imamura:2011su, Pestun:2007rz, Kim:2009wb}. As a special case of a more general formula derived in \cite{Imamura:2011su}, the superconformal index of ${\cal N}=2$ $U(N)$ CS theory with $g$ adjoint chiral multiplets is given by the following expression involving a sum over magnetic flux sectors on the $S^2$ and integration over the maximal torus of $U(N)$,
\ie
& {\cal I}_+(x,y) = {\rm Tr} \left[ (-)^F e^{-\beta (H-J-R)} x^{H+J} \prod_{I=1}^g y_I^{G_I} \right]
\\
&= \sum_{s_i\in {\mathbb Z}} x^{\epsilon_0(s)}\prod_{I=1}^G y_I^{q(s) G_I} {1\over N!} \prod_{i=1}^N \int_0^{2\pi}{da_i\over 2\pi} \, e^{-S_{CS}^{(0)}(s,a) }  \exp\left[
\sum_{m=1}^\infty {1\over m} f_{tot}(e^{ima},x^m,y_I^m) \right]
\\
& = \sum_{s_i\in {\mathbb Z}} \left[x^{g(1-h)-1}\prod_{I=1}^G y_I^{-G_I}\right]^{\sum_{i<j} |s_{ij}|} {1\over N!}\prod_{i=1}^N \int_0^{2\pi}{da_i\over 2\pi} \, e^{-ik\sum_i s_i a_i }  \exp\left[
\sum_{m=1}^\infty {1\over m} f_{tot}(e^{ima},x^m,y_I^m;s) \right]
\fe
where $S_{CS}^{(0)}(s,a)$ is the contribution from the supersymmetric CS action evaluated on the localized solutions, $\epsilon_0(s)$ and $q(s)G_I$ are the zero-point energy and global symmetry charges in the magnetic flux sector $s$. Their explicit expressions are given by
\ie
& S_{CS}^{(0)}(s,a) = {i k} \sum_{i=1}^N s_i a_i ,\\
& \epsilon_0(s) = \left[{g\over 2}(1-h) -{1\over 2}\right] \sum_{1\leq i,j\leq N} |s_i-s_j|, \\
& q(s) = - {1\over 2}\sum_{1\leq i,j\leq N} |s_i-s_j|.
\fe
The exponential involving $f_{tot}(e^{ia},x,y_I)$ is the $1$-loop determinant from the vector multiplet and the chiral multiplets. They are given by
\ie
& f_{tot}(e^{ia}, x, y_I;s) = f_{V}(e^{ia}, x;s) + g f_{\Phi}(e^{ia}, x, y_I;s),\\
& f_{V}(e^{ia}, x;s) = - \sum_{1\leq i\not=j\leq N} e^{i(a_i-a_j)} x^{|s_i-s_j|}, \\
& f_\Phi(e^{ia}, x, y_I;s) =  \sum_{1\leq i,j\leq N}e^{i(a_i-a_j)} x^{|s_i-s_j|}  \left[ {x^{h} \over 1-x^2}\prod_I y_I^{G_I} - {x^{2-h}\over 1-x^2} \prod_I y_I^{-G_I} \right]
\fe
In the large $k$ limit, the contribution to the index from operators with finite dimension (namely dimension that does not scale with $k$) comes from the $s=0$ sector only. This part of the index is given by the integral formula
\ie
{\cal I}^{s=0}(x,y) & ={1\over N!} \prod_{i=1}^N \int_0^{2\pi}{da_i\over 2\pi}  \exp\left[
\sum_{m=1}^\infty {1\over m} f_{tot}(e^{ima},x^m,y_I^m;0) \right]
\\
&= {2^{N(N-1)}\over N!} \prod_{i=1}^N \int_0^{2\pi}{da_i\over 2\pi} \left[ \prod_{i<j}\sin^2({a_{ij}\over 2})\right]
\exp\left[
g\sum_{m=1}^\infty {1\over m} f_{\Phi}(e^{ima},x^m,y_I^m;0) \right]
\fe
where
\ie
& f_\Phi(e^{ia}, x, y_I;0) =  \sum_{1\leq i,j\leq N}e^{ia_{ij}}  \left[ {x^{h} \over 1-x^2}\prod_I y_I^{G_I} - {x^{2-h}\over 1-x^2} \prod_I y_I^{-G_I} \right].
\fe
Let us define $z\equiv x^{h-1}\prod_I y_I^{G_I}$, then the index in the large $k$ limit depends on $x$ and $z$ only. We can write ${\cal I}^{s=0}(x,y)$ as
\ie
\widetilde {\cal I}(x|z) &= {2^{N(N-1)}\over N!}  \prod_{i=1}^N \int_0^{2\pi}{da_i\over 2\pi} \left[ \prod_{i<j}\sin^2({a_{ij}\over 2})\right]
\exp\left[
g\sum_{m=1}^\infty {1\over m} \sum_{i,j} e^{ima_{ij}} {x^m\over 1-x^{2m}} (z^m-z^{-m}) \right]
\\
&= {2^{N(N-1)}\over N!} \prod_{i=1}^N \int_0^{2\pi}{da_i\over 2\pi} \left[ \prod_{i<j}\sin^2({a_{ij}\over 2})\right]
\prod_{n=0}^\infty \prod_{i,j} \left(
{1-z^{-1} x^{2n+1} e^{ia_{ij}}\over 1-z \,x^{2n+1} e^{ia_{ij}}} \right)^g
\fe
One can verify that this indeed agrees with the index derived from the free limit of ${\cal N}=2$ $U(N)$ CS theory with $g$ adjoint chiral multiplets.

Let us focus on the $g=1$ example, and set the flavor charge $G=1/2$. In the theory with no superpotential, $W=0$, $h=h(\lambda)$ is the renormalized $R$-charge given by ${\cal Z}$-minimization when there are no accidental global symmetries. In the case $W=\alpha {\rm Tr}\Phi^4$, $h=1/2$, whereas in the case $W={\rm Tr}\Phi^3$ the renormalized $R$-charge is $h=2/3$. Let ${\cal I}_{free}(x,y)$ be the index of the $k=\infty$ $W=0$ theory. Then by writing ${\cal I}_{free}(x,y) = \widetilde {\cal I}_{free}(x| x^{-1} y)$, we can relate the superconformal indices of the three theories in the 't Hooft limit to ${\cal I}_{free}(x,y)$ as
\ie \label{relation}
& {\cal I}_{W=0}(x,y) ={\cal I}_{free}(x, x^{2h(\lambda)-1}y), \\
&{\cal I}_{W= {\rm Tr}\Phi^4}(x) = {\cal I}_{free}(x, 1), \\
& {\cal I}_{W= {\rm Tr}\Phi^3}(x) ={\cal I}_{free}(x, x^{{1\over 3}}).
\fe
where
\ie
{\cal I}_{free}(x,y) = \prod_{n=1}^\infty {1-x^{2n}\over (1-x^{n\over 2} y^{n\over 2}) (1+x^{3n\over 2} y^{-{n\over 2}})}.
\fe
In particular, we have
\ie
{\cal I}_{W= {\rm Tr}\Phi^3}(x) =  \prod_{n=1}^\infty {1-x^{2n}\over (1-x^{2n\over 3} ) (1+x^{4n\over 3} )}.
\fe
We can rewrite it in terms of single trace conformal primary contribution ${\cal I}^{st}_{W= {\rm Tr}\Phi^3}(x)$, through
\ie
{\cal I}_{W= {\rm Tr}\Phi^3}(x) = \exp\left[ \sum_{n=1}^\infty {{\cal I}^{st}_{W= {\rm Tr}\Phi^3}(x^n)\over n (1-x^{2n})} \right]
\fe
The explicit expression for ${\cal I}^{st}$ is
\ie \label{eqnindexA}
{\cal I}^{st}_{W= {\rm Tr}\Phi^3}(x) &= {1+x^{2\over 3}\over 2(1+x^{4\over 3})} - {1\over 2(1+x^{2\over 3})} + x^{4\over 3} 
\\
&=x^{2/3}+x^{10/3}-x^4+x^6-x^{20/3}+x^{26/3}-x^{28/3}+x^{34/3}-x^{12}+x^{14}+\cdots
\fe
The above equation is the same as the index for the free theory \eqref{compone} upon setting $y\rightarrow  x^{1/3}$.
The index is precisely reproduced by the conjectured BPS spectrum in the theory with ${\rm Tr}\,\Phi^3$ superpotential listed in Table (\ref{tab:table2adjsupphi3}).

As argued earlier, the ${\cal N}=2$ $g=1$ theory with no superpotential in the 't Hooft limit has a renormalized $R$-charge $h(\lambda)$ that approaches $1/4$ where ${\rm Tr}\, \Phi^2$ becomes a free field and decouples. Near this point, ${\rm Tr}\, \Phi^m$ for $m\leq 7$ are relative superpotential deformations, each of which gives rise to a strongly coupled critical point. More precisely, we have a fixed line parameterized by (a range of) $\lambda$ for each superpotential deformation $W={\rm Tr}\, \Phi^m$. At such a critical point, the $R$-charge is renormalized to $h = 2/m$. Its superconformal index is then given by
\ie
{\cal I}_m(x) = {\cal I}_{free}(x, x^{{4\over m}-1}) = \prod_{n=1}^\infty {1-x^{2n}\over (1-x^{2n\over m}) (1+x^{2n(1-{1\over m})})}.
\fe

\section{Discussion}
So, what did we learn about the gravity dual of these large $N$ Chern-Simons-matter theories?

Perhaps the ``nicest" theories we studied are the ${\cal N}=3$ theory with one adjoint hypermultiplet
and the ${\cal N}=2$ superpotential deformed theories with two adjoint chiral multiplets and with $U(1)$ or no flavor symmetry. We found that their supersymmetric spectrum consists of only operators of spin $\leq 2$, suggesting a possible supergravity dual in the strong coupling limit. In the ${\cal N}=3$ case, while part of the supersymmetric spectrum looks like the Kaluza-Klein spectrum of $7$-dimensional supergravity compactified on $S^3$, there is an additional tower of states in spectrum that do not seem to come from standard KK modes. In the ${\cal N}=2$ deformed theories, the spectrum contains states of arbitrarily high $U(1)$ charges, suggesting that they could come from KK modes of $S^1$-compactification of supergravity theories, but to identify their duals appears difficult due to some unusual features of the spectrum.

The ${\cal N}=2$ theories with one adjoint chiral multiplet are even more intriguing. With either $\Tr\,\Phi^4$ or $\Tr\,\Phi^3$ superpotential, there is a line of fixed points. At these fixed point theories, in the large $N$ limit, the supersymmetric spectrum involves a single tower of operators/states of arbitrarily high spin as well as $R$-charge. This rules out the possibility of a supergravity dual, but leaves open the possibility that the duals of the strongly coupled SCFTs are higher spin theories of gravity in $AdS_4$. 

The most mysterious case is the ${\cal N}=2$ theory with one adjoint chiral multiplet and no superpotential. The $R$-charge of this theory is renormalized and decreases monotonically with the 't Hooft coupling $\lambda$. At some point, when $\lambda=\lambda_2^f\approx 1.23$, the operator ${\rm Tr}\, \Phi^2$ becomes a free field and decouples from the theory. At this point, a new $U(1)$ global symmetry emerges and in principle the ${\cal Z}$-minimization prescription no longer determines the superconformal $R$-charge. If we assume that the naive ${\cal Z}$-minimization is still valid at large $N$ for $\lambda>\lambda_2^f$, then we find that the renormalized $R$-charge approaches zero asymptotically at strong coupling. If this is true, apart from the decoupled free fields, the BPS spectrum involves a discretum of states starting at dimension $\Delta=1/2$. While at general $\lambda$ the BPS spectrum consists of towers of states of arbitrarily high spin and $R$-charge, the $R$-charge form a discretum at strong coupling, suggesting that a new noncompact dimension emerges in the higher spin gravity dual.

Finally, in the cases with more than two adjoint flavours, the number of supersymmetric states grow exponentially with the dimension. It suggests that their dual theories are string theories in $AdS_4$ with an exponentially growing tower of {\it supersymmetric} string oscillator excitations. The superconformal index of these theories as a functional of the chemical potential undergoes a phase transition. After this phase transition, these theories are likely dual to supersymmetric black holes in the yet to be determined dual string theories in $AdS_4$.

Let us comment briefly on brane constructions for the ${\cal N}=2$ $U(N)$ Chern-Simons theory coupled to one adjoint chiral matter with no superpotential.\footnote{We thank Ofer Aharony and Daniel Jafferis for discussions on this point.} This theory can be embedded in type IIB string theory by suspending $N$ D3-branes between an NS5-brane and a $(1,k)$ $5$-brane. One takes the NS5-brane to extend in $012456$ directions, and take the $(1,k)$ $5$-brane to extend in $01245$ directions and at an angle in the $6-9$ plane in order to preserve ${\cal N}=2$ supersymmetry. The D3-branes extend in $0123$ directions, and are free to move in the $4-5$ plane.

To connect this brane configuration to the more familiar setup of \cite{Hanany-Witten}, one should deform it by rotating the $(1,k)$ $5$-brane in the $4-7$, $5-8$ and $6-9$ planes in such a way that the ${\cal N}=2$ supersymmetry is preserved; a superpotential mass term is then generated for the adjoint chiral matter multiplet. The $s$-rule \cite{Hanany-Witten} would indicate that supersymmetry is spontaneously broken if $N>k$, i.e. $\lambda>1$. It has been observed in \cite{Jafferis:2011ns} that this is consistent with supersymmetry being preserved by the undeformed theory.

Alternatively, the ${\cal N}=2$, $W=0$ theory may also be embedded as the world volume theory of $N$ M5-branes wrapped on a special Lagrangian lens space $S^3/\mathbb{Z}_k$ in a Calabi-Yau $3$-fold \cite{Gaiotto:2007qi}. The M5-brane extends in an $\mathbb{R}^{1,2}$ in the $\mathbb{R}^{1,4}$. It has been noted in \cite{Gaiotto:2007qi} that, however, finding the gravity dual by taking the decoupling limit from this brane construction is difficult.

\acknowledgments

We are especially grateful to J. Bhattacharya for collaboration during the initial stages of this project.
We also thank S. Trivedi for insights and comments.
We would like to acknowledge useful discussions and correspondences with 
D. Jafferis, Z. Komargodski, A. Mukherjee and S. Wadia. We thank H. M. Antia and S. Datta for advice on the numerics. We wish to thank Ofer Aharony for
comments on the preliminary draft of this work.
S.M. would like to acknowledge the hospitality of IUCAA,
Indian Strings Meeting 2011 
and ICTP Trieste. X.Y. would like to thank the organizers of Indian Strings Meeting 2011, Tata Institute of Fundamental Research, Berkeley Center for Theoretical Physics, and Simons Center for Geometry and Physics, for their hospitality. The work 
of S.M. was supported in part by a Swarnajayanti Fellowship. The work of X.Y. was supported by 
the Fundamental Laws Initiative Fund at Harvard University and in part by NSF Award PHY-0847457.
S.M., P.N., T.S. and U.V. 
would also like to acknowledge our debt to the people of India for 
their generous and steady support to research in the basic sciences.

\appendix
\appendixpage
\addappheadtotoc

\section{Details of numerics and plots} \label{numerics}

In this appendix we briefly describe the numerical technique we have 
used to determine the $R$-charge,  $h(\lambda)$, of the 
chiral multiplets in the theory with $g$ chiral multiplets and no 
superpotential. 

As we have explained above, the function $h(\lambda)$ is 
determined by the solution of the equations \eqref{eq:saddlept} and 
\eqref{eq:deltaeqn}. The basic idea is to determine $h(\lambda)$
by solving those equations numerically. This procedure has two possible 
pitfalls

\begin{itemize}
\item The equations \eqref{eq:saddlept} and \eqref{eq:deltaeqn}
correctly determine $h(\lambda)$ only in the large $N$ limit. Numerically, 
however, it is feasible to solve these equations only at finite $N$. 
It is important to check that our results do not change substantially 
upon increasing $N$.
\item The equations \eqref{eq:saddlept} and \eqref{eq:deltaeqn}
could admit multiple solutions; we need to help the numerical solving 
procedure to focus on the correct solution. We achieved this as follows. 
At small $\lambda$ we used as an input guess the results of our perturbative 
computation as our initial guess for the equation solving technique, and 
then increased $\lambda$ in small steps. At every subsequent step we used 
the result of the previous step as our input guess. We ensure, by this 
procedure, that we always zoom into the correct saddle point, atleast in 
a finite neighbourhood of $\lambda=0$. It is of course possible that 
Jafferis' matrix integral (and hence the field theory) 
undergoes a large $N$ phase transition at finite $\lambda=\lambda_c$. 
If this indeed does happen then all results of this paper are valid 
only for $\lambda<\lambda_c$. We leave the investigation of possible 
phase transitions in this path integral to future work. 
\end{itemize}

In actual practice we found it easier to solve \eqref{eq:saddlept} but 
not \eqref{eq:deltaeqn} numerically. This solution determines $|{\cal Z}|$ as 
a function of $h(\lambda)$. We actually proceeded to evaluate $|{\cal Z}|$ 
for 60 closely spaced trial values of $h(\lambda)$ and then estimated $h(\lambda)$ by 
the value (of our 60 trial point) that minimizes $|{\cal Z}|$, thus effectively 
solving \eqref{eq:deltaeqn}. We performed all our numerics using Mathematica).

\begin{figure}[h]
\begin{center}
\includegraphics[width=100mm]{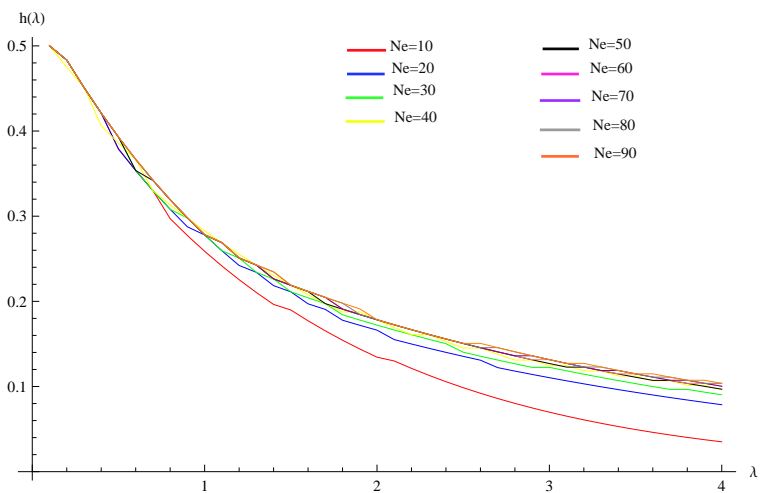}
\end{center}
\caption{Numeric plot of $h$ vs. $\lambda$ for $\lambda$ upto 4, at $g=1$.}\label{fig:numeric_res}
\end{figure}

In the rest of this appendix we will present evidence that the results of 
our numerical routine are reliable. To start with, in Fig. \ref{fig:numeric_res}
we present a plot of $h(\lambda)$ versus $\lambda$, obtained from our numerical
routine, with $Ne=(10,20,30,...,100)$ in the range $\lambda \in (0, 4)$. 
As is apparent from the Fig \ref{fig:numeric_res}, the result changes 
substantially
from $Ne=10$ (the lowest graph) to $Ne=20$ (the second lowest graph), but 
appears to converge to a limit curve for $Ne\geq 30$ or so. The lesson of 
this exercise is that numerics with $Ne \geq 30$ are rather reliable for  $\lambda<4$. 

In order to understand the convergence of $h(\lambda)$ as $N$ is taken to $\infty$, we present a plot
of $h(\lambda)$ vs. $Ne$ at $\lambda = 4.0$ and we also best fit our data to
$$h = a + \frac{b}{c+N^2}.$$
Note that the best fit seems to agree rather well with the data indicating that the error in the
$N \rightarrow \infty$ limit scales as $1/N^2$. As we have explained in the appendix, we have performed 
a similar best fit of our data (as a function of $N$) for all values of $\lambda$, and have  
used this best fit value to generate the curves presented in Section \ref{smallglargelmda}. This bestfitting 
procedure appears to work rather well for every $\lambda \in (1, 10)$.

\begin{figure}[h]
\begin{center}
\includegraphics[width=100mm]{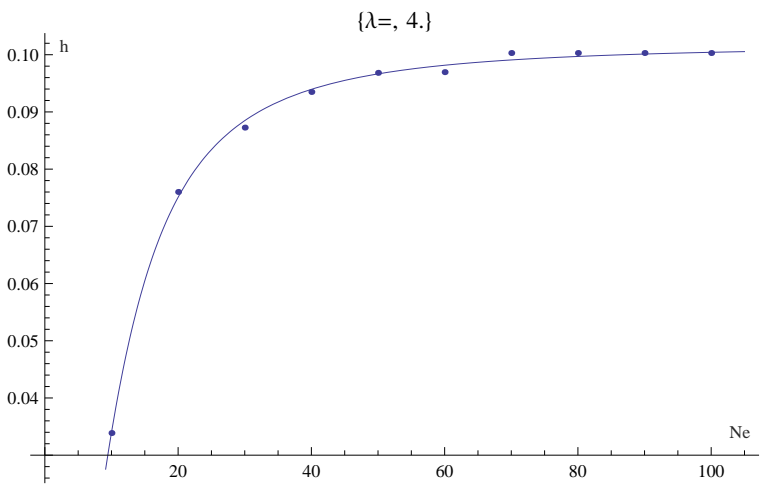}
\end{center}
\caption{$h(\lambda)$ vs. $Ne$ at $\lambda=4.0$. Data best fit to $a + b/(c+N^2)$, $a$, $b$ and $c$ were found to be $0.102518$, -$16.9934$
and $239.4509$ respectively.
Note that the fit seems rather good.}\label{fig:hvsNelmdasmall}
\end{figure}

\begin{figure}[h]
\begin{center}
\includegraphics[width=100mm]{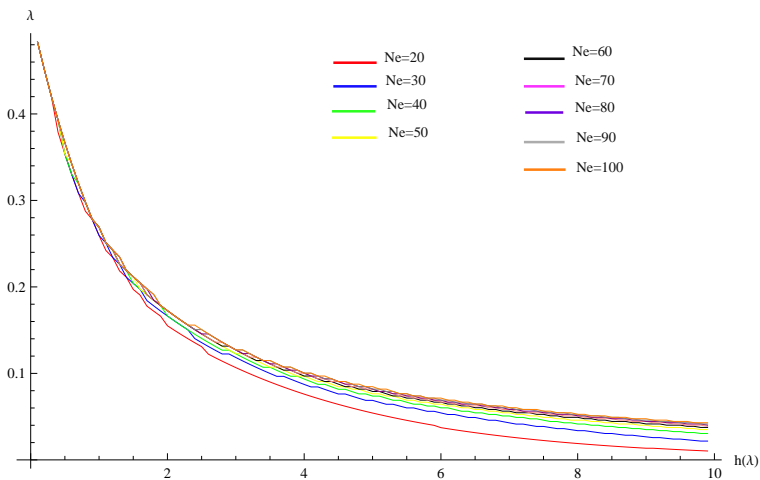}
\end{center}
\caption{Numeric plot of $h$ vs. $\lambda$ for $\lambda$ upto 10, at $g=1$,with different $Ne$.}\label{fig:largelmdadiffN}
\end{figure}

\begin{figure}[h]
\begin{center}
\includegraphics[width=100mm]{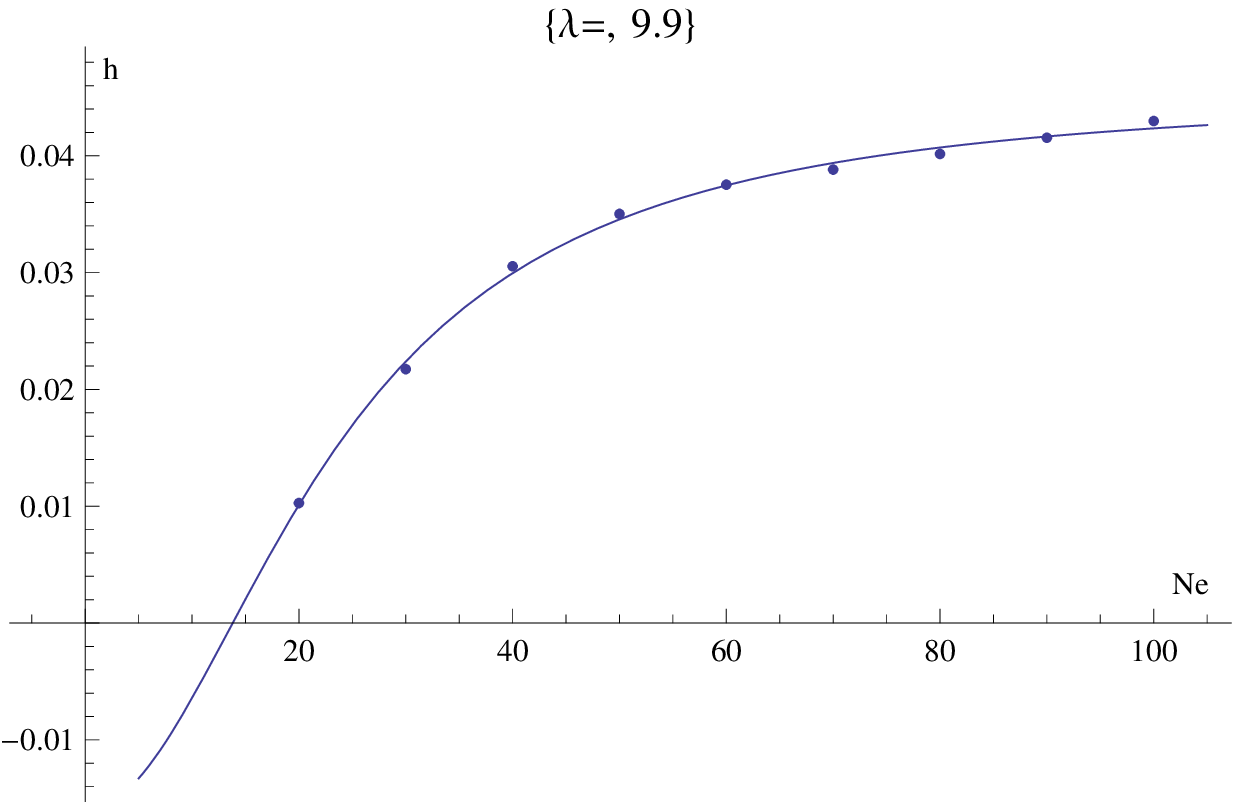}
\end{center}
\caption{$h(\lambda)$ vs. $Ne$ at $\lambda=9.9$. Data best fit to $a + b/(c+N^2)$, $a$, $b$ and $c$ were found to be $0.0455$, -$33.3054$
and $541.42$ respectively.
Note again that the fit seems rather good.}
\label{fig:hvsNelmdalarge}
\end{figure}

It is important that generating accurate results at large $\lambda$ requires larger values of $Ne$. This is seen in
Fig.\ref{fig:largelmdadiffN} where $h(\lambda)$ is plotted against $\lambda$ for $Ne=20,30,...100$. 
Compairing Fig(\ref{fig:hvsNelmdasmall}) and Fig(\ref{fig:hvsNelmdalarge}) that while for $\lambda \sim 4$ Ne=30 seems good enough, 
for $\lambda \sim 10$ one has to go to $Ne \ge 60$ for results reliable upto a few percent of accuracy. This is also apparent from 
the curves displayed in Fig. \ref{fig:numeric_res} and \ref{fig:largelmdadiffN}.

\begin{figure}[h]
\begin{center}
\includegraphics[width=100mm]{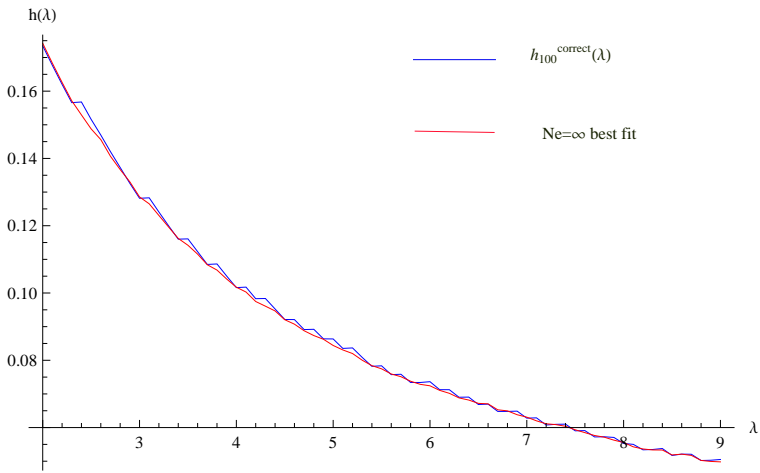}
\end{center}
\caption{$h_{100}^{corr}(\lambda)$ data(Blue curve) and $h_{\infty}(\lambda)$(red curve) best fit as a function of $\lambda$}\label{fig:correctedN100}
\end{figure}

\begin{figure}[h]
\begin{center}
\includegraphics[width=100mm]{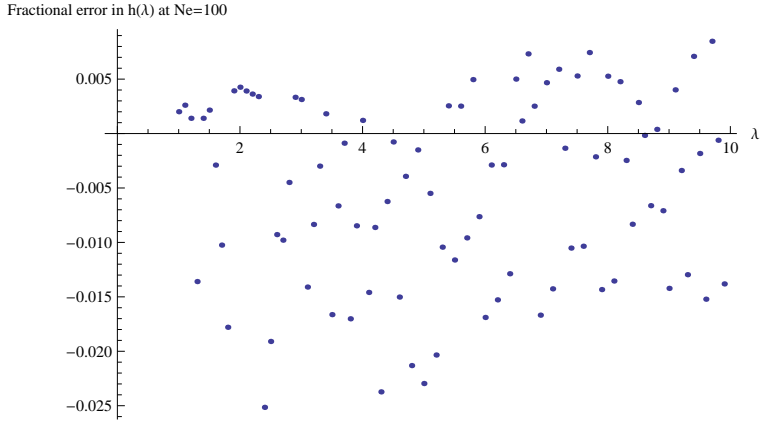}
\end{center}
\caption{Fractional error in $Ne=100$ data as a function of $\lambda$}\label{fig:N100error}
\end{figure}

Though we have not performed a serious estimate of errors in these calculations, we can crudely estimate the errors in our procedure as follows. 
Let us define
\begin{equation} \label{corrected} 
 h_{Ne}^{corr}(\lambda)=
 h_{Ne}(\lambda)-\frac{a(\lambda)}{c^2(\lambda)+100^2}
\end{equation}
Here $h_{ne}(\lambda)$ is the raw data for $h$ obtained from a numerical run with $N_e$ eigenvalues, and the subtraction represents the best fit correction for 
finite $N$ effects. $h^{corr}_{Ne}(\lambda)$ differs from $h(\lambda)$, the value of $h$ obtained from bestfitting our results at $Ne=20 \ldots 100$. 
The difference between $h_{100}^{corr}(\lambda)$ and $h_{\lambda}$ may be taken as a crude estimate of the errors in our results. In Fig \ref{fig:correctedN100} we  present a plot of 
these two functions versus $\lambda$. Note that they agree very closely for $\lambda \in (1, 10)$. More quantitatively, in Fig. \ref{fig:N100error} 
we have plotted the fractional error. 
$$\frac{h(\lambda)-h^{corr}_{100}(\lambda)}{h(\lambda)}.$$
Note that all errors lie within three percent. This is the basis of our belief that our results for $h(\lambda)$ are accurate to within a few percent. 

\begin{figure}[h]
\begin{center}
\includegraphics[width=100mm]{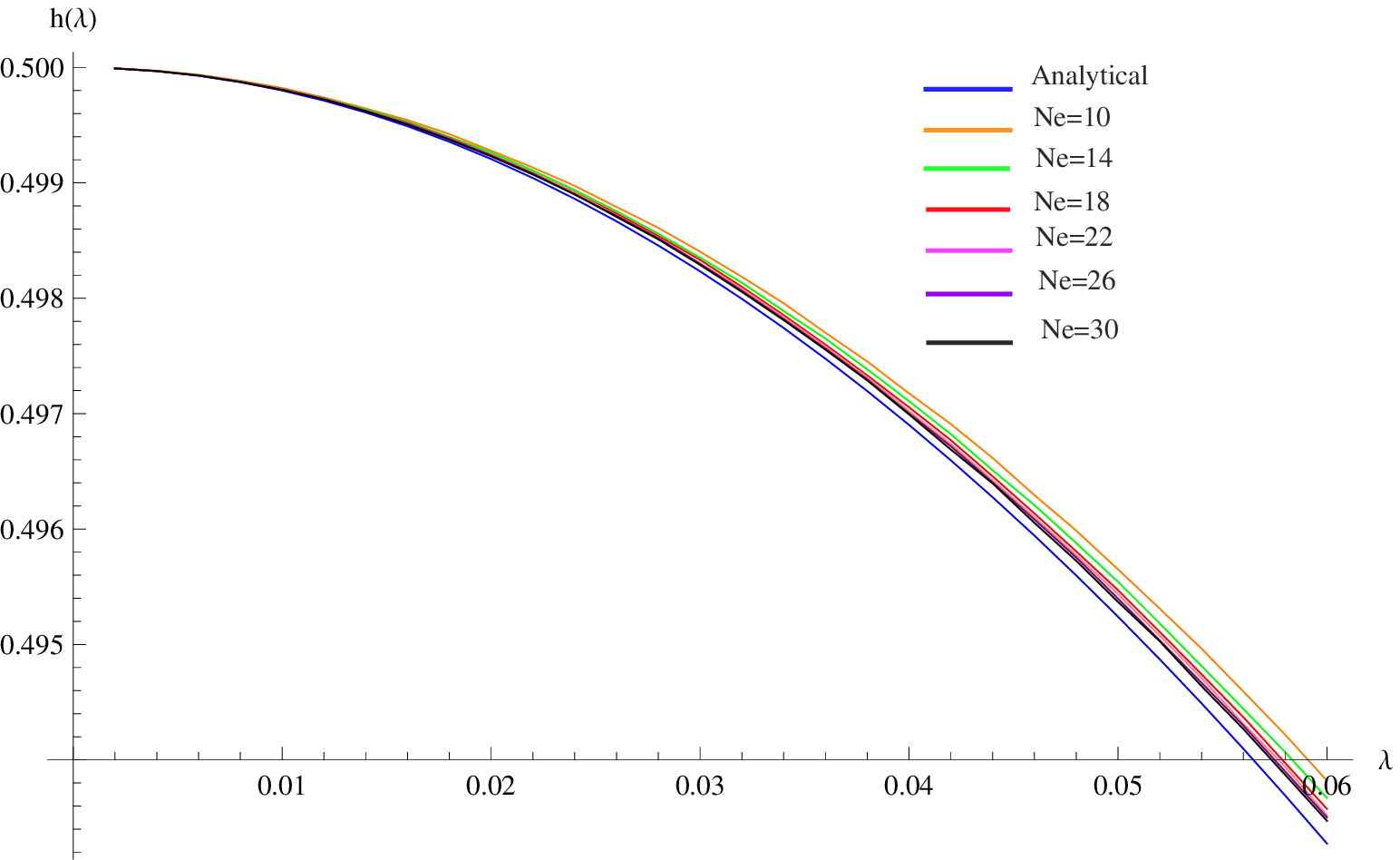}
\end{center}
\caption{Comparison of perturbative result and numeric results (different $Ne$) for small $\lambda$.}\label{fig:compare}
\end{figure}

As another comparison of our numerical results versus those of perturbation 
theory, in Fig. \ref{fig:compare} we have plotted $h(\lambda)$ (both numerical
and perturbative) against $\lambda$ at small $\lambda$. We have plotted 
our numerical results at $Ne=10, 14, 18, 22, 26, 30$. As is apparent from the 
graph, the numerical results converge towards the perturbative values 
at large $Ne$; the agreement with perturbation theory is already rather 
good at $Ne=20$. 

\begin{figure}[h]
\begin{center}
\includegraphics[width=100mm]{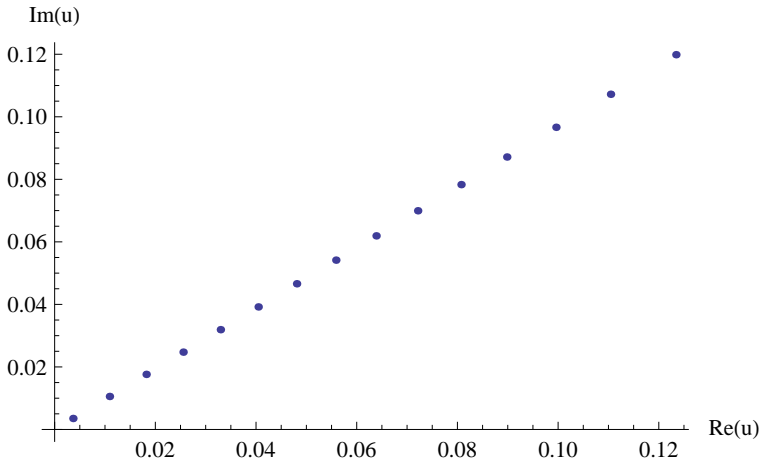}
\end{center}
\caption{Scatter plot (on the complex plane) of the eigenvalue 
distribution obtained at $Ne=30$ and $\lambda = 0.06$.}\label{fig:numeric_distro}
\end{figure}

As a more sensitive test of our numerics we next compare the {\it eigenvalue distributions}
obtained from perturbation theory to those obtained from our numerics.  A scatter plot  
of the eigenvalues (on the complex plane) 
generated by the numerics for $Ne=30$ at $\lambda = 0.06$ 
is presented in Fig \ref{fig:numeric_distro}. As is visually apparent, 
the eigenvalues lie in a straight line. The angle of this cut turns out, 
numerically, to be  $\frac{\pi}{4}-0.014$ radians and its magnitude (crudely estimated by 
the distance of the largest eigenvalue from the origin plus a rough correction \footnote{We estimate the correction 
as follows. Given 30 eigenvalues distributed according to the Wigner distribution. The last eigenvalue 
in such a distribution will not be located at $x=a$ by instead, most probably, at $x=a-\frac{y}{2}$, 
where $y$ is the solution to the equation 
$$\int_y^a \rho(y)=\frac{1}{30}.$$
In the situation at hand the distance of the largest eigenvalue form the origin was approximately $0.17$ while 
we estimated the shift by } is numerically given by approximately $0.17$ (this value is obtained by fitting the observed 
eigenvalue density function to the Weigner form). This compares reasonably well with the 
perturbative prediction of the angle of the line 
($\frac{\pi}{4}-0.016$) and magnitude ($\sqrt{\frac{2}{\pi}\lambda}=0.19$). 
The crude comparison reported above can be improved by bestfitting 
the results from various different value of $Ne$; we will not pause to do so here. 

As a final check on our numerics, we have used our numerical routine to 
compute $h(\infty)$ at $g=10$ with $Ne=20$. Numerically we found 
$h=0.458$. This compares rather well with the prediction of our large 
$g$ perturbative expansion, $h=0.46$.


\section{Cohomology calculation}
In this appendix we explain the calculation of $Q$ cohomology in more detail.
For definiteness, consider the calculation of $Q$ cohomology in the theory with a
$U(N)$ gauge group with a single chiral adjoint matter with a superpotential $\Tr\, \Phi^4$ as given in section \ref{superphi4}.
Since $Q$ carries quantum numbers $(\Delta,j,h)=({1 \over 2},-{1 \over 2},1)$, action of $Q$ does not change
the value of $\Delta + j$. Hence cohomology can be calculated independently for each $\Delta + j$ sector.
For a fixed $\Delta + j$, operators can be arranged into ``levels''. The level of an operator is just twice the angular momentum.
Level $0$ operators are made only of $\phi$ and are of the form $\Tr\, (\phi^n)$. Level $1$ operators are of the form $\Tr\, (\phi^n \bar{\psi})$ and so on.

By using Mathematica one can construct all the states at a given $\Delta +j$ and level. Also, by using Mathematica a state at level $k+1$
say $|\xi\rangle$ can be acted upon by $Q$ and decomposed into a linear combination of states at level $k$. Thus a matrix $Q_{k+1,k}$ can 
be constructed whose rows correspond to level $k+1$ operators and columns correspond to operators at level $k$. Then the number of states
in $Q$ cohomology at level $k+1$ will be naively given by 
$N_1(k+1,\Delta+j) = (\# \text{ of states at level }k+1) -{\rm rank}(Q_{k+1,k})-{\rm rank}(Q_{k+2,k+1})$. But to remove the conformal descendant operators, i.e 
operators $\xi$ which are of the form $D_{++} |\xi'\rangle = |\xi\rangle$, one should further subtract by all allowed $|\xi'\rangle$ states. Therefore the total
number of states in cohomology at level $k+1$, at a given $\Delta+j$ is $N_\text{cohomology}(k+1,\Delta+j) = N_1(k+1,\Delta+j) -  N_1(k-1,\Delta+j-2)$. This is because, in this theory $D$
increases the level by $2$ and has  $\Delta = 1,j=1$. 

\bibliographystyle{JHEP}
\bibliography{CSreference}

\end{document}